\documentclass[12pt]{article}

\usepackage{amsmath}
\usepackage{appendix,epic,eepic,amsmath,amsthm,amssymb,epsfig,cite,indentfirst,oldgerm}
\usepackage{multicol,boxedminipage}

\newcommand{\EQ}{\begin{equation}}
\newcommand{\EN}{\end{equation}}

\newcommand{\bear}{\begin{eqnarray}}
\newcommand{\ear}{\end{eqnarray}}
\newcommand{\bt} { \begin{tabular} }
\newcommand{\et}{ \end{tabular} }
\newcommand{\bc} { \begin{center} }
\newcommand{\ec}{ \end{center} }

\newcommand{\btb} { \begin{table} }
\newcommand{\etb}{ \end{table} }

\begin{document}

\topmargin 0pt
\oddsidemargin 5mm
\newcommand{\NP}[1]{Nucl.\ Phys.\ {\bf #1}}
\newcommand{\PL}[1]{Phys.\ Lett.\ {\bf #1}}
\newcommand{\NC}[1]{Nuovo Cimento {\bf #1}}
\newcommand{\CMP}[1]{Comm.\ Math.\ Phys.\ {\bf #1}}
\newcommand{\PR}[1]{Phys.\ Rev.\ {\bf #1}}
\newcommand{\PRL}[1]{Phys.\ Rev.\ Lett.\ {\bf #1}}
\newcommand{\MPL}[1]{Mod.\ Phys.\ Lett.\ {\bf #1}}
\newcommand{\JETP}[1]{Sov.\ Phys.\ JETP {\bf #1}}
\newcommand{\TMP}[1]{Teor.\ Mat.\ Fiz.\ {\bf #1}}

\renewcommand{\thefootnote}{\fnsymbol{footnote}}

\newpage
\setcounter{page}{0}
\begin{titlepage}
\begin{flushright}

\end{flushright}
\vspace{0.5cm}
\begin{center}
{\large The monodromy matrix in the F-basis for} \\
{\large arbitrary six-vertex models }\\
\vspace{1cm}
{\large M. J. Martins and M. Zuparic} \\
\vspace{0.15cm}
{\em Universidade Federal de S\~ao Carlos\\
Departamento de F\'{\i}sica \\
C.P. 676, 13565-905, S\~ao Carlos(SP), Brazil\\
E-mail Address: {\tt martins, zuparic@df.ufscar.br}}\\
\vspace{0.35cm}
\end{center}
\vspace{0.5cm}

\begin{abstract}
We present the expressions for the monodromy matrix elements of the six-vertex model in the F-basis for arbitrary Boltzmann weights. The results rely solely on the property of unitarity and Yang-Baxter relations, avoiding any specific parameterization of the weights. This allows us to write complete algebraic expressions for the inner products and the underlying domain wall partition functions in the case of arbitrary rapidities. We then apply our results for the trigonometric six-vertex model in the presence of inhomogeneous electric fields and obtain a determinant formula for the respective on-shell scalar products.

\end{abstract}

\vspace{.15cm} \centerline{}
\vspace{.1cm} \centerline{Keywords: Six-Vertex Model, F-Basis, Monodromy Matrix}
\vspace{.15cm} \centerline{June 2011}

\end{titlepage}


\pagestyle{empty}

\newpage

\pagestyle{plain}
\pagenumbering{arabic}

\renewcommand{\thefootnote}{\arabic{footnote}}
\newtheorem{proposition}{Proposition}
\newtheorem{pr}{Proposition}
\newtheorem{remark}{Remark}
\newtheorem{re}{Remark}
\newtheorem{theorem}{Theorem}
\newtheorem{theo}{Theorem}

\def\ll{\left\lgroup}
\def\rr{\right\rgroup}

\newtheorem{Theorem}{Theorem}[section]
\newtheorem{Corollary}[Theorem]{Corollary}
\newtheorem{Proposition}[Theorem]{Proposition}
\newtheorem{Conjecture}[Theorem]{Conjecture}
\newtheorem{Lemma}[Theorem]{Lemma}
\newtheorem{Example}[Theorem]{Example}
\newtheorem{Note}[Theorem]{Note}
\newtheorem{Definition}[Theorem]{Definition}

\section{Introduction}
The quantum inverse scattering method is a powerful tool used to solve integrable two-dimensional models from a unified algebraic perspective \cite{FA,KO}. One fundamental object in this approach is a set of quadratic relations called the \textit{Yang-Baxter algebra} which can be written as follows,
\begin{equation}
R_{ab}(\mu,\nu) \mathcal{T}_{a,1\dots L}(\mu) \mathcal{T}_{b, 1\dots L}(\nu) =  \mathcal{T}_{b, 1\dots L}(\nu)\mathcal{T}_{a,1\dots L}(\mu)R_{ab}(\mu,\nu),
\label{one}\end{equation}
where $\mu,\nu$ are complex parameters and $L$ is an integer related to the two-dimensional volume $L \times L$.

The monodromy operator $\mathcal{T}_{a,1\dots L}(\mu)$ is usually viewed as a matrix on a given auxiliary space $\mathcal{A}_a$ and its elements act on the standard Hilbert space, $V_1 \otimes \dots \otimes V_L$, of a quantum chain system. The respective degrees of freedom per site are then encoded in the space $V_i$. The $R$-matrix $R_{ab}(\mu,\nu)$ acts on the auxiliary tensor product space $\mathcal{A}_a \otimes \mathcal{A}_b$, and is required to satisfy the famous Yang-Baxter relation \cite{BA},
\begin{equation}
R_{12}(\lambda_1,\lambda_2)R_{13}(\lambda_1,\lambda_3)R_{23}(\lambda_2,\lambda_3) = R_{23}(\lambda_2,\lambda_3)R_{13}(\lambda_1,\lambda_3)R_{12}(\lambda_1,\lambda_2),
\label{two}\end{equation}
for the complex parameters $\lambda_i$.

The trace of the monodromy matrix on the auxiliary space becomes the generator of the commuting integrals of motion. This is seen by taking the trace of Eq. (\ref{one}) on the tensor space $\mathcal{A}_a \otimes \mathcal{A}_b$ under the requirement that the $R$-matrix is invertible. This latter property is assured for instance, when the $R$-matrix satisfies the so-called unitarity relation,
\begin{equation}
 R_{12}(\lambda_1,\lambda_2) R_{21}(\lambda_2,\lambda_1) = \mathcal{I}_1 \otimes \mathcal{I}_2,
\label{three}\end{equation}
where $\mathcal{I}_a$ denotes the identity matrix in $\mathcal{A}_a$.

The spectrum of the integrals of motion can then be formulated in terms of the action of some off-diagonal monodromy matrix elements, named creation operators, on a suitable reference state. The necessary commutation rules required to uncover the spectrum can be obtained from Eq. (\ref{one}).

Possessing the eigenstates, one can in principle compute their scalar product within this algebraic framework. This aforementioned problem has additional complications however as it involves the combinatorial analysis of commuting an arbitrary number of creation operators \cite{KO1,IK}. It was in this context that the concept of partition functions with domain wall boundary conditions was introduced in \cite{KO1}, and further elaborated in \cite{IZ,KMT}.

It has been argued that the combinatorial complexity surrounding the scalar product has been circumvented with the introduction of a new basis for the monodromy matrix elements, commonly referred to as the $F$-basis \cite{MS}. Its origin goes back to the notion of Drinfel'd twists associated with the $R$-matrix of the symmetrical six-vertex model \cite{DRI}. By means of a similarity transformation the elements of the monodromy matrix can be written in a symmetrical form with respect to an arbitrary permutation of the indices $1\dots L$. This achievement helps to simplify the combinatorial problem regarding the computation of scalar products of Bethe states of the symmetrical six-vertex model - and is also important in the formulation of correlation functions of local spin operators, see \cite{KMTS} for a review. We remark that the explicit construction of the $F$-basis has also been undertaken for other solvable vertex models. Examples include higher spin and elliptic extensions of the six-vertex model \cite{TE,BOS}, as well as systems solved by the nested Bethe ansatz \cite{BOS1,CH1,CH2}. We note that the Boltzmann weights in all the aforementioned works contain explicit parameterizations to build up the corresponding $F$-basis.

However, it is reasonable to think that the $F$-basis formulation should be independent of the $R$-matrix parameterization and rely solely on the basic algebraic relations (\ref{one}-\ref{three}). This approach has been shown to work at 
least in the context of the construction of the Bethe eigenvectors of $U(1)$ invariant 
$N$-state vertex models \cite{MM}. As far as the theory of integrable models is concerned 
it seems important to extend this strategy to tackle the scalar product of 
the respective algebraic eigenvector given its fundamental role in calculating correlation functions, 
see for instance \cite{Kita,GER1,GER2,GER3} and references therein.

The main purpose of this paper is to consider this problem for the simplest vertex system satisfying the ice-rule: the fully asymmetric six-vertex model. We shall detail how the $F$-basis of an arbitrary six-vertex model can indeed be constructed relying solely on the algebraic relations (\ref{one}-\ref{three}). Interestingly enough, exactly half of the commutation rules that come from Eq. (\ref{one}) are needed to demonstrate the main results for the simplified monodromy matrix elements. This highlights the expectation that the algebra (\ref{one}) contains all the base ingredients necessary for the exact solution of a given integrable model.

We organize this paper as follows. In section \textbf{2} we introduce the six vertex model with arbitrary Boltzmann weights and give the core algebraic relations (\ref{one}-\ref{three}) that are integral to this work. In \textbf{3} we define the $F$-basis and detail exactly how one needs to adapt past results in order to obtain a complete definition of the factorizing $F$-matrix for the arbitrary six-vertex model. In \textbf{4} we provide explicit expressions for the elements of the monodromy matrix transformed through the $F$-basis and present a complete algebraic proof based on relations (\ref{one}-\ref{three}) using arbitrary Boltzmann weights. In \textbf{5} and \textbf{6} we apply the $F$-basis and obtain complete algebraic expressions for the scalar product and domain wall partition functions of the arbitrary six-vertex model. In \textbf{7} we apply the results of this paper and consider an explicit asymmetric parameterization of the six-vertex model \cite{BA,BA2}, obtaining determinant forms for the domain wall partition functions and the scalar product. In \textbf{8} we offer some concluding remarks and in the appendices we summarize a number of technical details omitted in the main text.
\section{The arbitrary six-vertex model}
In this section we recall the definitions of the $R$-matrix of the asymmetric six-vertex model. We summarize the relations which arise from Eqs. (\ref{one}-\ref{three}) that will be used in the computations of this work. We also give necessary details on the concept of permuting $R$-matrices and required results on the symmetric group $S_L$.
\subsection{The asymmetric R-matrix, unitarity and Yang-Baxter equations} The $R$-matrix of the asymmetric six-vertex model has in general six distinct entries associated to the statistical configurations of a two state vertex model satisfying the ice-rule \cite{BA}. Because of the unitarity relation (\ref{three}) it is possible to normalize the $R$-matrix by one of its weights. We shall represent the corresponding $R$-matrix by,
\begin{equation*}
R_{12}(\lambda_1,\lambda_2) = \left( \begin{array}{cccc}
1 & 0 & 0 & 0 \\
0 & b_+(\lambda_1,\lambda_2) & c_+(\lambda_1,\lambda_2) & 0 \\
0 & c_-(\lambda_1,\lambda_2) & b_-(\lambda_1,\lambda_2) & 0 \\
0 & 0 & 0 & a_-(\lambda_1,\lambda_2)
\end{array}\right)_{12}.
\end{equation*}
Here we recall that the complex numbers $\lambda_i$ denote the set of spectral parameters necessary to parameterize the Boltzmann weights. It turns out that the unitarity condition (\ref{three}) generates the following set of relations amongst the Boltzmann weights:
\begin{eqnarray}
b_-(\lambda_{1},\lambda_{2})b_+(\lambda_{2},\lambda_{1})+c_-(\lambda_{1},\lambda_{2})c_-(\lambda_{2},\lambda_{1}) &=&1\label{uni1}\\
b_-(\lambda_{1},\lambda_{2})b_+(\lambda_{2},\lambda_{1})+c_+(\lambda_{1},\lambda_{2})c_+(\lambda_{2},\lambda_{1}) &=&1\label{uni2}\\
c_+(\lambda_{1},\lambda_{2})b_-(\lambda_{2},\lambda_{1})+b_-(\lambda_{1},\lambda_{2})c_-(\lambda_{2},\lambda_{1}) &=&0\label{uni3}\\
c_-(\lambda_{1},\lambda_{2})b_+(\lambda_{2},\lambda_{1})+b_+(\lambda_{1},\lambda_{2})c_+(\lambda_{2},\lambda_{1}) &=&0\label{uni4}\\
a_-(\lambda_{1},\lambda_{2})a_-(\lambda_{2},\lambda_{1}) &=&1\label{uni5}.
\end{eqnarray}
In addition to unitarity the $R$-matrix also satisfies the Yang-Baxter equation (\ref{two}), leading to the weights obeying thirteen separate relations. We explicitly list the following eight as we rely solely on these in this work:
\begin{eqnarray}
c_{+}(\lambda_{1},\lambda_{2})b_-(\lambda_{2},\lambda_{3})+b_-(\lambda_{1},\lambda_{2})c_{+}(\lambda_{1},\lambda_{3})c_{-}(\lambda_{2},\lambda_{3}) &=&\label{YB1} \\
c_{+}(\lambda_{1},\lambda_{2})b_-(\lambda_{1},\lambda_{3})\nonumber\\
c_{-}(\lambda_{1},\lambda_{2})b_-(\lambda_{2},\lambda_{3})+b_-(\lambda_{1},\lambda_{2})c_{-}(\lambda_{1},\lambda_{3})c_{+}(\lambda_{2},\lambda_{3}) &=&\label{YB2} \\
c_{-}(\lambda_{1},\lambda_{2})b_-(\lambda_{1},\lambda_{3})\nonumber\\
b_+(\lambda_{1},\lambda_{2})c_{-}(\lambda_{2},\lambda_{3})+c_{+}(\lambda_{1},\lambda_{2})c_{-}(\lambda_{1},\lambda_{3})b_+(\lambda_{2},\lambda_{3}) &=&\label{YB3} \\
b_+(\lambda_{1},\lambda_{3})c_{-}(\lambda_{2},\lambda_{3})\nonumber\\
b_+(\lambda_{1},\lambda_{2})c_{+}(\lambda_{1},\lambda_{3})b_-(\lambda_{2},\lambda_{3})+c_{+}(\lambda_{1},\lambda_{2})a_-(\lambda_{1},\lambda_{3}) c_{+} (\lambda_{2},\lambda_{3})&=&\label{YB4}\\
 a_-(\lambda_{1},\lambda_{2}) c_{+}(\lambda_{1},\lambda_{3})a_-(\lambda_{2},\lambda_{3})\nonumber
\end{eqnarray}
\begin{eqnarray}
b_+(\lambda_{1},\lambda_{2})c_{-}(\lambda_{1},\lambda_{3})b_-(\lambda_{2},\lambda_{3})+c_{-}(\lambda_{1},\lambda_{2})a_-(\lambda_{1},\lambda_{3}) c_{-} (\lambda_{2},\lambda_{3})&=&\label{YB5}\\
 a_-(\lambda_{1},\lambda_{2}) c_{-}(\lambda_{1},\lambda_{3})a_-(\lambda_{2},\lambda_{3})\nonumber\\
b_-(\lambda_{1},\lambda_{2})a_-(\lambda_{1},\lambda_{3})c_{-}(\lambda_{2},\lambda_{3})+c_{+}(\lambda_{1},\lambda_{2})c_{-}(\lambda_{1},\lambda_{3})b_-(\lambda_{2},\lambda_{3})&=& \label{YB6}\\
 a_-(\lambda_{1},\lambda_{2}) b_-(\lambda_{1},\lambda_{3})c_{-}(\lambda_{2},\lambda_{3})\nonumber\\
c_{-}(\lambda_{1},\lambda_{2})c_{+}(\lambda_{1},\lambda_{3})c_{-}(\lambda_{2},\lambda_{3}) &=& \label{YB7}\\
c_{+}(\lambda_{1},\lambda_{2})c_{-}(\lambda_{1},\lambda_{3})c_{+}(\lambda_{2},\lambda_{3}) \nonumber\\
b_+(\lambda_{1},\lambda_{2})c_+(\lambda_{1},\lambda_{3})c_{-}(\lambda_{2},\lambda_{3})+c_{+}(\lambda_{1},\lambda_{2})a_{-}(\lambda_{1},\lambda_{3})b_+(\lambda_{2},\lambda_{3})&=& \label{YB8}\\
 c_+(\lambda_{1},\lambda_{2}) b_+(\lambda_{1},\lambda_{3})a_{-}(\lambda_{2},\lambda_{3}).\nonumber
\end{eqnarray}
\subsubsection{The co-cycle relation for R-matrices}
Throughout the remainder of this work we use the following \textit{left-handed} notation,
\begin{equation*}
R_{1,2\dots L} (\lambda_1| \lambda_2,\dots,\lambda_L) = R_{1L}(\lambda_1,\lambda_L) R_{1(L-1)}(\lambda_1,\lambda_{L-1}) \dots R_{12}(\lambda_1,\lambda_2),
\end{equation*}
and \textit{right-handed} notation,
\begin{equation*}
R_{1\dots (L-1),L} (\lambda_1,\dots,\lambda_{L-1}|\lambda_L) = R_{1L}(\lambda_1,\lambda_L) R_{2 L} (\lambda_2,\lambda_L)\dots R_{(L-1)L}(\lambda_{L-1},\lambda_L),
\end{equation*}
for the ordered product of $R$-matrices. Using the above notation the $R$-matrices satisfy the following global unitarity condition,
\begin{equation}
R_{1,2\dots L}(\lambda_1| \lambda_2,\dots,\lambda_L)R_{2\dots L,1} (\lambda_2,\dots,\lambda_{L}|\lambda_1) =\mathcal{I}_{1\dots L},
\label{biguni}\end{equation}
and co-cycle relation,
\begin{equation}\begin{array}{ll}
&R_{1,2\dots (L-1)}(\lambda_1| \lambda_2,\dots,\lambda_{L-1})R_{1\dots (L-1),L}(\lambda_1,\dots,\lambda_{L-1}|\lambda_L)\\
 =&  R_{2\dots (L-1),L}(\lambda_2,\dots,\lambda_{L-1}|\lambda_L) R_{1, 2 \dots L}(\lambda_1| \lambda_2,\dots,\lambda_L).
\end{array}\label{Rcocycle}
\end{equation}
Both Eqs. (\ref{biguni}) and (\ref{Rcocycle}) can be verified using induction. 
\subsection{The monodromy matrix}
\subsubsection{Rapidities and inhomogeneities - notational conventions} Through the remainder of this work we shall label the rapidities by $\{\mu,\nu\}$, and we denote the corresponding inhomogeneities by $ \{\xi_1, \dots, \xi_L\}$. To this point in the paper we have used indices on operators to indicate what vector spaces the operators act on. In addition we will now use the following notational conventions:
\begin{itemize}
\item{To indicate that an operator acts on the auxiliary spaces $\mathcal{A}_a$ or $\mathcal{A}_b$, we shall always use the Latin based indices $a$ or $b$. Furthermore, associated with auxiliary space $\mathcal{A}_a$ is the rapidity $\mu$, and associated with auxiliary space $\mathcal{A}_b$ is the rapídity $\nu$.}
\item{To indicate that an operator acts on the vector space $V_{\alpha}$, $\alpha \in \{1,\dots, L\}$, we shall use the Greek based index $\alpha$, or the actual numbers $\{1,\dots,L\}$. Furthermore, associated with vector space $V_{\alpha}$ is the inhomogeneity $\xi_{\alpha}$.}
\end{itemize}
Hence under the above conventions we have the following examples of notational equivalences that we rely on heavily throughout the remainder of this work:
\begin{equation*}\begin{array}{lll}
X_{a,1\dots L} (\mu)  &\equiv& X_{a,1\dots L} (\mu|\xi_1, \dots, \xi_L)\\
Y_{b,\sigma(1\dots L)} (\nu) &\equiv&  Y_{b,\sigma(1\dots L)} (\nu|\xi_{\sigma(1)}, \dots, \xi_{\sigma(L)})\\
Z_{1,2\dots L} & \equiv & Z_{1,2\dots L} (\xi_1|\xi_2,\dots \xi_L)
\end{array} 
\end{equation*}
where $\sigma \in S_L$, $X_{a,1\dots L} \in \textrm{End}(\mathcal{A}_a \otimes V_1 \otimes \dots \otimes V_L)$, $Y_{b,\sigma(1\dots L)} \in \textrm{End}(\mathcal{A}_b \otimes V_{\sigma(1)} \otimes \dots \otimes V_{\sigma(L)})$ and $Z_{1,2\dots L} \in \textrm{End}(V_1 \otimes \dots \otimes V_L)$.
\subsubsection{The local $\mathcal{L}$-matrix and global $\mathcal{T}$-matrix} We now define the local $\mathcal{L}$-matrix as,
\begin{equation*}
\mathcal{L}_{a \alpha} (\mu) = R_{a \alpha} (\mu).
\end{equation*}
The global monodromy matrix,
\begin{equation*}
\mathcal{T}_{a,1\dots L}(\mu ) \in \textrm{End}(\mathcal{A}_a \otimes V_1 \otimes \dots \otimes V_L),
\end{equation*}
is constructed from the following ordered product of $\mathcal{L}$-matrices,
\begin{equation*}\begin{array}{lcl}
\mathcal{T}_{a,1 \dots L}(\mu) &=& \mathcal{L}_{aL}(\mu) \mathcal{L}_{a(L-1)}(\mu)\dots \mathcal{L}_{a1}(\mu)\\
&=& \left( \begin{array}{cc}
A_{1\dots L}(\mu)&  B_{1\dots L}(\mu) \\
C_{1\dots L}(\mu)&  D_{1\dots L}(\mu)
\end{array}\right)_{a}.
\end{array}
\end{equation*}
The operators,
\begin{equation*}
\left\{ A_{1\dots L}(\mu),  B_{1\dots L}(\mu) ,C_{1\dots L}(\mu),  D_{1\dots L}(\mu) \right\} \in \textrm{End} (V_1\otimes \dots \otimes V_L),
\end{equation*}
are referred to as the entries of the monodromy matrix. We give the following necessary result for the operators.
\begin{proposition}\label{propABCD}
The operators $A_{1\dots L}(\mu)$ and $C_{1\dots L}(\mu)$ are \textit{upper triangular} and the operators $B_{1\dots L}(\mu)$ and $D_{1\dots L}(\mu)$ are \textit{lower triangular}. Moreover, the diagonal entries of $B_{1\dots L}(\mu)$ and $C_{1\dots L}(\mu)$ are entirely zero, and the diagonal entries of $D_{1\dots L}(\mu)$ are given by,
\begin{equation*}
\textrm{diag}\left\{ D_{1\dots L}(\mu) \right\} = \otimes^L_{i=1} \left(\begin{array}{cc}
b_-(\mu,\xi_i) & 0\\
0 & a_-(\mu,\xi_i) \end{array} \right)_i.
\end{equation*}
Additionally, the entries of the final row of $ D_{1\dots L}(\mu) $ are given by,
\begin{equation*}\begin{array}{llcll}
\left( D_{1\dots L}(\mu)\right)_{(2^L)j} &=& 0 & \textrm{for} & j \in \{1,\dots, 2^L-1\}\\
\left( D_{1\dots L}(\mu)\right)_{(2^L)(2^L)} &=& {\displaystyle \prod^L_{i=1}}a_-(\mu,\xi_i).
\end{array}\end{equation*}
\end{proposition}
The verification of the above result can be obtained by induction through decomposing the monodromy matrix in the following way,
\begin{equation*}
 \mathcal{T}_{a,1\dots L}(\mu)= R_{a L}(\mu)R_{a, 1 \dots (L-1)}(\mu).
\end{equation*}
We offer the complete proof in appendix \ref{APP1}.
\subsection{The Yang-Baxter algebra} From the Yang-Baxter equation (\ref{two}) one can show that the monodromy matrix satisfies the \textit{global intertwining relation} given by Eq. (\ref{one}). Expanding Eq. (\ref{one}) as a $4 \times 4$ matrix equation in the auxiliary vector space $\mathcal{A}_a \otimes \mathcal{A}_b$, we obtain sixteen algebraic equations for the operators $A_{1\dots L},  B_{1\dots L} ,C_{1\dots L}$ and $D_{1\dots L}$. We list the eight relevant equations for this work below:
\begin{eqnarray}
A(\mu)D(\nu)-A(\nu)D(\mu)&=& \frac{b_+}{c_-}(\mu,\nu)B(\nu)C(\mu)-\frac{b_-}{c_-}(\mu,\nu)C(\mu)B(\nu)\label{eq6}\\
A(\mu)C(\nu)&=& \frac{1}{b_+}(\mu,\nu)C(\nu)A(\mu)-\frac{c_+}{b_+}(\mu,\nu)C(\mu)A(\nu)\label{eq9}\\
D(\nu)A(\mu)-A(\mu)D(\nu)&=& \frac{c_+}{b_+}(\mu,\nu) C(\mu)B(\nu)-\frac{c_-}{b_+}(\mu,\nu) C(\nu)B(\mu)\label{eq10}\\
D(\mu)A(\nu)-D(\nu)A(\mu)&=& \frac{b_-}{c_+}(\mu,\nu) C(\nu)B(\mu)-\frac{b_+}{c_+}(\mu,\nu) B(\mu)C(\nu)\label{eq11}\\
C(\nu)C(\mu)&=& a_-(\mu,\nu)C(\mu)C(\nu)\label{eq13}\\
D(\nu)C(\mu)&=& \frac{a_-}{b_+}(\mu,\nu)C(\mu)D(\nu)-\frac{c_-}{b_+}(\mu,\nu)C(\nu)D(\mu)\label{eq14}\\
C(\nu)D(\mu)&=& \frac{a_-}{b_-}(\mu,\nu)D(\mu)C(\nu)-\frac{c_+}{b_-}(\mu,\nu)D(\nu)C(\mu)\label{eq15}\\
D(\mu)D(\nu)-D(\nu)D(\mu)&=&0.\label{eq16}
\end{eqnarray}
\subsection{The permuted R-matrix} The concept of permuting the $R$-matrix is integral in defining the $F$-basis for this model. As such, we shall spend some time giving necessary definitions and results. For self-consistency we also give the necessary results regarding the symmetric group $S_L$ in Appendix \ref{APP3}.

For any given $\sigma \in S_L$ the permuted $R$-matrix,
\begin{equation*} 
R^{\{\sigma \}}_{1\dots L} \in \textrm{End}(V_1 \otimes \dots \otimes V_L),
\end{equation*}
is defined through the following product of auxiliary operators,
\begin{equation}
R^{\{\sigma\}}_{1\dots L} = P^{\left\{\sigma^{-1}\right\}}_{1\dots L} \hat{R}^{\{\sigma\}}_{1\dots L}.
\label{decom1}\end{equation}
Decomposing $\sigma$ in terms of adjacent permutations, 
\begin{equation*}
\sigma = \sigma_{\alpha_p (\alpha_p+1)}\dots \sigma_{\alpha_2 (\alpha_2+1)}\sigma_{\alpha_1 (\alpha_1+1)},
\end{equation*}
the auxiliary operators $P^{\left\{\sigma^{-1}\right\}}_{1\dots L}$ and $\hat{R}^{\{\sigma\}}_{1\dots L}$ decompose the following way,
\begin{equation}\begin{array}{lll}
P^{\left\{\sigma^{-1}\right\}}_{1\dots L} &=& P^{\left\{\sigma_{\alpha_1 (\alpha_1+1)}\right\}}_{1\dots L}  P^{\left\{\sigma_{\alpha_2 (\alpha_2 +1)}\right\}}_{1\dots L} \dots P^{\left\{\sigma_{\alpha_p (\alpha_p+1)}\right\}}_{1\dots L}\\
\hat{R}^{\{\sigma\}}_{1\dots L} &=& \hat{R}^{\left\{\sigma_{\alpha_p (\alpha_p+1)}\right\}}_{1\dots L} \dots \hat{R}^{\left\{\sigma_{\alpha_2 (\alpha_2 +1)}\right\}}_{1\dots L} \hat{R}^{\left\{\sigma_{\alpha_1 (\alpha_1+1)}\right\}}_{1\dots L},
\end{array}\label{decom2}\end{equation}
where,
\begin{equation}\begin{array}{lll}
P^{\left\{\sigma_{\alpha (\alpha+1)}\right\}}_{1\dots L} &=& P_{\alpha (\alpha+1)}\\
\hat{R}^{\left\{\sigma_{\alpha (\alpha+1)}\right\}}_{1\dots L} &=& P_{\alpha (\alpha+1)} R_{\alpha (\alpha+1)}.
\end{array}\label{decom3}\end{equation}
Given a general operator $X_{1 \dots L} \in \textrm{End}(V_1 \otimes \dots \otimes V_L)$ and the auxiliary operator $ P^{\left\{\sigma\right\}}_{1\dots L}$, we have the following useful relation,
\begin{equation}
X_{\sigma(1\dots L)} = P^{\left\{\sigma^{-1}\right\}}_{1\dots L} X_{1\dots L} P^{\left\{\sigma\right\}}_{1\dots L}.
\label{useful}\end{equation}
We also give the following necessary result.
\subsubsection{A necessary result - representations of the symmetric group}
\begin{proposition}\label{PROP2}
The auxiliary operators $P^{\left\{\sigma\right\}}_{1\dots L}$ and $\hat{R}^{\{\sigma\}}_{1\dots L}$ provide valid representations for the symmetric group $S_L$. 
\end{proposition}
\textbf{Proof.} We recall that the symmetric group $S_L$ can be generated through the set of $(L-1)$ adjacent permutations,
\begin{equation*}
\{\sigma_{12}, \sigma_{23}, \dots, \sigma_{(L-1)L}\}.
\end{equation*}
We note that each expression for $P^{\left\{\sigma_{\alpha(\alpha+1)}\right\}}_{1\dots L}$ and $\hat{R}^{\{\sigma_{\alpha(\alpha+1)}\}}_{1\dots L}$ is unique for each $\alpha \in \{1,\dots,(L-1) \}$. Hence, to show that $P^{\left\{\sigma\right\}}_{1\dots L}$ and $\hat{R}^{\{\sigma\}}_{1\dots L}$ provide valid representations for the symmetric group $S_L$, one is required to show that they obey\footnote{Refer to Appendix (\ref{APP3}) for more details.} Eqs. (\ref{rel1}-\ref{rel3}). This can be achieved using elementary matrix multiplication, unitarity (\ref{three}) and Yang-Baxter (\ref{two}). $\square$

We detailed this result because its application is crucial in the construction of the $F$-basis. 
\subsubsection{Important example - the cyclic permutation} The permuted $R$-operator for $\sigma_{c}$, given explicitly by Eq. (\ref{CYCLE}), has the following form:
\begin{equation}\begin{array}{lll}
 R^{\left\{\sigma_{c}\right\}}_{1\dots L} &=&  P^{\left\{\sigma^{-1}_{c}\right\}}_{1\dots L}\hat{R}^{\left\{\sigma_{c}\right\}}_{1\dots L}\\
&=& P^{\left\{\sigma_{12}\right\}}_{1\dots L} \dots P^{\left\{\sigma_{(L-1)L}\right\}}_{1\dots L}\hat{R}^{\left\{\sigma_{(L-1)L}\right\}}_{1\dots L} \dots \hat{R}^{\left\{\sigma_{12}\right\}}_{1\dots L} \\
&=& R_{1,2\dots L}.
\end{array}
\label{cyclicR}\end{equation}
\section{The F-basis}\label{theojust}
We now provide a detailed analysis of the factorizing $F$-matrix associated with the arbitrary six-vertex model. 
\subsection{The factorizing F-matrix - definition}\label{definI}
We begin by imposing the three following standard properties on the matrix $F_{1\dots L}$ first offered in \cite{MS}:
\begin{itemize}
\item{$F_{1\dots L}$ is lower triangular.} 
\item{$F^{-1}_{1\dots L}$ exists.}
\item{$F_{1\dots L}$ obeys the following factorization property for all $\sigma \in S_L$,}
\end{itemize}
\begin{equation}
 F_{\sigma(1\dots L)} R^{\{\sigma\}}_{1\dots L} =  F_{1\dots L}.
\label{defining}\end{equation}
\subsection{The $L=2$ example} As a relevant example, let us consider the case $L=2$. For this case, the only non-trivial permutation is the adjacent permutation $\sigma_{12}$. Hence Eq. (\ref{defining}) explicitly becomes,
\begin{equation*}
 F_{21}  R_{12}  = F_{12}.
\end{equation*}
Relying on the unitarity of the $R$-matrix (\ref{three}) we find that the solution to $F_{12}$ is given by,
\begin{eqnarray}
F_{12} &=& \mathcal{N}_{12}\left(e^{(11)}_1+e^{(22)}_1 R_{12}\right) \label{N=2a}\\
       &=& \mathcal{N}_{21} \left(e^{(11)}_2 R_{12}+ e^{(22)}_2\right) \label{N=2b},
\end{eqnarray}
where we use the notation $e^{(ij)}_{\alpha} \in \textrm{End}(V_{\alpha})$ for the $2 \times 2$ matrix with $1$ in entry $(i,j)$ and zero in all other entries. Additionally, 
$\mathcal{N}_{12}$ is the following diagonal matrix,
\begin{equation}
 \mathcal{N}_{12} = \left( \begin{array}{cccc}
                         1 & 0 & 0 & 0\\
			 0 & 1 & 0 & 0\\
			 0 & 0 & 1 & 0\\
			 0 & 0 & 0 & \frac{1}{\sqrt{a_-(\xi_1,\xi_2)}}
                        \end{array}\right).
\label{curlyN}\end{equation}
The solution to $F_{12}$ in Eq. (\ref{N=2a}) can be decomposed into two distinct parts - the diagonal matrix $\mathcal{N}_{12}$, and the remainder which we shall label,
\begin{equation*}
\mathcal{F}_{12} = e^{(11)}_1+e^{(22)}_1 R_{12}.
\end{equation*}
We note that the form of $\mathcal{F}_{12}$ coincides exactly with the solution originally provided in \cite{MS}.
\subsection{An ansatz for general L} Taking inspiration from Eq. (\ref{N=2a}), we provide the following ansatz for the form of $F_{1\dots L}$,
\begin{equation}
F_{1\dots L}  = \mathcal{N}_{1\dots L}\mathcal{F}_{1\dots L},
\label{ansatz}\end{equation}
where the definition of $\mathcal{N}_{1\dots L}$ is provided by the product of \textit{partial} $\mathcal{N}$-matrices,
\begin{equation}\begin{array}{lll}
\mathcal{N}_{1\dots L} &=& \mathcal{N}_{2\dots L}\mathcal{N}_{1,2\dots L}\\
&=& \mathcal{N}_{(L-1)L}\mathcal{N}_{L-2,(L-1)L} \dots \mathcal{N}_{1,2\dots L},
\end{array}\label{BIGN}\end{equation}
such that the partial $\mathcal{N}$-matrices $\mathcal{N}_{i,(i+1)\dots L}$, are given by,
\begin{equation}
\mathcal{N}_{i,(i+1)\dots L} = \mathcal{N}_{i L} \mathcal{N}_{i (L-1)} \dots \mathcal{N}_{i(i+1)}.
\label{BIGN2}\end{equation}
As with the $L=2$ case, the terms $\mathcal{F}_{1\dots L}$ coincide with the form of the solution given in \cite{MS}:
\begin{equation}\begin{array}{lll}
 \mathcal{F}_{1\dots L} &=& \mathcal{F}_{2\dots L}\mathcal{F}_{1,2\dots L}\\
&=&\mathcal{F}_{(L-1)L}\mathcal{F}_{(L-2),(L-1)L} \dots \mathcal{F}_{1,2\dots L},
\end{array}\label{ansatz1a}\end{equation}
where we use the following \textit{left-handed} notation for the \textit{partial} $\mathcal{F}$-matrices,
\begin{equation}\begin{array}{lll}
\mathcal{F}_{i,(i+1)\dots L} &=& e^{(11)}_i + e^{(22)}_i R_{i,(i+1)\dots L}\\
&=& \left( \begin{array}{cc}
            \mathcal{I}_{(i+1)\dots L} & 0\\
	    C_{(i+1)\dots L}(\xi_i) & D_{(i+1)\dots L}(\xi_i)
           \end{array}\right)_i.
\end{array}\label{lefthanded}\end{equation}
Hence $F_{1\dots L}$ is expressed as a product of partial left-handed matrices ($\mathcal{F}_{1\dots L}$) multiplied by a diagonal matrix ($\mathcal{N}_{1 \dots L}$). We can immediately notice that such a form is lower triangular (due to $D_{i+1\dots L}(\xi_i)$ being lower triangular), and since each diagonal entry is non zero, $F^{-1}_{1\dots L}$ is assured to exist. We are left with the task of verifying the factorization condition (\ref{defining}).
\subsubsection{Recasting the factorization condition - the twisted $R$-matrix}
In order to show the validity of the factorization condition (\ref{defining}) for all $\sigma \in S_L$ we proceed by first taking advantage of the decomposition of $F_{1\dots L}$ present in Eq. (\ref{ansatz}). In what follows we recast Eq. (\ref{defining}) as an equation involving $\mathcal{F}_{1 \dots L}$ and obtain a factorization expression in which the \textit{twisted} $R$-matrix is not proportional to the identity. We then adapt the procedure first devised in \cite{MS} to tackle the problem. To this end we substitute Eq. (\ref{ansatz}) into Eq. (\ref{defining}) to obtain,
\begin{equation*}\begin{array}{lll}
 \mathcal{F}_{\sigma(1\dots L)} R^{\{\sigma\}}_{1\dots L}  = \mathcal{N}^{-1}_{\sigma(1\dots L)}\mathcal{N}_{1\dots L} \mathcal{F}_{1\dots L} & \textrm{for all}& \sigma \in S_L.
\end{array}\end{equation*}
We now offer the following result.
\begin{proposition}\label{NandR}
\begin{equation}\begin{array}{lll}
\mathcal{N}^{-1}_{\sigma(1\dots L)}\mathcal{N}_{1\dots L}  = \mathcal{R}^{\{\sigma\}}_{1\dots L} &\textrm{for all}& \sigma \in S_L,
\end{array}\label{auxR}\end{equation}
where $\mathcal{R}^{\{\sigma\}}_{1\dots L}$ follows the same decomposition rules as $R^{\{\sigma\}}_{1\dots L}$, given by Eqs. (\ref{decom1}-\ref{decom3}), and $\mathcal{R}_{12}$ is explicitly given as,
\begin{equation}
 \mathcal{R}_{12}= \mathcal{N}^{-1}_{21}\mathcal{N}_{12}  = \left( \begin{array}{cccc}
                         1 & 0 & 0 & 0\\
			 0 & 1 & 0 & 0\\
			 0 & 0 & 1 & 0\\
			 0 & 0 & 0 & \frac{1}{a_-(\xi_1,\xi_2)}
                        \end{array}\right)_{12}.
\label{twistedRR}\end{equation}
\end{proposition}
Since the corresponding auxiliary operators $\hat{\mathcal{R}}^{\{\sigma\}}_{1\dots L}$ provide a valid representation for $S_L$, the above result can be verified by showing that Eq. (\ref{auxR}) holds for only two permutations: the adjacent permutation $\sigma_{12}$ and the cyclic permutation $\sigma_c$. We give the verification in Appendix \ref{APP4}. 

Hence we recast the factorization condition in the following form,
\begin{equation}\begin{array}{lll}
\mathcal{F}_{\sigma(1\dots L)} R^{\{\sigma\}}_{1\dots L}  = \mathcal{R}^{\{\sigma\}}_{1\dots L} \mathcal{F}_{1\dots L} & \textrm{for all} & \sigma \in S_L,
\end{array}\label{defining2}\end{equation}
where we note that the $\mathcal{R}$-matrices of this work are referred to as \textit{twisted} $R$-matrices in \cite{MS}.

We impose that the $\mathcal{R}$-matrices follow the same left-handed and right-handed convention as the $R$-matrices,
\begin{eqnarray*}
\mathcal{R}_{1,2\dots,N} &=& \mathcal{R}_{1N} \mathcal{R}_{1(N-1)} \dots \mathcal{R}_{12} \\
\mathcal{R}_{1\dots N-1,N} &=& \mathcal{R}_{1N} \mathcal{R}_{2 N} \dots \mathcal{R}_{(N-1)N}.
\end{eqnarray*}
Following the above convention the $\mathcal{R}$-matrices obey the same global unitarity condition as the $R$-matrices,
\begin{equation}
 \mathcal{R}_{1,2\dots L}\mathcal{R}_{2\dots L,1}  =\mathcal{I}_{1\dots L},
\label{unitcurlyR}\end{equation}
and being diagonal the $\mathcal{R}$-matrices commute amongst themselves.\\
\\
\textbf{Techniques employed to prove the factorization condition.} To complete the proof we proceed in a similar fashion to \cite{MS} - we introduce an appropriate \textit{right-handed} notation for the partial matrices $\mathcal{F}_{1\dots (L-1), L}$ which permits us to obtain the corresponding co-cycle relation for the factorizing $\mathcal{F}$-matrices. We then use the co-cycle relation to express the ansatz (\ref{ansatz1a}) as a product of partial right-handed matrices. After these ingredients are obtained the proof is almost automatic. 
\subsubsection{The co-cycle relation} Taking inspiration from Eq. (\ref{N=2b}), we introduce the following right-handed notation,
\begin{equation}
\mathcal{F}_{i \dots (L-1),L}  = \mathcal{R}_{L,i \dots (L-1)} \left( e^{(11)}_L R_{i \dots (L-1),L} +e^{(22)}_L \right) .
\label{righthanded}\end{equation}
With the aforementioned left-handed and right-handed partial $\mathcal{F}$-matrices, we give the following result:
\begin{proposition}{The co-cycle relation.}
\begin{equation}
\mathcal{R}_{L,1\dots (L-1)} \mathcal{F}_{1,2\dots (L-1)}\mathcal{R}_{1\dots (L-1),L} \mathcal{F}_{1\dots (L-1),L} = \mathcal{F}_{2 \dots (L-1),L} \mathcal{F}_{1,2 \dots L}
\label{co-cycleF}\end{equation}
\end{proposition}
\textbf{Proof.} The difference between the left-handed and right-handed partial $\mathcal{F}$-matrices is crucial for proving the validity of the corresponding co-cycle relation. Substituting the explicit form of the partial matrices on the left hand side of Eq. (\ref{co-cycleF}) we obtain,
\begin{equation*}\begin{array}{ll}
  & \mathcal{R}_{L,1\dots (L-1)} \overbrace{\mathcal{F}_{1,2\dots (L-1)}}^{\textrm{use Eq. (\ref{lefthanded})}}\overbrace{\mathcal{R}_{1\dots (L-1),L} \mathcal{F}_{1\dots (L-1),L}}^{\textrm{use Eq. (\ref{righthanded})}}\\
=& \mathcal{R}_{L,1\dots (L-1)} \left( e^{(11)}_1 e^{(22)}_L +e^{(11)}_1e^{(11)}_L R_{1\dots (L-1),L} \right. \\
&\left. + e^{(22)}_1 e^{(11)}_L R_{1,2\dots (L-1)}R_{1\dots (L-1),L} +e^{(22)}_1 e^{(22)}_L R_{1,2\dots (L-1)} \right) \\
=& \mathcal{R}_{L,2\dots (L-1)} \textrm{\huge(\normalsize}  \overbrace{\mathcal{R}_{L1}e^{(11)}_1  e^{(22)}_L}^{e^{(11)}_1 e^{(22)}_L} +\overbrace{\mathcal{R}_{L1}e^{(11)}_1e^{(11)}_L R_{1L}}^{ e^{(11)}_1e^{(11)}_L}R_{2\dots (L-1),L}  \\
& + \overbrace{\mathcal{R}_{L1}e^{(22)}_1 e^{(11)}_L}^{e^{(22)}_1 e^{(11)}_L} \overbrace{R_{1,2\dots (L-1)}R_{1\dots (L-1),L}}^{\textrm{use Eq. (\ref{Rcocycle})}} +\overbrace{\mathcal{R}_{L1}e^{(22)}_1 e^{(22)}_L}^{e^{(22)}_1 e^{(22)}_L R_{1L}} R_{1,2\dots (L-1)} \textrm{\huge)\normalsize}\\
=& \mathcal{R}_{L,2\dots (L-1)} \left(  e^{(11)}_1 e^{(22)}_L +e^{(11)}_1e^{(11)}_L R_{2\dots (L-1),L}  \right.\\
&\left.+ e^{(22)}_1 e^{(11)}_L R_{2\dots (L-1),L}R_{1,2\dots L} +e^{(22)}_1 e^{(22)}_L R_{1,2\dots L} \right). \textrm{  $\square$}
\end{array}
\end{equation*}
\subsubsection{Extending the ansatz}
With the application of the co-cycle relation, we can now express the ansatz (\ref{ansatz1a}) as the following product of right-handed partial $\mathcal{F}$-matrices:
\begin{proposition}
\begin{equation}
 \mathcal{F}_{1\dots L} = \mathcal{R}_{L,1\dots (L-1)} \mathcal{F}_{1\dots (L-1)}\mathcal{R}_{1\dots (L-1),L} \mathcal{F}_{1\dots (L-1),L}
\label{ansatz2}\end{equation}
\end{proposition}
\textbf{Proof.} We shall proceed using induction. Notice that the $L=3$ case produces the co-cycle relation (\ref{co-cycleF}). Assuming that Eq. (\ref{ansatz2}) is true for some $L$ we obtain the following from the $(L+1)$ case,
\begin{equation*}\begin{array}{lll}
\mathcal{F}_{1\dots (L+1)} &= & \overbrace{\mathcal{F}_{2\dots (L+1)}}^{\textrm{use Eq. (\ref{ansatz2})} } \mathcal{F}_{1,2\dots (L+1)}\\
&=&\mathcal{R}_{(L+1),2\dots L} \mathcal{F}_{2\dots L}\mathcal{R}_{2\dots L,(L+1)}\overbrace{ \mathcal{F}_{2\dots L,(L+1)} \mathcal{F}_{1,2\dots (L+1)}}^{\textrm{use Eq. (\ref{co-cycleF})}}\\
&=& \mathcal{R}_{(L+1),1\dots L} \mathcal{F}_{2\dots L}\overbrace{\mathcal{R}_{2\dots L,(L+1)}\mathcal{R}_{(L+1),2\dots L}}^{\textrm{use Eq. (\ref{unitcurlyR})}} \mathcal{F}_{1,2\dots L} \mathcal{R}_{1\dots L,(L+1)}\mathcal{F}_{1\dots L,(L+1)}\\
&=& \mathcal{R}_{(L+1),1\dots L} \overbrace{\mathcal{F}_{2\dots L}\mathcal{F}_{1,2\dots L}}^{\textrm{use Eq. (\ref{ansatz1a})}} \mathcal{R}_{1\dots L,(L+1)}\mathcal{F}_{1\dots L, (L+1)}
\end{array}\end{equation*}
\begin{equation*}\begin{array}{lll}
&=&\mathcal{R}_{(L+1),1\dots L} \mathcal{F}_{1\dots L} \mathcal{R}_{1\dots L,(L+1)}\mathcal{F}_{1\dots L,(L+1)}. \textrm{  $\square$}
\end{array}\end{equation*}
We are now prepared to verify the factorization property.
\subsection{Verification of the factorization property} \label{PROOF}
We begin by applying Eq. (\ref{useful}) to Eq. (\ref{defining2}) to obtain,
\begin{equation}
 \hat{R}^{\{\sigma\}}_{1\dots L} =\mathcal{F}^{-1}_{1\dots L} \hat{\mathcal{R}}^{\left\{\sigma \right\}}_{1\dots L} \mathcal{F}_{1\dots L}.
\label{defining3}\end{equation} 
Since both $\hat{R}^{\{\sigma\}}_{1 \dots L}$ and $\hat{\mathcal{R}}^{\{\sigma\}}_{1\dots L}$ provide valid representations of $S_L$, we remark that Eq. (\ref{defining3}) is in a form that one can readily decompose the permutation $\sigma$. To illustrate this consider the permutation $\sigma = \sigma_1 \sigma_2$ on the left-hand side and right-hand side of Eq. (\ref{defining3}) respectively,
\begin{equation*}
\begin{array}{lll}
\hat{R}^{\{\sigma_1\sigma_2\}}_{1\dots L}&=& \hat{R}^{\{\sigma_1\}}_{1\dots L}\hat{R}^{\{\sigma_2\}}_{1\dots L}\\
\mathcal{F}^{-1}_{1\dots L} \hat{\mathcal{R}}^{\{\sigma_1\sigma_2\}}_{1 \dots L}\mathcal{F}_{1\dots L} &=& \mathcal{F}^{-1}_{1\dots L} \hat{\mathcal{R}}^{\{\sigma_1\}}_{1 \dots L}\mathcal{F}_{1\dots L}\mathcal{F}^{-1}_{1\dots L} \hat{\mathcal{R}}^{\{\sigma_2\}}_{1 \dots L}\mathcal{F}_{1\dots L}
\end{array}
\end{equation*}
where $\{\sigma_1,\sigma_2\} \in S_L$. Since $S_L$ can be constructed entirely from the adjacent permutation $\sigma_{12}$ and the cyclic permutation $\sigma_c$ (see Appendix \ref{APP3}), we need only verify Eq. (\ref{defining2}) for the aforementioned permutations to guarantee its validity for all $S_L$.
\subsubsection{The adjacent permutation $\sigma_{12}$} Beginning with $\sigma_{12}$ we proceed using induction. Having previously shown the validity of the $L=2$ case, we assume that the case is true up to some $L$. Hence for $(L+1)$, the left hand side of Eq. (\ref{defining}) becomes,
\begin{equation*}
 \overbrace{\mathcal{F}_{213 \dots (L+1)}}^{\textrm{use Eq. (\ref{ansatz2})}} R_{12} = \mathcal{R}_{(L+1),213\dots L} \mathcal{F}_{213 \dots L}\mathcal{R}_{213\dots L,(L+1)} \mathcal{F}_{213 \dots L,(L+1)}R_{12}
\end{equation*}
where,
\begin{equation*}\begin{array}{ll}
& \mathcal{R}_{213\dots L,(L+1)}\overbrace{ \mathcal{F}_{213 \dots L,(L+1)}}^{\textrm{use Eq. (\ref{righthanded})}} R_{12}\\
=& \overbrace{\mathcal{R}_{213\dots L,(L+1)}\mathcal{R}_{(L+1),213\dots L}}^{\textrm{use Eq. (\ref{unitcurlyR})}} \left( e^{(11)}_{L+1} R_{213 \dots L,(L+1)} + e^{(22)}_{L+1}\right) R_{12} \\
=&   e^{(11)}_{L+1} \overbrace{R_{2(L+1)}R_{1(L+1)}R_{12}}^{\textrm{use Eq. (\ref{two})}} R_{3 \dots L,(L+1)}   + R_{12}e^{(22)}_{L+1}  
\end{array}\end{equation*}
\begin{equation*}\begin{array}{ll}
=& R_{12}\left(  e^{(11)}_{L+1} R_{1 \dots L,(L+1)}  + e^{(22)}_{L+1} \right)\\
=& R_{12}\mathcal{R}_{1 \dots L,(L+1)}\overbrace{\mathcal{R}_{(L+1),1 \dots L}\left(  e^{(11)}_{L+1} R_{1 \dots L,(L+1)}  + e^{(22)}_{L+1} \right)}^{\textrm{use Eq. (\ref{righthanded})}}\\
=& R_{12} \mathcal{R}_{1 \dots L,(L+1)} \mathcal{F}_{1 \dots L,(L+1)}.
\end{array}
\end{equation*}
Hence,
\begin{eqnarray*}
\mathcal{F}_{213 \dots (L+1)}R_{12} &=&  \mathcal{R}_{(L+1),213\dots L} \mathcal{F}_{213 \dots L}R_{12} \mathcal{R}_{1 \dots L,(L+1)} \mathcal{F}_{1 \dots L,(L+1)}\\
&=&  \mathcal{R}_{(L+1),1\dots L}\overbrace{\mathcal{F}_{\sigma_{12}(1 \dots L)}R^{\left\{\sigma_{12} \right\}}_{1\dots L}}^{\textrm{use Eq. (\ref{defining2})}} \mathcal{R}_{1 \dots L,(L+1)} \mathcal{F}_{1 \dots L,(L+1)}\\
&=& \mathcal{R}^{\left\{\sigma_{12} \right\}}_{1\dots L} \overbrace{\mathcal{R}_{(L+1),1\dots L}\mathcal{F}_{1 \dots L} \mathcal{R}_{1 \dots L,(L+1)} \mathcal{F}_{1 \dots L,(L+1)}}^{\textrm{use Eq. (\ref{ansatz2})}}\\
&=& \mathcal{R}_{12} \mathcal{F}_{1 \dots (L+1)},
\end{eqnarray*}
which is the required expression. 
\subsubsection{The cyclic permutation $\sigma_c$} Moving on to $\sigma_{c}$ we can proceed without induction. Starting from the left hand side of Eq. (\ref{defining2}),
\begin{equation*}
\overbrace{\mathcal{F}_{2\dots L 1}}^{\textrm{use Eq. (\ref{ansatz2})}}R_{1,2\dots L } =\mathcal{R}_{1,2\dots L} \mathcal{F}_{2 \dots L}\mathcal{R}_{2\dots L,1} \mathcal{F}_{2 \dots L,1}R_{1,2\dots L},
\end{equation*}
where,
\begin{equation*}
\begin{array}{ll}
&\mathcal{R}_{2\dots L,1} \overbrace{\mathcal{F}_{2 \dots L,1}}^{\textrm{use Eq. (\ref{righthanded})}}R_{1,2\dots L}\\
=& \overbrace{\mathcal{R}_{2\dots L,1}\mathcal{R}_{1,2\dots L}}^{\textrm{use Eq. (\ref{unitcurlyR})}} \textrm{\huge(\normalsize} e^{(11)}_{1} \overbrace{R_{2\dots L,1}R_{1,2\dots L }}^{\textrm{use Eq. (\ref{biguni})}} + e^{(22)}_{1} R_{1,2\dots L } \textrm{\huge)\normalsize}\\
=& \mathcal{F}_{1,2\dots L }.
\end{array}
\end{equation*}
Hence,
\begin{eqnarray*}
 \mathcal{F}_{2\dots L 1}R_{1,2\dots L } &=&  \mathcal{R}_{1,2\dots L} \overbrace{\mathcal{F}_{2 \dots L} \mathcal{F}_{1,2\dots L }}^{\textrm{use Eq. (\ref{ansatz1a})}}\\
&=&  \mathcal{R}_{1,2\dots L} \mathcal{F}_{1 \dots L},
\end{eqnarray*}
which is the required expression - thus verifying the factorization condition for all $\sigma \in S_L$.\\
\\
\textbf{Remark.} In \cite{WHE} a trigonometric free fermion six-vertex model was considered and a similar form of $F_{1\dots L}$ to that in \cite{MS} was proposed. This lead to a simplified form of the creation operators $B_{1\dots L}$ and $C_{1 \dots L}$. However, one can explicitly verify that the chosen form for $F_{1 \dots L}$ in the aforementioned work does not satisfy Eq. (\ref{defining}). The above analysis of this work offers a theoretical justification as to why the $F$-basis analysis in \cite{WHE} lead to positive results.
\section{The twisted monodromy operators}
\subsection{Similarity transforms and twisted operators}
In past works \cite{MS,TE, BOS,BOS1,CH1,CH2} which applied factorizing $F$-matrices as similarity transforms to simplify the monodromy operators, the choice of the form of the factorizing matrix was obvious. In this work we have two such choices, $F_{1\dots L }$ and $\mathcal{F}_{1 \dots L}$, which are related via the multiplication of a diagonal matrix (\ref{ansatz}). Hence, the similarity transforms are related via the multiplication of diagonal matrices,
\begin{equation}
F_{1\dots L} X_{1 \dots L} F^{-1}_{1 \dots L} = \mathcal{N}_{1\dots L} \mathcal{F}_{1\dots L} X_{1 \dots L} \mathcal{F}^{-1}_{1 \dots L} \mathcal{N}^{-1}_{1\dots L},
\label{useL}\end{equation}
for $X_{1 \dots L} \in \textrm{End}(V_1 \otimes \dots \otimes V_L)$. 

Being related as such through diagonal matrices, it is enough to compute the following non-trivial \textit{twisted} operators,
\begin{equation}
\tilde{X}_{1 \dots L} = \mathcal{F}_{1 \dots L} X_{1 \dots L} \mathcal{F}^{-1}_{1\dots L}.
\label{twisting}\end{equation}
The purpose of this section is to provide explicit forms for the twisted monodromy operators $\tilde{A}_{1\dots L}(\mu)$, $\tilde{B}_{1\dots L}(\mu)$, $\tilde{C}_{1\dots L}(\mu)$ and $\tilde{D}_{1\dots L}(\mu)$, without relying on any specific parameterization of the Boltzmann weights - relying only on the algebraic relations provided in Eqs. (\ref{one}-\ref{three}).
\subsection{Explicit expressions for the twisted monodromy operators} 
\begin{eqnarray}
\tilde{D}_{1\dots L}(\mu) &=& \otimes^L_{i=1} \left( \begin{array}{cc}
b_-(\mu,\xi_i)&  0\\
0&  a_-(\mu,\xi_i)
\end{array}\right)_{i}\label{D}\\
\tilde{C}_{1\dots L}(\mu) &=& \sum^L_{l=1}c_-(\mu,\xi_l)\otimes^{l-1}_{i=1} \left(  \begin{array}{cc} 
\frac{b_-(\mu,\xi_i)}{b_-(\xi_l,\xi_i)}  & 0  \\
 0  & a_-(\mu,\xi_i)
\end{array}\right)_i  e^{(12)}_l\label{C}\\
&& \otimes^L_{j=l+1}\left(  \begin{array}{cc} 
\frac{b_-(\mu,\xi_j)}{b_-(\xi_l,\xi_j)}  & 0  \\
 0  & \frac{a_-(\mu,\xi_j)}{a_-(\xi_l,\xi_j)} 
\end{array}\right)_j \nonumber\\
\tilde{B}_{1\dots L}(\mu) &=& \sum^L_{l=1}c_+(\mu,\xi_l)\otimes^{l-1}_{i=1} \left(  \begin{array}{cc} 
b_-(\mu,\xi_i)  & 0  \\
 0  & \frac{a_-(\mu,\xi_i)a_-(\xi_i,\xi_l)}{b_-(\xi_i,\xi_l)}
\end{array}\right)_i  e^{(21)}_l\label{B}\\
&& \otimes^L_{j=l+1}\left(  \begin{array}{cc} 
b_-(\mu,\xi_j)  & 0  \\
 0  & \frac{a_-(\mu,\xi_j)}{b_-(\xi_j,\xi_l)} 
\end{array}\right)_j \nonumber\\
\tilde{A}_{1\dots L}(\mu) &=& \otimes^L_{i=1} \left( \begin{array}{cc}
1&  0\\
0&  \frac{1}{b_-(\xi_i,\mu)}
\end{array}\right)_{i}+\tilde{B}_{1\dots L}(\mu)\tilde{D}^{-1}_{1\dots L}(\mu)\tilde{C}_{1\dots L}(\mu) \label{A}
\end{eqnarray}
To verify the above expressions we proceed slowly as there are many technical details, which we provide for the sake of completeness. The verification is executed via induction, as the form of $\mathcal{F}_{1\dots L}$ given in Eq. (\ref{ansatz1a}) is well suited for this method. Relying on induction, we begin by explicitly verifying that the $L=2$ case holds for each of the twisted operators. 
\subsection{The case L=2} Calculating the twisted monodromy operators directly using Eq. (\ref{twisting}) we obtain,
\begin{equation*}\begin{array}{lll}
\tilde{D}_{12}(\mu) &=& \left( \begin{array}{cc}
b_-(\mu,\xi_1)&  0\\
0&  a_-(\mu,\xi_1)
\end{array}\right)_{1} \otimes \left( \begin{array}{cc}
b_-(\mu,\xi_2)&  0\\
0&  a_-(\mu,\xi_2)
\end{array}\right)_{2}\\
&&+ \left( \begin{array}{cccc}
0&0&  0&0\\
0& 0 & 0 &0\\
0& \kappa^{(D)}_1 & 0 & 0\\
0 & 0 & 0 &0 \end{array}\right)_{12}\\
\textrm{where},\\
\kappa^{(D)}_1&=& c_+(\mu,\xi_1)c_-(\mu,\xi_2)b_-(\xi_1,\xi_2)+b_-(\mu,\xi_1)a_-(\mu,\xi_2)c_-(\xi_1,\xi_2)\\
&&- a_-(\mu,\xi_1)b_-(\mu,\xi_2)c_-(\xi_1,\xi_2),\\
\tilde{C}_{12}(\mu)& =& \left( \begin{array}{cccc}
0&\kappa^{(C)}_1& \frac{c_-(\mu,\xi_1)b_-(\mu,\xi_2)}{b_-(\xi_1,\xi_2)}&  0\\
0& 0 & 0 & \frac{c_-(\mu,\xi_1)a_-(\mu,\xi_2)}{a_-(\xi_1,\xi_2)}\\
0& 0 & 0 & \kappa^{(C)}_2\\
0 & 0 & 0 &0
\end{array}\right)_{12}\\
\textrm{where},\\
 \kappa^{(C)}_1 &=& c_-(\mu,\xi_2) - \frac{c_-(\mu,\xi_1)b_-(\mu,\xi_2) c_-(\xi_1,\xi_2)}{b_-(\xi_1,\xi_2)}\\
 \kappa^{(C)}_2 &=&\frac{c_-(\mu,\xi_1)a_-(\mu,\xi_2) c_-(\xi_1,\xi_2)+b_+(\mu,\xi_1)c_-(\mu,\xi_2) b_-(\xi_1,\xi_2) }{a_-(\xi_1,\xi_2)},\\
\tilde{B}_{12}(\mu)& =& \left( \begin{array}{cccc}
0&0& 0&  0\\
b_-(\mu,\xi_1)c_+(\mu,\xi_2)& 0 & 0 & 0\\
\kappa^{(B)}_1& 0 & 0 & 0\\
0 & \kappa^{(B)}_2 & \frac{a_-(\mu,\xi_1)c_+(\mu,\xi_2)a_-(\xi_1,\xi_2)}{b_-(\xi_1,\xi_2)} &0
\end{array}\right)_{12}\\
\textrm{where},\\
 \kappa^{(B)}_1 &=&b_-(\mu,\xi_1)c_+(\mu,\xi_2)c_-(\xi_1,\xi_2)+c_+(\mu,\xi_1)b_-(\xi_1,\xi_2) \\
 \kappa^{(B)}_2 &=& c_+(\mu,\xi_1)b_+(\mu,\xi_2)a_-(\xi_1,\xi_2)-\frac{a_-(\mu,\xi_1)c_+(\mu,\xi_2)a_-(\xi_1,\xi_2)c_-(\xi_1,\xi_2)}{b_-(\xi_1,\xi_2)},
\end{array}
\end{equation*}
\begin{equation*}\begin{array}{lll}
\tilde{A}_{12}(\mu)& =& \left( \begin{array}{cccc}
1&0& 0&  0\\
0&  \kappa^{(A)}_1 & \frac{c_-(\mu,\xi_1)c_+(\mu,\xi_2)}{b_-(\xi_1,\xi_2)} & 0\\
0& \kappa^{(A)}_2 & \kappa^{(A)}_3 & 0\\
0 & 0 & 0 &\kappa^{(A)}_4 
\end{array}\right)_{12}\\
\textrm{where},\\
 \kappa^{(A)}_1 &=& b_+(\mu,\xi_2) + \frac{c_-(\mu,\xi_1)c_+(\mu,\xi_2) c_-(\xi_2,\xi_1)}{ b_-(\xi_2,\xi_1)}\\
\kappa^{(A)}_2 &=& b_+(\mu,\xi_2)c_-(\xi_1,\xi_2)-b_+(\mu,\xi_1)c_-(\xi_1,\xi_2) - \frac{c_-(\mu,\xi_1)c_+(\mu,\xi_2) c^2_-(\xi_1,\xi_2)}{ b_-(\xi_1,\xi_2)}\\
 \kappa^{(A)}_3 &=& b_+(\mu,\xi_1) + \frac{c_-(\mu,\xi_1)c_+(\mu,\xi_2) c_-(\xi_1,\xi_2)}{ b_-(\xi_1,\xi_2)}\\
\kappa^{(A)}_4&=& b_+(\mu,\xi_1)b_+(\mu,\xi_2).
\end{array}
\end{equation*}
All the $\kappa$ entries of the above matrices do not immediately match the corresponding entries calculated from the $L=2$ expressions of Eqs. (\ref{D}-\ref{A}). In Table (\ref{tab:template}) we provide a summary  of the unitarity and Yang-Baxter relations that are required to simplify such non-trivial entries in order to bring the operators $\tilde{A}_{12}$, $\tilde{B}_{12}$, $\tilde{C}_{12}$ and $\tilde{D}_{12}$ in the form given by Eqs. (\ref{D}-\ref{A}). The technical details are quite cumbersome and thus have been deferred to Appendix \ref{APP5}. 

Having verified the case $L=2$, we now assume that Eqs. (\ref{D}-\ref{A}) hold up to some $(L-1)$, and we devote the remainder of this section verifying that the case for general $L$ holds. To facilitate this we are now required to construct recurrence relations for each of the four monodromy operators to proceed with the induction.
\begin{table}\begin{center}
\begin{tabular}{| c | c | c |}
  \hline                       
  Operator &  Y-B equations used & Unitarity equations used \\\hline
  $\tilde{A}_{12}$  & (\ref{YB1})(\ref{YB2}) (\ref{YB3}) (\ref{YB4})(\ref{YB5}) (\ref{YB6}) (\ref{YB7})&(\ref{uni1}) (\ref{uni2}) (\ref{uni3}) \\\hline
  $\tilde{B}_{12}$ & (\ref{YB1}) (\ref{YB4}) &(\ref{uni3}) (\ref{uni5}) \\\hline
 $\tilde{C}_{12}$  & (\ref{YB2}) (\ref{YB5})  & (\ref{uni3}) \\\hline
 $\tilde{D}_{12}$  & (\ref{YB6}) &  \\\hline 
\end{tabular}
\caption{Required Yang-Baxter and unitarity relations for $L=2$.}
\label{tab:template}\end{center}\end{table}
\subsection{Recurrence relations for the monodromy operators} We proceed by expressing the following decomposition of the monodromy matrix,
\begin{equation*}
\mathcal{T}_{a,1\dots L}(\mu) =  R_{a,2 \dots L}(\mu)R_{a1}(\mu),
\end{equation*}
as a matrix equation in auxiliary vector space $\mathcal{A}_a$ to obtain,
\begin{equation*}\begin{array}{ll}
& \left( \begin{array}{cc}
A_{1\dots L}(\mu)&  B_{1\dots L}(\mu) \\
C_{1\dots L}(\mu)&  D_{1\dots L}(\mu)
\end{array}\right)_{a}\\
=& \left(  \begin{array}{cc} 
\mathcal{I}_1\otimes A_{2\dots L}(\mu) & \mathcal{I}_1\otimes B_{2\dots L}(\mu)  \\
\mathcal{I}_1\otimes C_{2\dots L}(\mu) & \mathcal{I}_1\otimes D_{2\dots L}(\mu) 
\end{array}\right)_a\\
& \times \left( \begin{array}{cc}
 \left(\begin{array}{cc} 1 & 0 \\ 0 & b_+(\mu,\xi_1) \end{array}\right)_1 \otimes \mathcal{I}_{2\dots L} &  
 c_+(\mu,\xi_1) e^{(21)}_1 \otimes \mathcal{I}_{2\dots L}\\
 c_-(\mu,\xi_1) e^{(12)}_1 \otimes \mathcal{I}_{2\dots L} &  \ \left(\begin{array}{cc} b_-(\mu,\xi_1) & 0 \\ 0 & a_-(\mu,\xi_1) \end{array}\right)_1\otimes \mathcal{I}_{2\dots L}
\end{array}\right)_{a}.
\end{array}
\end{equation*}
Expanding each entry of the monodromy matrix in vector space $V_1$, we obtain the four recurrence relations,
\begin{eqnarray}
D_{1\dots L}(\mu) &=& \left(  \begin{array}{cc} 
b_-(\mu,\xi_1) D_{2\dots L}(\mu) & 0  \\
c_+(\mu,\xi_1) C_{2\dots L}(\mu) & a_-(\mu,\xi_1) D_{2\dots L}(\mu)
\end{array}\right)_1 \label{master1}\\
C_{1\dots L}(\mu) &=& \left(  \begin{array}{cc} 
C_{2\dots L}(\mu) & c_-(\mu,\xi_1) D_{2\dots L}(\mu)  \\
 0& b_+(\mu,\xi_1) C_{2\dots L}(\mu) 
\end{array}\right)_1\label{master2}\\
B_{1\dots L}(\mu) &=& \left(  \begin{array}{cc} 
b_-(\mu,\xi_1) B_{2\dots L}(\mu) & 0  \\
c_+(\mu,\xi_1) A_{2\dots L}(\mu) & a_-(\mu,\xi_1) B_{2\dots L}(\mu)
\end{array}\right)_1 \label{master3}\\
A_{1\dots L}(\mu) &=& \left(  \begin{array}{cc} 
A_{2\dots L}(\mu) & c_-(\mu,\xi_1) B_{2\dots L}(\mu)  \\
 0& b_+(\mu,\xi_1) A_{2\dots L}(\mu) 
\end{array}\right)_1\label{master4},
\end{eqnarray}
which are true by construction. In the work below we use the following relations extensively,
\begin{equation}\begin{array}{lll}
\mathcal{F}_{1\dots L} \mathcal{F}^{-1}_{2\dots L}&=&\mathcal{F}_{2\dots L} \mathcal{F}_{1,2\dots L} \mathcal{F}^{-1}_{2\dots L}\\
&=& \mathcal{F}_{2\dots L} \left(  \begin{array}{cc} 
\mathcal{I}_{2\dots L} & 0  \\
 C_{2\dots L}(\xi_1)&D_{2\dots L}(\xi_1) 
\end{array}\right)_1 \mathcal{F}^{-1}_{2\dots L}\\
&=&\left(  \begin{array}{cc} 
\mathcal{I}_{2\dots L}  & 0  \\
 \tilde{C}_{2\dots L}(\xi_1)& \tilde{D}_{2\dots L}(\xi_1) 
\end{array}\right)_1 ,
\end{array}
\label{FF1}\end{equation}
and,
\begin{equation}
\left( \mathcal{F}_{1\dots L} \mathcal{F}^{-1}_{2\dots L}\right)^{-1}=  \left(  \begin{array}{cc} 
\mathcal{I}_{2\dots L}  & 0  \\
-\tilde{D}^{-1}_{2\dots L}(\xi_1) \tilde{C}_{2\dots L}(\xi_1)&\tilde{D}^{-1}_{2\dots L}(\xi_1) 
\end{array}\right)_1 .
\label{FF2}\end{equation}
We now act on each of the recurrence relations (\ref{master1}-\ref{master4}) on the left with $\mathcal{F}_{1\dots L }$ and on the right with $\mathcal{F}^{-1}_{2\dots L}$. In doing so Eqs. (\ref{master1}-\ref{master4}) become the twisted recurrence relations,
\begin{eqnarray}
&& \tilde{D}_{1\dots L}(\mu) \mathcal{F}_{1\dots L} \mathcal{F}^{-1}_{2\dots L} \label{master1b} \\
&=& \left(  \begin{array}{cc} 
b_-(\mu,\xi_1) \tilde{D}_{2\dots L}(\mu) & 0  \\
\begin{array}{c} b_-(\mu,\xi_1) \tilde{C}_{2\dots L}(\xi_1)\tilde{D}_{2\dots L}(\mu) \\+ c_+(\mu,\xi_1)\tilde{D}_{2\dots L}(\xi_1) \tilde{C}_{2\dots L}(\mu) \end{array} & a_-(\mu,\xi_1) \tilde{D}_{2\dots L}(\xi_1)\tilde{D}_{2\dots L}(\mu)
\end{array}\right)_1 \nonumber \\
&& \tilde{C}_{1\dots L}(\mu) \mathcal{F}_{1\dots L} \mathcal{F}^{-1}_{2\dots L} \label{master2b} \\
&=& \left(  \begin{array}{cc} 
\tilde{C}_{2\dots L}(\mu) & c_-(\mu,\xi_1) \tilde{D}_{2\dots L}(\mu)  \\
 \tilde{C}_{2\dots L}(\xi_1)\tilde{C}_{2\dots L}(\mu) & \begin{array}{c} c_-(\mu,\xi_1) \tilde{C}_{2\dots L}(\xi_1)\tilde{D}_{2\dots L}(\mu)\\
+ b_+(\mu,\xi_1) \tilde{D}_{2\dots L}(\xi_1)\tilde{C}_{2\dots L}(\mu) \end{array}
\end{array}\right)_1 \nonumber\\
&& \tilde{B}_{1\dots L}(\mu) \mathcal{F}_{1\dots L} \mathcal{F}^{-1}_{2\dots L} \label{master3b} \\
&=& \left(  \begin{array}{cc} 
b_-(\mu,\xi_1)  \tilde{B}_{2\dots L}(\mu) & 0  \\
\begin{array}{c} b_-(\mu,\xi_1) \tilde{C}_{2\dots L}(\xi_1) \tilde{B}_{2\dots L}(\mu)\\+c_+(\mu,\xi_1)  \tilde{D}_{2\dots L}(\xi_1) \tilde{A}_{2\dots L}(\mu)\end{array} & a_-(\mu,\xi_1) \tilde{D}_{2\dots L}(\xi_1) \tilde{B}_{2\dots L}(\mu)
\end{array}\right)_1 \nonumber\\
&& \tilde{A}_{1\dots L}(\mu) \mathcal{F}_{1\dots L} \mathcal{F}^{-1}_{2\dots L}\label{master4b}  \\
&=& \left(  \begin{array}{cc} 
\tilde{A}_{2\dots L}(\mu) & c_-(\mu,\xi_1) \tilde{B}_{2\dots L}(\mu)  \\
 \tilde{C}_{2\dots L}(\xi_1) \tilde{A}_{2\dots L}(\mu) & \begin{array}{c} c_-(\mu,\xi_1) \tilde{C}_{2\dots L}(\xi_1) \tilde{B}_{2\dots L}(\mu)\\+b_+(\mu,\xi_1)  \tilde{D}_{2\dots L}(\xi_1) \tilde{A}_{2\dots L}(\mu)\end{array} 
\end{array}\right)_1.  \nonumber
\end{eqnarray}
Given these twisted recurrence relations we now apply them in our inductive proof. We begin with the operator $\tilde{D}_{1\dots L}$ as it is by far the easiest to verify.
\subsection{The operator $\tilde{D}_{1\dots L}$} 
\subsubsection{Applying algebraic operations} This is the simplest twisted operator one has to deal with. We begin by applying the Yang-Baxter algebra Eqs. (\ref{eq15}-\ref{eq16}) to Eq. (\ref{master1b}) in the following manner, 
\begin{equation*}\begin{array}{ll}
& \tilde{D}_{1\dots L}(\mu)\mathcal{F}_{1\dots L} \mathcal{F}^{-1}_{2\dots L}\\
=& \left(  \begin{array}{cc} 
b_-(\mu,\xi_1) \tilde{D}_{2\dots L}(\mu) & 0  \\
\underbrace{\begin{array}{c} b_-(\mu,\xi_1) \tilde{C}_{2\dots L}(\xi_1)\tilde{D}_{2\dots L}(\mu) \\+ c_+(\mu,\xi_1)\tilde{D}_{2\dots L}(\xi_1) \tilde{C}_{2\dots L}(\mu) \end{array}}_{\textrm{use Eq. (\ref{eq15})}} & a_-(\mu,\xi_1) \underbrace{\tilde{D}_{2\dots L}(\xi_1)\tilde{D}_{2\dots L}(\mu)}_{\textrm{use Eq. (\ref{eq16})}}
\end{array}\right)_1 \\
=& \left(  \begin{array}{cc} 
b_-(\mu,\xi_1) \tilde{D}_{2\dots L}(\mu) & 0  \\
 a_-(\mu,\xi_1) \tilde{D}_{2\dots L}(\mu)\tilde{C}_{2\dots L}(\xi_1)  & a_-(\mu,\xi_1) \tilde{D}_{2\dots L}(\mu)\tilde{D}_{2\dots L}(\xi_1)
\end{array}\right)_1,
\end{array}
\end{equation*}
and by taking into account Eq. (\ref{FF2}) we are able to rewrite the above expression as,
\begin{equation}\begin{array}{lll}
\tilde{D}_{1\dots L}(\mu)&=&   \left(  \begin{array}{cc} 
b_-(\mu,\xi_1) \tilde{D}_{2\dots L}(\mu) & 0  \\
0  & a_-(\mu,\xi_1) \tilde{D}_{2\dots L}(\mu)
\end{array}\right)_1 \\
&=&  \left(  \begin{array}{cc} 
b_-(\mu,\xi_1) & 0  \\
0  & a_-(\mu,\xi_1) 
\end{array}\right)_1 \otimes \tilde{D}_{2\dots L}(\mu). 
\end{array}
\label{twistrecD}\end{equation}
\subsubsection{Applying induction} 
If we assume that, 
\begin{equation*}
\tilde{D}_{2\dots L}(\mu) = \otimes^L_{i=2} \left( \begin{array}{cc}
b_-(\mu,\xi_i)&  0\\
0&  a_-(\mu,\xi_i)
\end{array}\right)_{i},
\end{equation*}
one can immediately show the equivalence between Eqs. (\ref{twistrecD}) and (\ref{D}) - hence completing the verification of Eq. (\ref{D}).
\subsection{The operator $\tilde{C}_{1\dots L}$} The remaining twisted operators require a great deal more work to verify. To facilitate the inductive process we are required to expand Eq. (\ref{C}) as an expression in vector space $V_1$ (as a $2 \times 2$ matrix) and perform algebraic manipulations on the four aforementioned matrix entries.
\subsubsection{Applying algebraic operations} We begin by applying the following necessary algebraic manipulations to Eq. (\ref{master2b}),
\begin{equation*}\begin{array}{lcl}
\tilde{C}_{1\dots L}(\mu)\mathcal{F}_{1\dots L} \mathcal{F}^{-1}_{2\dots L}&=& \left(  \begin{array}{cc} 
\tilde{C}_{2\dots L}(\mu) & c_-(\mu,\xi_1) \tilde{D}_{2\dots L}(\mu)  \\
 \tilde{C}_{2\dots L}(\xi_1)\tilde{C}_{2\dots L}(\mu) & \underbrace{\begin{array}{c} c_-(\mu,\xi_1) \tilde{C}_{2\dots L}(\xi_1)\tilde{D}_{2\dots L}(\mu)\\
+ b_+(\mu,\xi_1) \tilde{D}_{2\dots L}(\xi_1)\tilde{C}_{2\dots L}(\mu) \end{array}}_{\textrm{use Eq. (\ref{eq14})}}
\end{array}\right)_1 \\
&=&  \left(  \begin{array}{cc} 
\tilde{C}_{2\dots L}(\mu) & c_-(\mu,\xi_1) \tilde{D}_{2\dots L}(\mu)  \\
 \tilde{C}_{2\dots L}(\xi_1)\tilde{C}_{2\dots L}(\mu) & a_-(\mu,\xi_1) \tilde{C}_{2\dots L}(\mu)\tilde{D}_{2\dots L}(\xi_1)
\end{array}\right)_1 ,
\end{array}\end{equation*}
\begin{eqnarray}
\Rightarrow \tilde{C}_{1\dots L}(\mu)& =&\left(  \begin{array}{cc} 
\alpha^{(C)} & \beta^{(C)}  \\
 \gamma^{(C)} & \delta^{(C)}
\end{array}\right)_1\nonumber\\
\textrm{where},\nonumber\\
\alpha^{(C)} &=& \tilde{C}_{2\dots L}(\mu) - c_-(\mu,\xi_1)\tilde{D}_{2\dots L}(\mu)\tilde{D}^{-1}_{2\dots L}(\xi_1)\tilde{C}_{2\dots L}(\xi_1) \label{alphaC}\\
\beta^{(C)} &=&  c_-(\mu,\xi_1)\tilde{D}_{2\dots L}(\mu)\tilde{D}^{-1}_{2\dots L}(\xi_1)\label{betaC}\\
\gamma^{(C)} &=& \underbrace{\tilde{C}_{2\dots L}(\xi_1)\tilde{C}_{2\dots L}(\mu) - a_-(\mu,\xi_1)\tilde{C}_{2\dots L}(\mu)\tilde{C}_{2\dots L}(\xi_1)}_{\textrm{use Eq. (\ref{eq13})}} = 0\label{gammaC}\\
\delta^{(C)} &=&  a_-(\mu,\xi_1)\tilde{C}_{2\dots L}(\mu).\label{deltaC}
\end{eqnarray}
It is apparent that the above recurrence relation is much more complicated than what we obtained for $\tilde{D}_{1 \dots L}$. In order to verify the assumed form of $\tilde{C}_{1 \dots L}$ for general $L$ we expand Eq. (\ref{C}) as a $2\times 2$ matrix in vector space $V_1$, obtaining four expressions which are then shown to be equivalent to Eqs. (\ref{alphaC}-\ref{deltaC}).
\subsubsection{The assumed form of $\tilde{C}_{1\dots L}$ in vector space $V_1$}
\begin{eqnarray}
\alpha^{(C)}_a &=& \sum^L_{l=2}\frac{b_-(\mu,\xi_1)c_-(\mu,\xi_l)}{b_-(\xi_l,\xi_1)}\otimes^{l-1}_{i=2} \left(  \begin{array}{cc} 
\frac{b_-(\mu,\xi_i)}{b_-(\xi_l,\xi_i)}  & 0  \\
 0  & a_-(\mu,\xi_i)
\end{array}\right)_i  e^{(12)}_l\label{alphaCa}\\
&&\otimes^L_{j=l+1}\left(  \begin{array}{cc} 
\frac{b_-(\mu,\xi_j)}{b_-(\xi_l,\xi_j)}  & 0  \\
 0  & \frac{a_-(\mu,\xi_j)}{a_-(\xi_l,\xi_j)} 
\end{array}\right)_j \nonumber\\
\beta^{(C)}_a &=&  c_-(\mu,\xi_1)   \otimes^L_{j=2}\left(  \begin{array}{cc} 
\frac{b_-(\mu,\xi_j)}{b_-(\xi_1,\xi_j)}  & 0  \\
 0  & \frac{a_-(\mu,\xi_j)}{a_-(\xi_1,\xi_j)} ,
\end{array}\right)_j\label{betaCa}\\
\gamma^{(C)}_a &=& 0\label{gammaCa}\\
\delta^{(C)}_a &=&  \sum^L_{l=2}a_-(\mu,\xi_1)c_-(\mu ,\xi_l) \otimes^{l-1}_{i=2} \left(  \begin{array}{cc} 
\frac{b_-(\mu,\xi_i)}{b_-(\xi_l,\xi_i)}  & 0  \\
 0  & a_-(\mu,\xi_i)
\end{array}\right)_i  e^{(12)}_l\label{deltaCa}\\
&&\otimes^L_{j=l+1}\left(  \begin{array}{cc} 
\frac{b_-(\mu,\xi_j)}{b_-(\xi_l,\xi_j)}  & 0  \\
 0  & \frac{a_-(\mu,\xi_j)}{a_-(\xi_l,\xi_j)} 
\end{array}\right)_j \nonumber,
\end{eqnarray}
where the subscript $a$ in the above expressions stands for \textit{assumed}. For the derivation of these expressions (\ref{alphaCa}-\ref{deltaCa}) refer to Appendix \ref{APP6}. We now show the equivalence between the two sets of expressions, (\ref{alphaC}-\ref{deltaC}) and (\ref{alphaCa}-\ref{deltaCa}).
\subsubsection{Applying induction - further algebraic operations} 
To verify the equivalence of $\alpha^{(C)}$ and $\alpha^{(C)}_a$, we explicitly expand $\alpha^{(C)}$ as follows,
\begin{equation*}\begin{array}{lll}
\alpha^{(C)} & = & \tilde{C}_{2\dots L}(\mu)-c_-(\mu,\xi_1) \tilde{D}_{2\dots L}(\mu) \tilde{D}^{-1}_{2\dots L}(\xi_1) \tilde{C}_{2\dots L}(\xi_1)\\
&=& \sum^L_{l=2}\overbrace{\left\{c_-(\mu,\xi_l) - \frac{c_-(\mu,\xi_1) b_-(\mu,\xi_l)c_-(\xi_1,\xi_l) }{b_-(\xi_1,\xi_l) }\right\}}^{\textrm{use Eqs. (\ref{uni3}) and (\ref{YB2})}}  \\
&&\otimes^{l-1}_{i=2} \left(  \begin{array}{cc} 
\frac{b_-(\mu,\xi_i)}{b_-(\xi_l,\xi_i)}  & 0  \\
 0  & a_-(\mu,\xi_i)
\end{array}\right)_i  e^{(12)}_l \otimes^L_{j=l+1}\left(  \begin{array}{cc} 
\frac{b_-(\mu,\xi_j)}{b_-(\xi_l,\xi_j)}  & 0  \\
 0  & \frac{a_-(\mu,\xi_j)}{a_-(\xi_l,\xi_j)} 
\end{array}\right)_j \\
&=& \sum^L_{l=2}\frac{b_-(\mu,\xi_1)c_-(\mu,\xi_l)}{b_-(\xi_l,\xi_1)}  
\otimes^{l-1}_{i=2} \left(  \begin{array}{cc} 
\frac{b_-(\mu,\xi_i)}{b_-(\xi_l,\xi_i)}  & 0  \\
 0  & a_-(\mu,\xi_i)
\end{array}\right)_i e^{(12)}_l  \otimes^L_{j=l+1}\left(  \begin{array}{cc} 
\frac{b_-(\mu,\xi_j)}{b_-(\xi_l,\xi_j)}  & 0  \\
 0  & \frac{a_-(\mu,\xi_j)}{a_-(\xi_l,\xi_j)} 
\end{array}\right)_j,
\end{array}
\end{equation*}
thus showing the equivalence of $\alpha^{(C)}_a$ and $\alpha^{(C)}$.

Through elementary inspection one can show the equivalence between $\beta^{(C)}_a$ and $\beta^{(C)}$, $\gamma^{(C)}_a$ and $\gamma^{(C)}$ and $\delta^{(C)}_a$ and $\delta^{(C)}$ - thus verifying the validity of Eq. (\ref{C}). We now perform an exactly analogous analysis to the remaining monodromy operators.
\subsection{The operator $\tilde{B}_{1\dots L}$} 
\subsubsection{Applying algebraic operations} Applying the following necessary algebraic manipulations to Eq. (\ref{master3b}),
\begin{equation*}\begin{array}{ll}
& \tilde{B}_{1\dots L}(\mu) \mathcal{F}_{1\dots L}\mathcal{F}^{-1}_{2\dots L}\\
=& \left(  \begin{array}{cc} 
b_-(\mu,\xi_1)  \tilde{B}_{2\dots L}(\mu) & 0  \\
\underbrace{\begin{array}{c} b_-(\mu,\xi_1) \tilde{C}_{2\dots L}(\xi_1) \tilde{B}_{2\dots L}(\mu)\\+c_+(\mu,\xi_1)  \tilde{D}_{2\dots L}(\xi_1) \tilde{A}_{2\dots L}(\mu)\end{array}}_{\textrm{use Eq. (\ref{eq11})}} & a_-(\mu,\xi_1) \tilde{D}_{2\dots L}(\xi_1) \tilde{B}_{2\dots L}(\mu)
\end{array}\right)_1\\
=&  \left(  \begin{array}{cc} 
b_-(\mu,\xi_1)  \tilde{B}_{2\dots L}(\mu) & 0  \\
\begin{array}{c} b_+(\mu,\xi_1) \tilde{B}_{2\dots L}(\mu) \tilde{C}_{2\dots L}(\xi_1)\\+c_+(\mu,\xi_1)  \tilde{D}_{2\dots L}(\mu) \tilde{A}_{2\dots L}(\xi_1)\end{array} & a_-(\mu,\xi_1) \tilde{D}_{2\dots L}(\xi_1) \tilde{B}_{2\dots L}(\mu)
\end{array}\right)_1,
\end{array}\end{equation*}
\begin{eqnarray}
\Rightarrow \tilde{B}_{1\dots L}(\mu)&=&  \left(  \begin{array}{cc} 
\alpha^{(B)} & 0  \\
\gamma^{(B)} &  \delta^{(B)}
\end{array}\right)_1 \nonumber\\
\textrm{where}, \nonumber\\
\alpha^{(B)}&=& b_-(\mu,\xi_1)  \tilde{B}_{2\dots L}(\mu)\label{alphaB}\\
\gamma^{(B)} &=& b_+(\mu,\xi_1) \tilde{B}_{2\dots L}(\mu) \tilde{C}_{2\dots L}(\xi_1)+c_+(\mu,\xi_1)  \tilde{D}_{2\dots L}(\mu) \tilde{A}_{2\dots L}(\xi_1)\label{betaB}\\
 &&-a_-(\mu,\xi_1) \tilde{D}_{2\dots L}(\xi_1) \tilde{B}_{2\dots L}(\mu)\tilde{D}^{-1}_{2\dots L}(\xi_1)\tilde{C}_{2\dots L}(\xi_1) \nonumber\\
 \delta^{(B)} &=& a_-(\mu,\xi_1) \tilde{D}_{2\dots L}(\xi_1) \tilde{B}_{2\dots L}(\mu)\tilde{D}^{-1}_{2\dots L}(\xi_1)\label{gammaB}.
\end{eqnarray}
In order to verify the assumed form of $\tilde{B}_{1 \dots L}$ for general $L$ we expand Eq. (\ref{B}) as a $2\times 2$ matrix in vector space $V_1$, obtaining three non zero entries which are then shown to be equivalent to Eqs. (\ref{alphaB}-\ref{gammaB}).
\subsubsection{The assumed form of $\tilde{B}_{1\dots L}$ in vector space $V_1$}
\begin{eqnarray}
\alpha^{(B)}_a &=& \sum^L_{l=2}b_-(\mu,\xi_1)c_+(\mu,\xi_l)\otimes^{l-1}_{i=2} \left(  \begin{array}{cc} 
b_-(\mu,\xi_i)  & 0  \\
 0  & \frac{a_-(\mu,\xi_i)a_-(\xi_i,\xi_l)}{b_-(\xi_i,\xi_l)}
\end{array}\right)_i \label{alphaBa} \\
&&e^{(21)}_l \otimes^L_{j=l+1}\left(  \begin{array}{cc} 
b_-(\mu,\xi_j)  & 0  \\
 0  & \frac{a_-(\mu,\xi_j)}{b_-(\xi_j,\xi_l)} 
\end{array}\right)_j  \nonumber  \\
\gamma^{(B)}_a &=&  c_+(\mu,\xi_1)   \otimes^L_{j=2}\left(  \begin{array}{cc} 
b_-(\mu,\xi_j)  & 0  \\
 0  & \frac{a_-(\mu,\xi_j)}{b_-(\xi_j,\xi_l)} ,
\end{array}\right)_j \label{betaBa}\\
\delta^{(B)}_a &=& \sum^L_{l=2} \frac{a_-(\mu,\xi_1)c_+(\mu,\xi_l)a_-(\xi_1,\xi_l)}{b_-(\xi_1,\xi_l)}\otimes^{l-1}_{i=2} \left(  \begin{array}{cc} 
b_-(\mu,\xi_i)  & 0  \\
 0  & \frac{a_-(\mu,\xi_i)a_-(\xi_i,\xi_l)}{b_-(\xi_i,\xi_l)}
\end{array}\right)_i  \label{gammaBa}  \\
&&e^{(21)}_l \otimes^L_{j=l+1}\left(  \begin{array}{cc} 
b_-(\mu,\xi_j)  & 0  \\
 0  & \frac{a_-(\mu,\xi_j)}{b_-(\xi_j,\xi_l)} 
\end{array}\right)_j ,\nonumber
\end{eqnarray}
where the subscript $a$ in the above expressions stands for \textit{assumed}. For the derivation of these expressions (\ref{alphaBa}-\ref{gammaBa}) refer to Appendix \ref{APP7}. We now show the equivalence between the two sets of expressions, (\ref{alphaB}-\ref{gammaB}) and (\ref{alphaBa}-\ref{gammaBa}).
\subsubsection{Applying induction - further algebraic operations} 
Through elementary inspection one can show the equivalence between $\alpha^{(B)}_a$ and $\alpha^{(B)}$, and $\delta^{(B)}_a$ and $\delta^{(B)}$. For $\gamma^{(B)}_a$ and $\gamma^{(B)}$ we proceed by explicitly expanding the expression for $\gamma^{(B)}$. Assuming that,
\begin{equation*}
 \tilde{A}_{2 \dots L}(\xi_1) = \otimes^L_{i=2} \left( \begin{array}{cc}
1&  0\\
0&  \frac{1}{b_-(\xi_i,\xi_1)}
\end{array}\right)_{i}+\tilde{B}_{2 \dots L}(\xi_1)\tilde{D}^{-1}_{2 \dots L}(\xi_1)\tilde{C}_{2 \dots L}(\xi_1) ,
\end{equation*}
we obtain the following form for $\gamma^{(B)}$,
\begin{equation*}\begin{array}{l}
 \gamma^{(B)} = c_+(\mu,\xi_1)\otimes^L_{i=2} \left( \begin{array}{cc}
1&  0\\
0&  \frac{1}{b_-(\xi_i,\xi_1)}
\end{array}\right)_{i}  \tilde{D}_{2 \dots L}(\mu) +\left\{ b_+(\mu,\xi_1) \tilde{B}_{2 \dots L}(\mu) \right. \\
   +c_+(\mu,\xi_1)  \tilde{D}_{2 \dots L}(\mu) \tilde{B}_{2 \dots L}(\xi_1)\tilde{D}^{-1}_{2 \dots L}(\xi_1)\left.-a_-(\mu,\xi_1) \tilde{D}_{2 \dots L}(\xi_1) \tilde{B}_{2 \dots L}(\mu)\tilde{D}^{-1}_{2 \dots L}(\xi_1) \right\} \tilde{C}_{2 \dots L}(\xi_1),
\end{array}\end{equation*}
where,
\begin{equation*}\begin{array}{ll}
& b_+(\mu,\xi_1) \tilde{B}_{2 \dots L}(\mu) +c_+(\mu,\xi_1)  \tilde{D}_{2 \dots L}(\mu) \tilde{B}_{2 \dots L}(\xi_1)\tilde{D}^{-1}_{2 \dots L}(\xi_1)\\
&-a_-(\mu,\xi_1) \tilde{D}_{2 \dots L}(\xi_1) \tilde{B}_{2 \dots L}(\mu)\tilde{D}^{-1}_{2 \dots L}(\xi_1) \\
=& \sum^L_{l=2}\textrm{\huge\{ \normalsize} \overbrace{b_+(\mu,\xi_1)c_+(\mu,\xi_l)}^{\textrm{use Eq. (\ref{YB4})}} + \frac{c_+(\mu,\xi_1)a_-(\mu,\xi_l)c_+(\xi_1,\xi_l)}{b_-(\xi_1,\xi_l)} - \frac{a_-(\mu,\xi_1) c_+(\mu,\xi_l)a_-(\xi_1,\xi_l) }{b_-(\xi_1,\xi_l) }\textrm{\huge\} \normalsize} \\
& \otimes^{l-1}_{i=2} \left(  \begin{array}{cc} 
b_-(\mu,\xi_i)  & 0  \\
 0  & \frac{a_-(\mu,\xi_i) a_-(\xi_i,\xi_l)}{b_-(\xi_i,\xi_l)}
\end{array}\right)_i e^{(21)}_l \otimes^L_{j=l+1}\left(  \begin{array}{cc} 
b_-(\mu,\xi_j)  & 0  \\
 0  & \frac{a_-(\mu,\xi_j)}{b_-(\xi_j,\xi_l)} 
\end{array}\right)_j\\
=&0.
\end{array}
\end{equation*}
Hence,
\begin{equation*}
 \gamma^{(B)} = c_+(\mu,\xi_1)   \otimes^L_{i=2}\left(  \begin{array}{cc} 
b_-(\mu,\xi_i)  & 0  \\
 0  & \frac{a_-(\mu,\xi_i)}{b_-(\xi_i,\xi_l)} ,
\end{array}\right)_i,
\end{equation*}
ultimately displaying the validity of Eq. (\ref{B}).
\subsection{The operator $\tilde{A}_{1\dots L}$}
\subsubsection{Applying algebraic operations}
This twisted operator deserves special mention as it involves a number of complicated steps which easily outweigh anything witnessed from the previous operators. We start by applying the following necessary algebraic manipulations to Eq. (\ref{master4b}),
\begin{equation*}\begin{array}{rll} 
\tilde{A}_{1\dots L}(\mu) \mathcal{F}_{1\dots L} \mathcal{F}^{-1}_{2\dots L}&=& \left(  \begin{array}{cc} 
\tilde{A}_{2\dots L}(\mu) & c_-(\mu,\xi_1) \tilde{B}_{2\dots L}(\mu)  \\
 \tilde{C}_{2\dots L}(\xi_1) \tilde{A}_{2\dots L}(\mu) & \underbrace{\begin{array}{c} c_-(\mu,\xi_1) \tilde{C}_{2\dots L}(\xi_1) \tilde{B}_{2\dots L}(\mu)\\+b_+(\mu,\xi_1)  \tilde{D}_{2\dots L}(\xi_1) \tilde{A}_{2\dots L} (\mu) \end{array}}_{\textrm{use Eq. (\ref{eq10})}} 
\end{array}\right)_1\\
&=&  \left(  \begin{array}{cc} 
\tilde{A}_{2\dots L}(\mu) & c_-(\mu,\xi_1) \tilde{B}_{2\dots L}(\mu)  \\
 \tilde{C}_{2\dots L}(\xi_1) \tilde{A}_{2\dots L}(\mu) & \begin{array}{c} b_+(\mu,\xi_1) \tilde{A}_{2\dots L}(\mu) \tilde{D}_{2\dots L}(\xi_1)\\+c_+(\mu,\xi_1)  \tilde{C}_{2\dots L}(\mu) \tilde{B}_{2\dots L}(\xi_1)\end{array}
\end{array}\right)_1,\\
\Rightarrow \tilde{A}_{1\dots L}(\mu)&=& \left(  \begin{array}{cc} 
\alpha^{(A)} & \beta^{(A)}  \\
\gamma^{(A)} &  \delta^{(A)}
\end{array}\right)_1
\end{array}\end{equation*}
\begin{eqnarray}
\textrm{where},\nonumber\\
\alpha^{(A)}&=& \tilde{A}_{2\dots L}(\mu) - c_-(\mu,\xi_1)  \tilde{B}_{2\dots L} (\mu) \tilde{D}^{-1}_{2\dots L}(\xi_1)\tilde{C}_{2\dots L}(\xi_1)\label{alphaA}\\
\beta^{(A)} &=& c_-(\mu,\xi_1)  \tilde{B}_{2\dots L}(\mu)\tilde{D}^{-1}_{2\dots L}(\xi_1) \label{betaA}\\
 \gamma^{(A)} &=&\tilde{C}_{2\dots L}(\xi_1) \tilde{A}_{2\dots L}(\mu)- b_+(\mu,\xi_1) \tilde{A}_{2\dots L}(\mu)\tilde{C}_{2\dots L}(\xi_1)\label{gammaA}\\
&& - c_+(\mu,\xi_1)\tilde{C}_{2\dots L}(\mu)\tilde{B}_{2\dots L}(\xi_1) \tilde{D}^{-1}_{2\dots L} (\xi_1) \tilde{C}_{2\dots L}(\xi_1)\nonumber\\
\delta^{(A)} &=& b_+(\mu,\xi_1)\tilde{A}_{2\dots L}(\mu) +c_+(\mu,\xi_1)\tilde{C}_{2\dots L}(\mu)\tilde{B}_{2\dots L} (\xi_1)\tilde{D}^{-1}_{2\dots L}(\xi_1)\label{deltaA}.
\end{eqnarray}
\subsubsection{The assumed form of $\tilde{A}_{1\dots L}$ in vector space $V_1$} As the expressions associated with $\tilde{A}_{1 \dots L}$ are quite long, it is wise to introduce the following auxiliary notation:
\begin{equation*}\begin{array}{lll}
\Omega_0(\mu) &=&  \otimes^L_{i=2} \left( \begin{array}{cc}
1&  0\\
0&  \frac{1}{b_-(\xi_i,\mu)}
\end{array}\right)_{i} \label{Omega0}\\
\Omega_1 (\mu)&=& \otimes^L_{i=2}\left(  \begin{array}{cc} 
 \frac{b_-(\mu,\xi_{i})}{b_-(\xi_{1},\xi_{i})}  & 0  \\
 0  & \frac{a_-(\mu,\xi_{i})}{a_-(\xi_{1},\xi_{i})b_-(\xi_{i},\xi_{1})} \end{array}\right)_{i}\\
\Omega^{(l)}_2(\mu) &=& \otimes^{l-1}_{i=2} \left(  \begin{array}{cc} 
 \frac{b_-(\mu,\xi_{i})}{b_-(\xi_{l},\xi_{i})} & 0  \\
 0  & \frac{a_-(\mu,\xi_{i})a_-(\xi_{i},\xi_{l})}{b_-(\xi_{i},\xi_{l})}
\end{array}\right)_{i} e^{(22)}_{l} \otimes^L_{j=l+1}\left(  \begin{array}{cc} 
 \frac{b_-(\mu,\xi_{j})}{b_-(\xi_{l},\xi_{j})}  & 0  \\
 0  & \frac{a_-(\mu,\xi_{j})}{a_-(\xi_{l},\xi_{j})b_-(\xi_{j},\xi_{l})} \end{array}\right)_{j}\\
 \Omega^{(l)}_3(\mu) &=& \otimes^{l-1}_{i=2} \left(  \begin{array}{cc} 
\frac{b_-(\mu,\xi_{i})}{b_-(\xi_{1},\xi_{i})}  & 0  \\
 0  & \frac{a_-(\mu,\xi_{i})a_-(\xi_i,\xi_{l})}{a_-(\xi_{1},\xi_{i})b_-(\xi_i,\xi_{l})}
\end{array}\right)_{i}  e^{(21)}_{l}  \otimes^L_{j=l+1}\left(  \begin{array}{cc} 
\frac{b_-(\mu,\xi_{j})}{b_-(\xi_{1},\xi_{j})}  & 0  \\
 0  & \frac{a_-(\mu,\xi_{j})}{a_-(\xi_{1},\xi_{j})b_-(\xi_{j},\xi_{l})} 
\end{array}\right)_{j}\\
\Omega^{(l)}_4(\mu) &=&  \otimes^{l-1}_{i=2} \left(  \begin{array}{cc} 
\frac{b_-(\mu,\xi_{i})}{b_-(\xi_{l},\xi_{i})}  & 0  \\
 0  & \frac{a_-(\mu,\xi_{i})}{b_-(\xi_{i},\xi_{1})}
\end{array}\right)_{i}  e^{(12)}_{l} \otimes^L_{j=l+1}\left(  \begin{array}{cc} 
\frac{b_-(\mu,\xi_{j})}{b_-(\xi_{l},\xi_{j})}  & 0  \\
 0  & \frac{a_-(\mu,\xi_{j})}{a_-(\xi_{l},\xi_{j})b_-(\xi_{j},\xi_{1})} 
\end{array}\right)_{j}\\
\Omega^{(l_1 l_2)}_5(\mu) &=& \otimes^{l_2-1}_{i=2} \left(  \begin{array}{cc} 
\frac{b_-(\mu,\xi_i)}{b_-(\xi_{l_2},\xi_i)}  & 0  \\
 0  & \frac{a_-(\mu,\xi_i)a_-(\xi_{i},\xi_{l_1})}{b_-(\xi_{i},\xi_{l_1})}
\end{array}\right)_{i}  e^{(12)}_{l_2} \otimes^{l_1-1}_{j=l_2+1} \left(  \begin{array}{cc} 
\frac{b_-(\mu,\xi_{j})}{b_-(\xi_{l_2},\xi_{j})}  & 0  \\
 0  & \frac{a_-(\mu,\xi_{j})a_-(\xi_j,\xi_{l_1})}{a_-(\xi_{l_2},\xi_{j})b_-(\xi_j,\xi_{l_1})}
\end{array}\right)_{j}  \\
&& e^{(21)}_{l_1} \otimes^L_{k=l_1+1}\left(  \begin{array}{cc} 
\frac{b_-(\mu,\xi_{k})}{b_-(\xi_{l_2},\xi_{k})}  & 0  \\
 0  & \frac{a_-(\mu,\xi_{k})}{a_-(\xi_{l_2},\xi_{k})b_-(\xi_{k},\xi_{l_1})} 
\end{array}\right)_{k}
\end{array}\end{equation*}
\begin{equation*}\begin{array}{lll}
 \Omega^{(l_1 l_2)}_6(\mu) &=& \otimes^{l_1-1}_{i=2} \left(  \begin{array}{cc} 
\frac{b_-(\mu,\xi_i)}{b_-(\xi_{l_2},\xi_i)}  & 0  \\
 0  & \frac{a_-(\mu,\xi_i)a_-(\xi_{i},\xi_{l_1})}{b_-(\xi_{i},\xi_{l_1})}
\end{array}\right)_{i} e^{(21)}_{l_1} \otimes^{l_2-1}_{j=l_1+1} \left(  \begin{array}{cc} 
\frac{b_-(\mu,\xi_{j})}{b_-(\xi_{l_2},\xi_{j})}  & 0  \\
 0  & \frac{a_-(\mu,\xi_{j})}{b_-(\xi_{j},\xi_{l_1})}
\end{array}\right)_{j}  \\
&&e^{(12)}_{l_2} \otimes^L_{k=l_2+1}\left(  \begin{array}{cc} 
\frac{b_-(\mu,\xi_{k})}{b_-(\xi_{l_2},\xi_{k})}  & 0  \\
 0  & \frac{a_-(\mu,\xi_{k})}{a_-(\xi_{l_2},\xi_{k})b_-(\xi_{k},\xi_{l_1})} 
\end{array}\right)_{k} \\
\mathcal{K}^{(l)}_{1} &=& \frac{b_-(\mu,\xi_{1})c_+(\mu,\xi_l)c_-(\mu,\xi_l)}{b_-(\xi_{l},\xi_{1})b_-(\mu,\xi_l)}\\
\mathcal{K}^{(l)}_{2} &=& \frac{a_-(\mu,\xi_{1})a_-(\xi_{1},\xi_{l})c_+(\mu,\xi_l)c_-(\mu,\xi_l)}{b_-(\xi_{1},\xi_{l})b_-(\mu,\xi_l)}\\
\mathcal{K}^{(l_1 l_2)}_{1} &=& \frac{b_-(\mu,\xi_1)c_+(\mu,\xi_{l_1}) c_-(\mu,\xi_{l_2})}{b_-(\xi_{l_2} ,\xi_1)b_-(\xi_{l_2},\xi_{l_1})}\\
\mathcal{K}^{(l_1 l_2)}_{2} &=& \frac{a_-(\mu,\xi_1)a_-(\xi_1,\xi_{l_1})c_+(\mu,\xi_{l_1}) c_-(\mu,\xi_{l_2})}{b_-(\xi_{1},\xi_{l_1})b_-(\xi_{l_2},\xi_{l_1})}.
\end{array}\end{equation*}
Using the above notation we obtain,
\begin{eqnarray}
 \alpha^{(A)}_a &=& \Omega_0 (\mu)+ \sum^L_{l=2}\mathcal{K}^{(l)}_1 \Omega^{(l)}_2(\mu) +  \sum_{2 \le l_2 < l_1 \le L}  \mathcal{K}^{(l_1 l_2)}_{1} \Omega^{(l_1 l_2)}_5(\mu)   +   \sum_{2 \le l_1 < l_2 \le L}  \mathcal{K}^{(l_1 l_2)}_{1} \Omega^{(l_1 l_2)}_6(\mu)    \label{alphaAa} \\
&&\nonumber\\
\beta^{(A)}_a &=&  \sum^L_{l=2}\frac{c_-(\mu,\xi_{1})c_+(\mu,\xi_{l})}{b_-(\xi_{1},\xi_{l})} \Omega^{(l)}_3(\mu) \label{betaAa}\\
 \gamma^{(A)}_a &=&  \sum^L_{l=2}\frac{c_+(\mu,\xi_{1})c_-(\mu,\xi_{l})}{b_-(\xi_{l},\xi_{1})} \Omega^{(l)}_4 (\mu)\label{gammaAa}\\
 \delta^{(A)}_a &=&  \frac{1}{b_-(\xi_1,\mu)}\Omega_0(\mu)+ \frac{c_+(\mu,\xi_1)c_-(\mu,\xi_1)}{b_-(\mu,\xi_1)}\Omega_1(\mu)+ \sum^L_{l=2}\mathcal{K}^{(l)}_2 \Omega^{(l)}_2(\mu) \label{deltaAa} \\
&&+   \sum_{2 \le l_2 < l_1 \le L} \mathcal{K}^{(l_1 l_2)}_{2} \Omega^{(l_1 l_2)}_5(\mu)+  \sum_{2 \le l_1 < l_2 \le L} \mathcal{K}^{(l_1 l_2)}_{2} \Omega^{(l_1 l_2)}_6 (\mu),\nonumber
\end{eqnarray}
where the subscript $a$ in the above expressions stands for \textit{assumed}. For the derivation of these expressions (\ref{alphaAa}-\ref{deltaAa}) refer to Appendix \ref{APP8}. We now show the equivalence between the two sets of expressions, (\ref{alphaA}-\ref{deltaA}) and (\ref{alphaAa}-\ref{deltaAa}).
\subsubsection{Applying induction - further algebraic operations} 
We begin this section by showing the equivalence of $\alpha^{(A)}$ and $\alpha^{(A)}_a$. From Eq. (\ref{alphaA}) we have,
\begin{equation*}
\alpha^{(A)}=\Omega_0(\mu) + \sum^L_{l_1 l_2 = 1} \mathcal{G}^{(l_1 l_2)}_{\alpha},
\end{equation*}
where,
\begin{equation}
\sum^L_{l_1 l_2 = 1} \mathcal{G}^{(l_1 l_2)}_{\alpha}=\tilde{B}_{2 \dots L}(\mu)\left\{ \tilde{D}^{-1}_{2 \dots L}(\mu)\tilde{C}_{2 \dots L}(\mu)  - c_-(\mu,\xi_1)  \tilde{D}^{-1}_{2 \dots L}(\xi_1)\tilde{C}_{2 \dots L}(\xi_1)\right\} .
\label{BC}
\end{equation}
We immediately notice the term $\Omega_0(\mu)$ in the above form of $\alpha^{(A)}$ matches with that in $\alpha^{(A)}_a$. The multiplication of the $\tilde{B}_{2\dots L}$ and $\tilde{C}_{2\dots L}$ operators in Eq. (\ref{BC}) results in a double summation, $\sum^L_{l_1 l_2 = 1}$, where the relevant sets of values of $l_1$ and $l_2$ are $l_1 = l_2 = l$, $l_1 > l_2$ and $l_1 < l_2$. Beginning with the case $l_1 = l_2 = l$ we obtain,
\begin{equation*}
\begin{array}{lll}
\sum^L_{ \genfrac{}{}{0pt}{}{l_1 l_2 = 1}{l_1 = l_2}} \mathcal{G}^{(l_1 l_2)}_{\alpha}&=& \sum^L_{l=2} c_+(\mu,\xi_{l})\overbrace{\left\{\frac{c_-(\mu,\xi_l)}{b_-(\mu,\xi_l)} - \frac{c_-(\mu,\xi_1)c_-(\xi_1,\xi_l)}{b_-(\xi_1,\xi_l)} \right\} }^{\textrm{use Eq. (\ref{uni3}) and (\ref{YB2})}}\Omega^{(l)}_2(\mu)\\
&=&  \sum^L_{l=2} \frac{b_-(\mu,\xi_1)c_+(\mu,\xi_l)c_-(\mu,\xi_l)}{b_-(\xi_l,\xi_1)b_-(\mu,\xi_l)} \Omega^{(l)}_2(\mu),
\end{array}
\end{equation*}
which verifies that the $l_1 = l_2 = l$ cases of both $\alpha^{(A)}$ and $\alpha^{(A)}_a$ are equivalent.

Moving on to the $l_1 > l_2$ case in Eq. (\ref{BC}) we obtain,
\begin{equation*}
\begin{array}{lll}
\sum^L_{ \genfrac{}{}{0pt}{}{l_1 l_2 = 1}{l_1 > l_2}} \mathcal{G}^{(l_1 l_2)}_{\alpha}
&=&  \sum_{2 \le l_2 < l_1 \le L}\frac{c_+(\mu,\xi_{l_1})}{b_-(\xi_{l_2},\xi_{l_1})} \overbrace{  \left\{c_-(\mu,\xi_{l_2})-\frac{c_-(\mu,\xi_{1}) b_-(\mu,\xi_{l_2}) c_-(\xi_1,\xi_{l_2})}{b_-(\xi_1,\xi_{l_2})}  \right\}}^{\textrm{use Eq. (\ref{uni3}) and (\ref{YB2})}} \Omega^{(l_1 l_2)}_5(\mu)\\
&=&  \sum_{2 \le l_2 < l_1 \le L}\frac{c_+(\mu,\xi_{l_1})b_-(\mu,\xi_1)c_-(\mu,\xi_{l_2})}{b_-(\xi_{l_2},\xi_{1})b_-(\xi_{l_2},\xi_{1_1})}   \Omega^{(l_1 l_2)}_5(\mu),
\end{array}
\end{equation*}
which verifies that the $l_1 > l_2$ cases of both $\alpha^{(A)}$ and $\alpha^{(A)}_a$ are equivalent.

Verifying the equivalence between the $l_1 < l_2$ terms of $\alpha^{(A)}$ and $\alpha^{(A)}_a$ is exactly analogous to verifying the $l_1 > l_2$ case, the only difference being that $\Omega^{(l_1 l_2)}_5(\mu)$ is replaced by $\Omega^{(l_1 l_2)}_6(\mu)$ in the calculations. Hence the overall expressions for both $\alpha^{(A)}$ and $\alpha^{(A)}_a$ are shown to be equivalent.

Considering $\beta^{(A)}$ and $\beta^{(A)}_a$, one can show their equivalence through elementary inspection without additional algebraic manipulation. For $\gamma^{(A)}$ and $\gamma^{(A)}_a$ we consider Eq. (\ref{gammaA}),
\begin{equation*}
 \begin{array}{lll}
   \gamma^{(A)} &=&\overbrace{\tilde{C}_{2 \dots L}(\xi_1) \tilde{A}_{2 \dots L}(\mu)- b_+(\mu,\xi_1) \tilde{A}_{2 \dots L} (\mu)\tilde{C}_{2 \dots L} (\xi_1)}^{\textrm{use Eq. (\ref{eq9})}} \\
&&- c_+(\mu,\xi_1)\tilde{C}_{2 \dots L}(\mu)\tilde{B}_{2 \dots L}(\xi_1) \tilde{D}^{-1}_{2 \dots L}(\xi_1)\tilde{C}_{2 \dots L}(\xi_1)\\
&=& c_+(\mu,\xi_1)\tilde{C}_{2 \dots L}(\mu) \left\{\tilde{A}_{2 \dots L}(\xi_1) - \tilde{B}_{2 \dots L}(\xi_1) \tilde{D}^{-1}_{2 \dots L}(\xi_1) \tilde{C}_{2 \dots L}(\xi_1)  \right\}\\
&=& c_+(\mu,\xi_1)\tilde{C}_{2 \dots L}(\mu) \Omega_0(\xi_1)\\
&=&  \sum^N_{l=2}\frac{c_+(\mu,\xi_1)c_-(\mu,\xi_l)}{b_-(\xi_l,\xi_1)}\Omega^{(l)}_4(\mu),
 \end{array}
\end{equation*}
thus verifying the equivalence of $\gamma^{(A)}$ and $\gamma^{(A)}_a$. Finally, for $\delta^{(A)}$ and $\delta^{(A)}_a$, we consider Eq. (\ref{deltaA}),
\begin{equation*}
 \begin{array}{lll}
  \delta^{(A)} &=& b_+(\mu,\xi_1)\tilde{A}_{2 \dots L}(\mu) +c_+(\mu,\xi_1)\overbrace{\tilde{C}_{2 \dots L}(\mu)\tilde{B}_{2 \dots L}(\xi_1)}^{\textrm{use Eq. (\ref{eq6})}}\tilde{D}^{-1}_{2 \dots L}(\xi_1)\\
&=&\overbrace{\left\{ b_+(\mu,\xi_1) - \frac{c_+(\mu,\xi_1)c_-(\mu,\xi_1)}{b_-(\mu,\xi_1)}\right\}}^{\textrm{use Eq. (\ref{uni1}) and (\ref{uni3})}} \Omega_0(\mu)+ \frac{c_+(\mu,\xi_1)c_-(\mu,\xi_1)}{b_-(\mu,\xi_1)} \overbrace{\Omega_0(\xi_1)\tilde{D}_{2 \dots L}(\mu)\tilde{D}^{-1}_{2 \dots L}(\xi_1)}^{\Omega_1(\mu)}\\
&&+ \frac{c_+(\mu,\xi_1)b_+(\mu,\xi_1)}{b_-(\mu,\xi_1)}\tilde{B}_{2 \dots L}(\xi_1)\tilde{C}_{2 \dots L}(\mu)\tilde{D}^{-1}_{2 \dots L}(\xi_1)\\
&& +\frac{c_+(\mu,\xi_1)c_-(\mu,\xi_1)}{b_-(\mu,\xi_1)} \tilde{B}_{2 \dots L} (\xi_1) \tilde{D}^{-1}_{2 \dots L}(\xi_1) \tilde{C}_{2 \dots L}(\xi_1) \tilde{D}_{2 \dots L}(\mu) \tilde{D}^{-1}_{2 \dots L}(\xi_1) \\
&&+ \left\{ b_+(\mu,\xi_1) - \frac{c_+(\mu,\xi_1)c_-(\mu,\xi_1)}{b_-(\mu,\xi_1)} \right\}\tilde{B}_{2 \dots L}(\mu)\tilde{D}^{-1}_{2 \dots L}(\mu)\tilde{C}_{2 \dots L}(\mu)\\
&=&\frac{1}{b_-(\xi_1,\mu)}\Omega_0(\mu)+ \frac{c_+(\mu,\xi_1)c_-(\mu,\xi_1)}{b_-(\mu,\xi_1)}\Omega_1(\mu)+ \sum^L_{l_1 l_2 = 1} \mathcal{G}^{(l_1 l_2)}_{\delta},
 \end{array}
\end{equation*}
where,
\begin{equation}\begin{array}{lll}
 \sum^L_{l_1 l_2 = 1} \mathcal{G}^{(l_1 l_2)}_{\delta}
&=&\frac{c_+(\mu,\xi_1)c_-(\mu,\xi_1)}{b_-(\mu,\xi_1)} \tilde{B}_{2 \dots L}(\xi_1) \tilde{D}^{-1}_{2 \dots L}(\xi_1) \tilde{C}_{2 \dots L}(\xi_1) \tilde{D}_{2 \dots L} (\mu)\tilde{D}^{-1}_{2 \dots L}(\xi_1) \\
&&+ \left\{ b_+(\mu,\xi_1) - \frac{c_+(\mu,\xi_1)c_-(\mu,\xi_1)}{b_-(\mu,\xi_1)} \right\}\tilde{B}_{2 \dots L}(\mu)\tilde{D}^{-1}_{2 \dots L}(\mu)\tilde{C}_{2 \dots L}(\mu)\\
&&+ \frac{c_+(\mu,\xi_1)b_+(\mu,\xi_1)}{b_-(\mu,\xi_1)}\tilde{B}_{2 \dots L}(\xi_1)\tilde{C}_{2 \dots L}(\mu)\tilde{D}^{-1}_{2 \dots L}(\xi_1).
\end{array}\label{BC2}
\end{equation}
We notice that the first two terms (containing $\Omega_0(\mu)$ and $\Omega_1(\mu)$) in the above expression for $\delta^{(A)}$ are equal to the first two terms in the expression for $\delta^{(A)}_a$. Concentrating on the remaining terms, we notice that the multiplication of the $\tilde{B}_{2\dots L}$ and $\tilde{C}_{2\dots L}$ operators in Eq. (\ref{BC2}) results in a double summation, $\sum^L_{l_1 l_2 = 1}$, where the relevant sets of values of $l_1$ and $l_2$ are $l_1 = l_2 = l$, $l_1 > l_2$ and $l_1 < l_2$. Beginning with the case $l_1 = l_2 = l$ we obtain,
\begin{equation*}\begin{array}{lll}
\sum^L_{ \genfrac{}{}{0pt}{}{l_1 l_2 = 1}{l_1 = l_2}} \mathcal{G}^{(l_1 l_2)}_{\delta}
&=& \sum^L_{l=2}\textrm{\huge\{\normalsize}  \frac{c_+(\mu,\xi_1)c_+(\xi_1,\xi_l)}{b_-(\mu,\xi_1)a_-(\xi_1,\xi_l)} \overbrace{\textrm{\huge(\normalsize} \frac{c_-(\mu,\xi_1) a_-(\mu,\xi_l)c_-(\xi_1,\xi_l)}{ b_-(\xi_1,\xi_l)}+ b_+(\mu,\xi_1)  c_-(\mu,\xi_l)\textrm{\huge)\normalsize}}^{\textrm{use Eq. (\ref{YB5}) and (\ref{uni3})}} \\
&& - \frac{c_+(\mu,\xi_1)c_-(\mu,\xi_1)c_+(\mu,\xi_l) c_-(\mu,\xi_l) }{ b_-(\mu,\xi_l)b_-(\mu,\xi_1)}+ \frac{c_+(\mu,\xi_l)c_-(\mu,\xi_l)b_+(\mu,\xi_1)}{b_-(\mu,\xi_l)}   \textrm{\huge\}\normalsize} \Omega^{(l)}_2(\mu) \\
&=& \sum^L_{l=2} \textrm{\huge\{\normalsize}- \frac{c_+(\mu,\xi_1)c_-(\mu,\xi_l)}{b_-(\mu,\xi_1)}\overbrace{ \textrm{\huge(\normalsize} \frac{a_-(\mu,\xi_1)c_-(\xi_l,\xi_1) }{b_-(\xi_l,\xi_1)}+ \frac{c_+(\mu,\xi_l)c_-(\mu,\xi_1)}{b_-(\mu,\xi_l)}\textrm{\huge)\normalsize}}^{\textrm{use Eq. (\ref{YB6}) and (\ref{uni3})}}\\
& +& \frac{c_+(\mu,\xi_l)c_-(\mu,\xi_l)b_+(\mu,\xi_1)}{b_-(\mu,\xi_l)} \textrm{\huge\}\normalsize}  \Omega^{(l)}_2(\mu)\\
&=& \sum^L_{l=2} \frac{c_-(\mu,\xi_l)}{b_-(\mu,\xi_l)} \overbrace{\textrm{\huge\{\normalsize} \frac{c_+(\mu,\xi_1)a_-(\mu,\xi_l) c_+(\xi_1,\xi_1)}{b_-(\xi_1,\xi_l)} +b_+(\mu,\xi_1)c_+(\mu,\xi_l)  \textrm{\huge\}\normalsize}}^{\textrm{use Eq. (\ref{YB4})}} \Omega^{(l)}_2(\mu)\\
&=& \sum^L_{l=2}\frac{a_-(\mu,\xi_{1})a_-(\xi_{1},\xi_{l})c_+(\mu,\xi_l)c_-(\mu,\xi_l)}{b_-(\xi_{1},\xi_{l})b_-(\mu,\xi_l)} \Omega^{(l)}_2(\mu),
\end{array}
\end{equation*}
thus showing the equivalence between the $l_1 = l_2 = l$ terms of $\delta^{(A)}$ and $\delta^{(A)}_a$. 

For $l_1>l_2$ we obtain,
\begin{equation*}\begin{array}{lll}
\sum^L_{ \genfrac{}{}{0pt}{}{l_1 l_2 = 1}{l_1 > l_2}} \mathcal{G}^{(l_1 l_2)}_{\delta}
&=&\sum_{2\le l_2 < l_1 \le L}\textrm{\huge\{\normalsize}\frac{c_+(\mu,\xi_1)b_-(\mu,\xi_{l_1})c_+(\xi_1,\xi_{l_1})}{b_-(\mu,\xi_1) b_-(\xi_1,\xi_{l_1}) b_-(\xi_{l_2},\xi_{l_1}) a_-(\xi_1,\xi_{l_2})}  \\
&& \times \overbrace{\textrm{\huge(\normalsize}  c_-(\mu,\xi_1) a_-(\mu, \xi_{l_2})c_-(\xi_1,\xi_{l_2})+b_+(\mu,\xi_1) c_-(\mu, \xi_{l_2})b_-(\xi_1,\xi_{l_2}) \textrm{\huge)\normalsize}}^{\textrm{use Eq. (\ref{YB5}) and (\ref{uni3})}} \\
&&- \frac{c_+(\mu,\xi_1)c_-(\mu,\xi_1)c_+(\mu,\xi_{l_1})c_-(\mu,\xi_{l_2})}{b_-(\mu,\xi_1)b_-(\xi_{l_2},\xi_{l_1})} + \frac{c_-(\mu,\xi_{l_2})b_+(\mu,\xi_1)c_+(\mu,\xi_{l_1})}{b_-(\xi_{l_2},\xi_{l_1})}    \textrm{\huge\}\normalsize} \Omega^{(l_1 l_2)}_5(\mu) \\
&=& \sum_{2\le l_2 < l_1 \le L}\textrm{\huge\{\normalsize}\frac{c_-(\mu,\xi_{l_2})b_+(\mu,\xi_1)c_+(\mu,\xi_{l_1})}{b_-(\xi_{l_2},\xi_{l_1})}- \frac{c_+(\mu,\xi_1)c_-(\mu,\xi_{l_2})}{b_-(\mu,\xi_1)b_-(\xi_{l_2},\xi_{l_1})}\\ 
&& \times \overbrace{\textrm{\huge(\normalsize} \frac{b_-(\mu,\xi_{l_1})a_-(\mu,\xi_1)c_-(\xi_{l_1},\xi_1)}{b_-(\xi_{l_1},\xi_1)} + c_+(\mu,\xi_{l_1})c_-(\mu,\xi_1)\textrm{\huge)\normalsize}}^{\textrm{use Eq. (\ref{YB6}) and (\ref{uni3})}} \textrm{\huge\}\normalsize}        \Omega^{(l_1 l_2)}_5(\mu)\\
&=&  \sum_{2\le l_2 < l_1 \le L} \frac{c_-(\mu,\xi_{l_2})}{b_-(\xi_{l_2},\xi_{l_1})} \overbrace{ \textrm{\huge\{\normalsize} \frac{c_+(\mu,\xi_{1})a_-(\mu,\xi_{l_1})c_+(\xi_1,\xi_{l_1})}{b_-(\xi_{1},\xi_{l_1})} +b_+(\mu,\xi_{1})c_+(\mu,\xi_{l_1}) \textrm{\huge\}\normalsize} }^{\textrm{use Eq. (\ref{YB4})}} \\
&&\times  \Omega^{(l_1 l_2)}_5(\mu)\\
&=&  \sum_{2 \le l_2 < l_1 \le L} \frac{a_-(\mu,\xi_1)a_-(\xi_1,\xi_{l_1})c_+(\mu,\xi_{l_1})c_-(\mu,\xi_{l_2})}{b_-(\xi_{1},\xi_{l_1})b_-(\xi_{l_2},\xi_{l_1})} \Omega^{(l_1 l_2)}_5(\mu),
\end{array}
\end{equation*}
hence showing the equivalence between the $l_1 > l_2$ terms of $\delta^{(A)}$ and $\delta^{(A)}_a$.

Verifying the equivalence between the $l_1 < l_2$ terms of $\delta^{(A)}$ and $\delta^{(A)}_a$ is exactly analogous to verifying the $l_1 > l_2$ case, the only difference being that $\Omega^{(l_1 l_2)}_5(\mu)$ is replaced by $\Omega^{(l_1 l_2)}_6(\mu)$ in the calculations. 

Finally we have verified the validity of the explicit expressions of the twisted monodromy matrices (\ref{D}-\ref{A}), solely relying on the information granted to us by the algebraic relations (\ref{one}-\ref{three}), forgoing any form of explicit parameterization of the Boltzmann weights. In Table (\ref{tab:table2}) we provide a summary of the information needed from Eqs. (\ref{one}-\ref{three}) in order to complete the verification.\\
\\
\textbf{Remark.} We shall conclude this section with the following comment. From Table (\ref{tab:table2}) we see that out of the distinct sixteen Yang-Baxter algebra relations generated from Eq. (\ref{one}), we are required to use exactly half in order complete the general proof of the explicit form of the twisted operators. This is because we have started with a proposal for the $F$-matrix that is lower-triangular. It is natural to conjecture that if our initial choice was for an upper-triangular $F$-matrix then we would require the exclusive use of the remaining half of the the Yang-Baxter algebra relations to prove the expressions for the corresponding twisted monodromy operators. Given a positive answer regarding the aforementioned conjecture we would find the self-consistent role played by the Yang-Baxter algebra relations in the formulation of the $F$-basis as rather remarkable.

\begin{table}\begin{center}
\begin{tabular}{| c | c | c | c |}
  \hline                       
  Operator & Y-B algebra used & Y-B equations used & Uni. equations used \\\hline
  $\tilde{A}_{1\dots L}$ & (\ref{eq6}) (\ref{eq9}) (\ref{eq10}) & (\ref{YB2}) (\ref{YB4})(\ref{YB5}) (\ref{YB6}) &(\ref{uni1}) (\ref{uni3}) \\\hline
  $\tilde{B}_{1\dots L}$ & (\ref{eq11}) & (\ref{YB4}) & \\\hline
 $\tilde{C}_{1\dots L}$ & (\ref{eq13}) (\ref{eq14}) & (\ref{YB2})  & (\ref{uni3}) \\\hline
 $\tilde{D}_{1\dots L}$ & (\ref{eq15}) (\ref{eq16}) &  &  \\\hline 
\end{tabular}
\caption{Required Yang-Baxter and unitarity relations for the general case.}
\label{tab:table2}\end{center}\end{table}
\section{The scalar product}
\subsection{State vectors} In order to be consistent with the lower-triangularity of the factorizing $F$-matrix, we shall define our Bethe state vectors by the following ket,
\begin{equation}
  | \{\nu\} \rangle = | \nu_M \dots \nu_2 \nu_1 \rangle = C_{1\dots L}(\nu_M) \dots C_{1\dots L}(\nu_2) C_{1\dots L}(\nu_1) |1\rangle,
\label{ket}\end{equation}
where $M$ is an integer and $M \le L$. The reference state $|1 \rangle$ is the standard ferromagnetic state,
\begin{equation}
   |1\rangle = \otimes^L_{i=1}\left( \begin{array}{l} 0\\  1  \end{array} \right)_i.
\label{ferro}\end{equation}
By the same token we define the conjugate state vector as the following bra,
\begin{equation}
 \langle \{\mu\}| = \langle \mu_1 \mu_2 \dots \mu_M|= \langle 1 | B_{1\dots L}(\mu_1) B_{1\dots L}(\mu_2)\dots B_{1\dots L}(\mu_M),
\label{bra}\end{equation}
where the transpose state $\langle 1 |$ is,
\begin{equation}
   \langle 1 | = \otimes^L_{i=1}\left( 0,1 \right)_i.
\label{ferroT}\end{equation}
The scalar product is then defined as the expectation value of the state vector with its conjugate,
\begin{equation*}\begin{array}{lll}
 \langle \{\mu\}| \{\nu\} \rangle &=& \mathcal{S}(\{\mu\}, \{\nu\})\\
&=& \langle 1 | B_{1\dots L}(\mu_1)\dots B_{1\dots L}(\mu_M) C_{1\dots L}(\nu_M) \dots C_{1\dots L}(\nu_1) |1\rangle.
\end{array}\end{equation*}
\subsection{On-shell states and the Bethe equations} The transfer matrix $T_{1\dots L}(\lambda)$ is defined as the trace of the monodromy matrix over the auxiliary space $\mathcal{A}_a$,
\begin{equation*}
T_{1 \dots L}(\lambda) = \textrm{Tr}_a\left[ \mathcal{T}_{a,1\dots L}(\lambda) \right] = A_{1\dots L}(\lambda) + D_{1 \dots L} (\lambda).
\end{equation*}
It can be diagonalized by imposing that the state $|\{\nu\} \rangle$ is the corresponding eigenvector. Continuing on the theme of the previous sections of this work, this eigenvalue problem can be solved by relying solely on the information granted to us from Eqs. (\ref{one}-\ref{three}) without any explicit weight parameterization. Following \cite{MM}, one can see that this task is undertaken with the help of the commutation rules (\ref{eq9}), (\ref{eq13}) and (\ref{eq14}), the unitarity properties (\ref{uni4}) and (\ref{uni5}), and the Yang-Baxter relations (\ref{YB3}) and (\ref{YB8}). The final results for the action of the operators $A_{1\dots L}(\lambda)$ and $D_{1 \dots L}(\lambda)$ on the state $|\{ \nu\} \rangle$ are,
\begin{eqnarray}
&&A_{1\dots L}(\lambda)|\nu_M \dots \nu_1 \rangle = \prod^L_{i=1}b_+(\lambda,\xi_i) \prod^M_{j=1}\frac{1}{b_+(\lambda,\nu_j)}|\nu_M \dots \nu_1 \rangle \label{eigA} \\
&&- \sum^M_{j=1}\frac{c_+(\lambda,\nu_j)}{b_+(\lambda,\nu_j)} \prod^L_{i=1}b_+(\nu_j,\xi_i) \prod^M_{\genfrac{}{}{0pt}{}{k=1}{ \ne j}}\frac{\theta(\nu_j,\nu_k)}{b_+(\nu_j,\nu_k)}C_{1\dots L}(\lambda)| \nu_M \dots \nu_{j+1} \nu_{j-1} \dots \nu_1 \rangle \nonumber \\
&&D_{1\dots L}(\lambda)|\nu_M \dots \nu_1 \rangle = \prod^L_{i=1}a_-(\lambda,\xi_i)\prod^M_{j=1}\frac{a_-(\nu_j,\lambda)}{b_+(\nu_j,\lambda)}|\nu_M \dots \nu_1 \rangle \label{eigD} \\
&&- \sum^M_{j=1}\frac{c_-(\nu_j,\lambda)}{b_+(\nu_j,\lambda)}\prod^L_{i=1}a_-(\nu_j,\xi_i)\prod^M_{\genfrac{}{}{0pt}{}{k=1}{ \ne j}} \frac{\theta(\nu_k,\nu_j)}{b_+(\nu_k,\nu_j)}C_{1\dots L}(\lambda)| \nu_M \dots \nu_{j+1} \nu_{j-1} \dots \nu_1 \rangle  \nonumber,
\end{eqnarray}
where the function $\theta(\nu_j.\nu_k)$ is defined as,
\begin{equation}
 \theta(\nu_j,\nu_k) = \left\{ \begin{array}{cll}
                                  a_-(\nu_j,\nu_k)& \textrm{if} & j < k\\
				  1 & \textrm{if}& j \ge k
                                 \end{array} \right. .
\label{theta}\end{equation}
In order for the states $|\{\nu\}\rangle$ to be eigenvectors of the transfer matrix one is required to cancel the undesirable terms $C_{1\dots L}(\lambda)| \nu_M \dots \nu_{j+1} \nu_{j-1} \dots \nu_1 \rangle$. All of these terms are eliminated by imposing that the rapidities $\{\nu\}$ satisfy the following set of \textit{Bethe equations},
\begin{equation}\begin{array}{lll}
{\displaystyle \prod^M_{\genfrac{}{}{0pt}{}{k=1}{ \ne j}} \frac{b_+(\nu_k,\nu_j)a_-(\nu_j,\nu_k)}{b_+(\nu_j,\nu_k)}= \prod^L_{i=1}\frac{a_-(\nu_j,\xi_i)}{b_+(\nu_j,\xi_i)}} &\textrm{for}& j \in \{1,\dots,M\}.
\end{array}\label{BETHEGEN}\end{equation}
From now on we shall denote the \textit{on-shell states}, i.e. states whose rapidities satisfy the Bethe equations (\ref{BETHEGEN}) by $\left.\left| \{\nu\}_{\beta}\right.\right\rangle $. Equivalently, we can refer to states $|\{\nu \} \rangle$ where the rapidities are free as \textit{off-shell}.
\subsection{Applying the F-basis} We now concentrate on applying the factorizing $\mathcal{F}$-matrices on the scalar product to obtain a complete algebraic expression in terms of Boltzmann weights. We begin by giving the following two necessary results:
\begin{proposition}\label{gurg}
 \begin{equation}
\begin{array}{lll}
\langle 1 | \mathcal{F}_{1\dots L} = {\displaystyle \prod_{1 \le i < j \le L}} a_-(\xi_i,\xi_j)  \langle 1 | &,& \mathcal{F}^{-1}_{1\dots L}|1\rangle =  {\displaystyle \prod_{1 \le i < j \le L}} a_-(\xi_j,\xi_i) |1\rangle
\end{array}\label{Fwith0}\end{equation}
\end{proposition}
The verification of the above relations can be obtained through careful consideration of the form of $\mathcal{F}_{1\dots L}$ given by Eq. (\ref{ansatz1a}). We offer the complete proof in Appendix \ref{APPFIRST}.

Applying the above results we obtain the following form for the scalar product in terms of the twisted monodromy operators,
\begin{equation}
 \mathcal{S}( \{\mu\}, \{\nu\} )= \langle 1| \tilde{B}_{1\dots L}(\mu_1)\dots \tilde{B}_{1\dots L}(\mu_M) \tilde{C}_{1\dots L}(\nu_M) \dots \tilde{C}_{1\dots L}(\nu_1) |1\rangle.
\label{twistedscalar}\end{equation}
\subsubsection{Complete set of states} The complete set of states for the Hilbert space $V_1 \otimes \dots \otimes V_L$ is given by,
\begin{equation*}
\sum^L_{p=0}\sum_{1 \le l_1 < \dots < l_p \le L} | l_1, \dots, l_p \rangle \langle l_1, \dots, l_p | ,
\end{equation*}
where we have used the following labels,
\begin{equation}\begin{array}{lll}
| l_1, \dots, l_p \rangle &=&  \otimes^L_{\genfrac{}{}{0pt}{}{i = 1}{\ne l_1, \dots, l_p}} \left( \begin{array}{c}0\\1 \end{array}\right)_i  \otimes^p_{j = 1 } \left( \begin{array}{c}1\\0 \end{array}\right)_{l_j}\\
\langle l_1, \dots, l_p | &=& \otimes^L_{\genfrac{}{}{0pt}{}{i = 1}{\ne l_1, \dots, l_p}} (0,1)_i  \otimes^p_{j = 1 } (1,0)_{l_j}.
\end{array}
\label{NICE}\end{equation}
Inserting a complete set of states into the expression for the scalar product (\ref{twistedscalar}) as follows,
\begin{equation}\begin{array}{lll}
\mathcal{S}( \{\mu\}, \{\nu\} ) &=& {\displaystyle \sum^L_{p=0}\sum_{1 \le l_1 < \dots < l_p \le L}} \langle 1 | \tilde{B}_{1 \dots L}(\mu_1)\dots \tilde{B}_{1 \dots L}(\mu_M)| l_1, \dots, l_p \rangle\\
 &&\times  \langle l_1, \dots, l_p |\tilde{C}_{1 \dots L}(\nu_M) \dots \tilde{C}_{1 \dots L}(\nu_1) |1\rangle,
\end{array}\label{raw}
\end{equation}
and considering the action of $e^{(12)}$ and $e^{(21)}$ on the orthonormal basis vectors we can immediately verify that the only value of $p$ in Eq. (\ref{raw}) that does not equal zero is $p=M$, hence,
\begin{equation}\begin{array}{lll}
\mathcal{S}( \{\mu\}, \{\nu\} ) &=& {\displaystyle \sum_{1 \le l_1 < \dots < l_M \le L}} \langle 1 | \tilde{B}_{1 \dots L}(\mu_1)\dots \tilde{B}_{1 \dots L} (\mu_M)| l_1, \dots, l_M \rangle\\
 &&\times  \langle l_1, \dots, l_M |\tilde{C}_{1 \dots L}(\nu_M) \dots \tilde{C}_{1 \dots L}(\nu_1) |1\rangle.
\end{array}\label{raw2}
\end{equation}
Reiterating that $\tilde{B}_{1 \dots L}(\mu)$ and $\tilde{C}_{1 \dots L}(\nu)$ are elements of $\textrm{End}(V_1 \otimes \dots \otimes V_L)$, in the following section we shall decrease the number of relevant vector spaces in the Hilbert space from $L$ to $M$. For the sake of clarity in the proceeding sections we note that with the application of the $\theta$-function, defined in Eq. (\ref{theta}), the twisted $B$ (\ref{B}) and $C$ (\ref{C}) operators can be expressed as,
\begin{equation*}\begin{array}{lll}
 \tilde{B}_{1 \dots L}(\mu) &=& \sum^L_{l=1}c_+(\mu,\xi_{l})e^{(21)}_{l}\otimes^{L}_{\genfrac{}{}{0pt}{}{i = 1}{\ne l}} \left(  \begin{array}{cc} 
b_-(\mu,\xi_{i})  & 0  \\
 0  & \frac{a_-(\mu,\xi_{i})\theta(\xi_{i},\xi_{l})}{b_-(\xi_{i},\xi_{l})}
\end{array}\right)_{i}\\
 \tilde{C}_{1 \dots L}(\nu) &=&\sum^L_{l=1}c_-(\nu,\xi_{l}) e^{(12)}_{l}\otimes^{L}_{\genfrac{}{}{0pt}{}{i = 1}{\ne l}} \left(  \begin{array}{cc} 
\frac{b_-(\nu,\xi_{i})}{b_-(\xi_{l},\xi_{i})}  & 0  \\
 0  & \frac{a_-(\nu,\xi_{i})}{\theta(\xi_{l},\xi_{i})}
\end{array}\right)_{i} .
\end{array}\end{equation*}
\subsection{Decreasing the number of relevant vector spaces}
In this section we discuss the steps that are necessary to express the scalar product (\ref{raw2}) as a weighted bilinear sum of domain wall partition functions. Here we follow the procedure originally devised in \cite{Kita}, but introduce a number of subtle adaptations that are necessary to deal with the generality of the Boltzmann weights.
\subsubsection{Simplification - I}\label{simp1}
Since the action of $e^{(21)}$ on the basis vectors of the ferromagnetic states (\ref{ferro}) is zero, one can use matrix multiplication on the expression $\langle 1 | \tilde{B}_{1 \dots L}(\mu_1)\dots \tilde{B}_{1 \dots L} (\mu_M)| l_1, \dots, l_M \rangle$ to decrease the number of relevant vector spaces the twisted $B$-operator acts on as follows, 
\begin{equation}\begin{array}{ll}
&  \langle 1 | \tilde{B}_{1\dots L }(\mu_{1}) \dots \tilde{B}_{1 \dots L}(\mu_M)| l_1, \dots, l_M \rangle \\
=&  \sum^M_{\genfrac{}{}{0pt}{}{q_1\dots q_M = 1}{q_1 \ne \dots \ne q_M }} \prod^M_{m=1} \prod^L_{\genfrac{}{}{0pt}{}{p = 1}{\ne l_1, \dots, l_M}} \frac{a_-(\mu_m,\xi_p)\theta(\xi_p,\xi_{l_{q_m}})}{b_-(\xi_p,\xi_{l_{q_m}})} \langle 1 | \tilde{B}^{(q_1)}_{l_1\dots l_M }(\mu_{1}) \dots \tilde{B}^{(q_M)}_{l_1 \dots l_M}(\mu_M)| l_1, \dots, l_M \rangle,
\end{array}\label{hj1}\end{equation}
where,
\begin{equation*}
  \tilde{B}^{(q)}_{l_1 \dots l_M}(\mu) = c_+(\mu,\xi_{l_q})e^{(21)}_{l_q}\otimes^{M}_{\genfrac{}{}{0pt}{}{i = 1}{\ne q}} \left(  \begin{array}{cc} 
b_-(\mu,\xi_{l_i})  & 0  \\
 0  & \frac{a_-(\mu,\xi_{l_i})\theta(\xi_{l_i},\xi_{l_q})}{b_-(\xi_{l_i},\xi_{l_q})}
\end{array}\right)_{l_i}.
\end{equation*}
Eq. (\ref{hj1}) is worthy of some remarks. Firstly, the coefficient $\prod^M_{m=1} \prod^L_{\genfrac{}{}{0pt}{}{p = 1}{\ne l_1, \dots, l_M}} \frac{a_-(\mu_m,\xi_p)\theta(\xi_p,\xi_{l_{q_m}})}{b_-(\xi_p,\xi_{l_{q_m}})}$ is independent of the value of $(q_1,\dots,q_M)$ in the sum due to the condition $q_1 \ne \dots \ne q_M$. Secondly, the operator $\tilde{B}^{(q)}_{l_1 \dots l_M}(\mu)$ acts trivially on the vector spaces $V_{\alpha}$, for $\alpha \ne (l_1,\dots,l_M)$, meaning we can decrease the number of relevant vector spaces in the reference states from $L$ to $M$ as follows,
\begin{equation*}
 \begin{array}{lll}
\langle 1 | \rightarrow  \,_{(l_1 \dots l_M)}\langle 1 | \equiv \otimes^M_{i=1}\left( 0,1 \right)_{l_i} & , & | l_1, \dots, l_M \rangle \rightarrow |0 \rangle_{(l_1 \dots l_M)} \equiv \otimes^M_{i=1}\left( \begin{array}{l} 1\\  0  \end{array} \right)_{l_i}.
\end{array}
\end{equation*}
Thirdly, when the aforementioned coefficient is taken out of the sum over $q_i$, we obtain the following simplification,
\begin{equation*}\begin{array}{ll}
& \sum^M_{\genfrac{}{}{0pt}{}{q_1\dots q_M = 1}{q_1 \ne \dots \ne q_M }}
\,_{(l_1 \dots l_M)}\langle 1 | \tilde{B}^{(q_1)}_{l_1\dots l_M }(\mu_{1}) \dots \tilde{B}^{(q_M)}_{l_1 \dots l_M}(\mu_M)| 0 \rangle_{(l_1 \dots l_M)}\\
=& \,_{(l_1 \dots l_M)}\langle 1 | \tilde{B}_{l_1\dots l_M }(\mu_{1}) \dots \tilde{B}_{l_1 \dots l_M}(\mu_M)|0 \rangle_{(l_1 \dots l_M)},
\end{array}\end{equation*}
where,
\begin{equation*}
 \tilde{B}_{l_1 \dots l_M}(\mu) = \sum^M_{q=1}c_+(\mu,\xi_{l_q})e^{(21)}_{l_q}\otimes^{M}_{\genfrac{}{}{0pt}{}{i = 1}{\ne q}} \left(  \begin{array}{cc} 
b_-(\mu,\xi_{l_i})  & 0  \\
 0  & \frac{a_-(\mu,\xi_{l_i})\theta(\xi_{l_i},\xi_{l_q})}{b_-(\xi_{l_i},\xi_{l_q})}
\end{array}\right)_{l_i}  .
\end{equation*}
Hence Eq. (\ref{hj1}) ultimately simplifies to the expression, 
\begin{equation*}
 \begin{array}{ll}
&  \langle 1 | \tilde{B}_{1\dots L }(\mu_{1}) \dots \tilde{B}_{1 \dots L}(\mu_M)| l_1, \dots, l_M \rangle\\
=&   \prod^M_{m=1} \prod^L_{\genfrac{}{}{0pt}{}{p = 1}{\ne l_1, \dots, l_M}} \frac{a_-(\mu_m,\xi_p)\theta(\xi_p,\xi_{l_m})}{b_-(\xi_p,\xi_{l_m})} \,_{(l_1 \dots l_M)}\langle 1 | \tilde{B}_{l_1 \dots l_M}(\mu_{1}) \dots \tilde{B}_{l_1 \dots l_M}(\mu_M) |0 \rangle_{(l_1 \dots l_M)}.
 \end{array}
\end{equation*}
\subsubsection{Simplification - II}\label{simp2} In a completely analogous method to Section (\ref{simp1}), since the action of $e^{(12)}$ on the basis vectors of the transpose ferromagnetic states (\ref{ferroT}) is zero, one can use matrix multiplication on the expression $ \langle l_1, \dots, l_M |\tilde{C}_{1 \dots L}(\nu_M) \dots \tilde{C}_{1 \dots L}(\nu_1) |1\rangle$ to decrease the number of relevant vector spaces the twisted $C$-operator acts on as follows, 
\begin{equation}\begin{array}{ll}
&  \langle l_1, \dots,  l_M|\tilde{C}_{1 \dots L}(\nu_M) \dots \tilde{C}_{1 \dots L}(\nu_1) |1\rangle \\
=& \sum^M_{\genfrac{}{}{0pt}{}{q_1\dots q_M = 1}{q_1 \ne \dots \ne q_M }} \prod^M_{m=1} \prod^L_{\genfrac{}{}{0pt}{}{p=1}{\ne l_1, \dots, l_M}}  \frac{a_-(\nu_m,\xi_p)}{\theta(\xi_{l_{q_m}},\xi_p)} \langle l_1, \dots,l_M|\tilde{C}^{(q_M)}_{l_1 \dots l_M}(\nu_M) \dots \tilde{C}^{(q_1)}_{l_1 \dots l_M}(\nu_1) |1\rangle,
 \end{array}\label{hj2}\end{equation}
where, 
\begin{equation*}
 \tilde{C}^{(q)}_{l_1 \dots l_M}(\nu) =c_-(\nu,\xi_{l_q}) e^{(12)}_{l_q}\otimes^{M}_{\genfrac{}{}{0pt}{}{i = 1}{\ne q}} \left(  \begin{array}{cc} 
\frac{b_-(\nu,\xi_{l_i})}{b_-(\xi_{l_q},\xi_{l_i})}  & 0  \\
 0  & \frac{a_-(\nu,\xi_{l_i})}{\theta(\xi_{l_q},\xi_{l_i})}
\end{array}\right)_{l_i}.
\end{equation*}
We offer the following remarks on Eq. (\ref{hj2}). Firstly, the coefficient $\prod^M_{m=1} \prod^L_{\genfrac{}{}{0pt}{}{p=1}{\ne l_1, \dots, l_M}}  \frac{a_-(\nu_m,\xi_p)}{\theta(\xi_{l_{q_m}},\xi_p)}$ is independent of the value of $(q_1,\dots,q_M)$ in the sum due to the condition $q_1 \ne \dots \ne q_M$. Secondly, the operator $\tilde{C}^{(q)}_{l_1 \dots l_M}(\nu)$ acts trivially on the vector spaces $V_{\alpha}$, for $\alpha \ne (l_1,\dots,l_M)$, meaning we can decrease the number of relevant vector spaces in the reference states from $L$ to $M$ as follows,
\begin{equation*}
 \begin{array}{lll}
\langle l_1,\dots,l_M | \rightarrow  \,_{(l_1 \dots l_M)}\langle 0 | \equiv \otimes^M_{i=1}\left( 1,0 \right)_{l_i} & , & | 1 \rangle \rightarrow |1 \rangle_{(l_1 \dots l_M)} \equiv \otimes^M_{i=1}\left( \begin{array}{l} 0\\  1  \end{array} \right)_{l_i}.
\end{array}
\end{equation*}
Thirdly, when the aforementioned coefficient is taken out of the sum over $q_i$, we obtain the following simplification,
\begin{equation*}\begin{array}{ll}
&\sum^M_{\genfrac{}{}{0pt}{}{q_1\dots q_M = 1}{q_1 \ne \dots \ne q_M }}   \,_{(l_1 \dots l_M)}\langle 0|\tilde{C}^{(q_M)}_{l_1 \dots l_M}(\nu_M) \dots \tilde{C}^{(q_1)}_{l_1 \dots l_M}(\nu_1) |1\rangle_{(l_1 \dots l_M)}\\
=& \,_{(l_1 \dots l_M)}\langle 0|\tilde{C}_{l_1 \dots l_M}(\nu_M) \dots \tilde{C}_{l_1 \dots l_M}(\nu_1) |1\rangle_{(l_1 \dots l_M)},
\end{array}\end{equation*}
where,
\begin{equation*}
 \tilde{C}_{l_1 \dots l_M}(\nu) =\sum^M_{q=1}c_-(\nu,\xi_{l_q}) e^{(12)}_{l_q}\otimes^{M}_{\genfrac{}{}{0pt}{}{i = 1}{\ne q}} \left(  \begin{array}{cc} 
\frac{b_-(\nu,\xi_{l_i})}{b_-(\xi_{l_q},\xi_{l_i})}  & 0  \\
 0  & \frac{a_-(\nu,\xi_{l_i})}{\theta(\xi_{l_q},\xi_{l_i})}
\end{array}\right)_{l_i} .
\end{equation*}
Hence Eq. (\ref{hj2}) ultimately simplifies to the expression, 
\begin{equation*}
\begin{array}{ll}
&  \langle l_1, \dots, l_M |\tilde{C}_{1 \dots L}(\nu_M) \dots \tilde{C}_{1 \dots L}(\nu_1) |1\rangle \\
=&  \prod^M_{m=1} \prod^L_{\genfrac{}{}{0pt}{}{p=1}{\ne l_1, \dots, l_M}}  \frac{a_-(\nu_m,\xi_p)}{\theta(\xi_{l_m},\xi_p)}  \,_{(l_1 \dots l_M)}\langle 0 | \tilde{C}_{l_1 \dots l_M}(\nu_M) \dots \tilde{C}_{l_1 \dots l_M}(\nu_1)  |1\rangle_{(l_1 \dots l_M)}.
 \end{array}\end{equation*}
\subsubsection{The scalar product as a weighted bilinear sum}
Combining the results of Sections (\ref{simp1}) and (\ref{simp2}) we obtain the following form for the scalar product,
\begin{equation}
 \begin{array}{l}
 \mathcal{S}( \{\mu\}, \{\nu\} ) =  \sum_{1 \le l_1 < \dots < l_M \le L}  \prod^M_{m=1} \prod^L_{\genfrac{}{}{0pt}{}{p = 1}{\ne l_1, \dots, l_M}} \frac{a_-(\mu_m,\xi_p)a_-(\nu_m,\xi_p)}{b_-(\xi_{p},\xi_{l_m})a_-(\xi_{l_m},\xi_p)}  \\
 \times  \,_{(l_1 \dots l_M)}\langle 1 |\tilde{B}_{l_1 \dots l_M}(\mu_{1}) \dots \tilde{B}_{l_1 \dots l_M}(\mu_M) |0 \rangle_{(l_1 \dots l_M)} \,_{(l_1 \dots l_M)}\langle 0 | \tilde{C}_{l_1 \dots l_M}(\nu_M) \dots \tilde{C}_{l_1 \dots l_M}(\nu_1)  |1\rangle_{(l_1 \dots l_M)}.
 \end{array}\label{simpscal}
\end{equation}
In the above expression we have applied the following $\theta$-function relation,
\begin{equation*}
  \prod^M_{m=1} \prod^L_{\genfrac{}{}{0pt}{}{p=1}{\ne l_1, \dots, l_M}} \frac{\theta(\xi_p,\xi_{l_m})}{\theta(\xi_{l_m},\xi_p)}=  \prod^M_{m=1} \prod^L_{\genfrac{}{}{0pt}{}{p=1}{\ne l_1, \dots, l_M}} \frac{1}{a_-(\xi_{l_m},\xi_p)},
\end{equation*}
which can be obtained using unitarity condition (\ref{uni5}). We refer to\\
$\,_{(l_1 \dots l_M)}\langle 1 |\tilde{B}_{l_1 \dots l_M}(\mu_{1}) \dots \tilde{B}_{l_1 \dots l_M}(\mu_M) |0 \rangle_{(l_1 \dots l_M)}$ and $\,_{(l_1 \dots l_M)}\langle 0 | \tilde{C}_{l_1 \dots l_M}(\nu_M) \dots \tilde{C}_{l_1 \dots l_M}(\nu_1)  |1\rangle_{(l_1 \dots l_M)}$ as the \textit{twisted domain wall partition functions of type $B$ and $C$} respectively. In the next section we shall express them (and hence the scalar product) explicitly in terms of Boltzmann weights.
\section{The domain wall partition function (DWPF)}\label{DOMAINY} We focus on the twisted DWPF's of type $B$ and $C$ separately. Beginning with type $B$, we employ the simplicity of the twisted $B$ operator to derive a recurrence relation for the DWPF which we can solve explicitly through induction. We then perform an equivalent analysis for type $C$.
\subsection{DWPF of type B}
\subsubsection{Twisted recurrence relation} Focusing on the expression $\,_{(l_1 \dots l_M)}\langle 1 | \tilde{B}_{l_1 \dots l_M}(\mu_1) \dots \tilde{B}_{l_1 \dots l_M}(\mu_M)  |0\rangle_{(l_1 \dots l_M)}$, we insert a complete set of states between the operators $\tilde{B}_{l_1 \dots l_M}(\mu_1)$ and $\tilde{B}_{l_1 \dots l_M}(\mu_{2})$ to obtain,
\begin{equation*}
 \begin{array}{ll}
 & \,_{(l_1 \dots l_M)}\langle 1 | \tilde{B}_{l_1 \dots l_M}(\mu_1) \dots \tilde{B}_{l_1 \dots l_M}(\mu_M)  |0\rangle_{(l_1 \dots l_M)}\\
=&  \sum^M_{q=1}\,_{(l_1 \dots l_M)}\langle 1 | \tilde{B}_{l_1 \dots l_M}(\mu_1)| q \rangle_{(l_1 \dots l_M)} \,_{(l_1 \dots l_M)}\langle q |\tilde{B}_{l_1 \dots l_M}(\mu_{2}) \dots \tilde{B}_{l_1 \dots l_M}(\mu_M)  |0\rangle_{(l_1 \dots l_M)},
\end{array}
\end{equation*}
where we have used the notation,
\begin{equation}
 \begin{array}{lll}
  | q \rangle_{(l_1 \dots l_M)} &=& \otimes^M_{\genfrac{}{}{0pt}{}{i = 1}{\ne q}} \left( \begin{array}{c}0\\1 \end{array}\right)_{l_i} \otimes\left( \begin{array}{c}1\\0 \end{array}\right)_{l_q}  \\
 \,_{(l_1 \dots l_M)}\langle q | &=& \otimes^M_{\genfrac{}{}{0pt}{}{i = 1}{\ne q}} (0,1)_{l_i} \otimes(1,0)_{l_q}.
 \end{array}\label{style}
\end{equation}
Since the action of $e^{(21)}$ on the basis vectors of the ferromagnetic states (\ref{ferro}) is zero, one can use matrix multiplication to obtain the following expression for $\,_{(l_1 \dots l_M)}\langle 1 | \tilde{B}_{l_1 \dots l_M}(\mu_1)| q \rangle_{(l_1 \dots l_M)}$,
\begin{equation*}
\,_{(l_1 \dots l_M)}\langle 1 | \tilde{B}_{l_1 \dots l_M}(\mu_1)| q \rangle_{(l_1 \dots l_M)}=
c_+(\mu_1,\xi_{l_{q}}){\displaystyle \prod^{M}_{\genfrac{}{}{0pt}{}{j=1}{\ne q} }} \frac{a_-(\mu_1,\xi_{l_j})\theta(\xi_{l_j},\xi_{l_{q}})}{b_-(\xi_{l_j},\xi_{l_{q}})}.
\end{equation*}
Concentrating on $ \,_{(l_1 \dots l_M)}\langle q |\tilde{B}_{l_1 \dots l_M}(\mu_{2}) \dots \tilde{B}_{l_1 \dots l_M}(\mu_M)  |0\rangle_{(l_1 \dots l_M)}$, we apply a similar method to that detailed in Section (\ref{simp1}) to decrease the relevant number of vector spaces in the reference states from $M$ to $M-1$ to obtain,
\begin{equation*}
 \begin{array}{ll}
&  \,_{(l_1 \dots l_M)}\langle q |\tilde{B}_{l_1 \dots l_M}(\mu_{2}) \dots \tilde{B}_{l_1 \dots l_M}(\mu_M)  |0\rangle_{(l_1 \dots l_M)}\\
=&{\displaystyle \prod^{M}_{m=2}}b_-(\mu_m,\xi_{l_q})  \,_{(\genfrac{}{}{0pt}{}{l_1 \dots l_M}{\ne l_q})}\langle 1 | \tilde{B}_{\genfrac{}{}{0pt}{}{l_1 \dots l_M}{\ne l_q}}(\mu_{2})\dots \tilde{B}_{\genfrac{}{}{0pt}{}{l_1 \dots l_M}{\ne l_q}}(\mu_{M})| 0\rangle_{(\genfrac{}{}{0pt}{}{l_1 \dots l_M}{\ne l_q})},
 \end{array}
\end{equation*}
where,
\begin{equation*}
 \tilde{B}_{\genfrac{}{}{0pt}{}{l_1 \dots l_M}{\ne l_q}}(\mu) =\sum^M_{\genfrac{}{}{0pt}{}{p=1}{ \ne q}}c_+(\mu,\xi_{l_p})  e^{(21)}_{l_p} \otimes^{M}_{\genfrac{}{}{0pt}{}{i=1}{ \ne p,q}} \left(  \begin{array}{cc} 
b_-(\mu,\xi_{l_i})  & 0  \\
 0  & \frac{a_-(\mu,\xi_{l_i})\theta(\xi_{l_i},\xi_{l_p})}{b_-(\xi_{l_i},\xi_{l_p})}
\end{array}\right)_{l_i} ,
\end{equation*}
and, 
\begin{equation*}\begin{array}{lll}
  \,_{(\genfrac{}{}{0pt}{}{l_1 \dots l_M}{\ne l_q})}\langle 1 | = \otimes^M_{\genfrac{}{}{0pt}{}{i=1}{ \ne q}}\left( 0,1 \right)_{l_i} & , & |0 \rangle_{(\genfrac{}{}{0pt}{}{l_1 \dots l_M}{\ne l_q})} = \otimes^M_{\genfrac{}{}{0pt}{}{i=1}{ \ne q}}\left( \begin{array}{l} 1\\  0  \end{array} \right)_{l_i}.
\end{array}
\end{equation*}
Thus we obtain the following twisted recursion relation,
\begin{equation}
\begin{array}{r}
  \,_{(l_1 \dots l_M)}\langle 1 |\tilde{B}_{l_1 \dots l_M}(\mu_{1}) \dots \tilde{B}_{l_1 \dots l_M}(\mu_M)  |0\rangle_{(l_1 \dots l_M)} = {\displaystyle \sum^M_{q=1}}c_+(\mu_1,\xi_{l_{q}}) 
 {\displaystyle \prod^{M}_{\genfrac{}{}{0pt}{}{j=1}{\ne q} } } \frac{a_-(\mu_1,\xi_{l_j})\theta(\xi_{l_j},\xi_{l_{q}})}{b_-(\xi_{l_j},\xi_{l_{q}})} \\
\times  {\displaystyle \prod^{M}_{m=2}}b_-(\mu_m,\xi_{l_q}) \,_{(\genfrac{}{}{0pt}{}{l_1 \dots l_M}{\ne l_q})}\langle 1 | \tilde{B}_{\genfrac{}{}{0pt}{}{l_1 \dots l_M}{\ne l_q}}(\mu_{2})\dots \tilde{B}_{\genfrac{}{}{0pt}{}{l_1 \dots l_M}{\ne l_q}}(\mu_{M})| 0\rangle_{(\genfrac{}{}{0pt}{}{l_1 \dots l_M}{\ne l_q})}.
\end{array}
\label{twistedrecursion}\end{equation}
We now focus on untwisting the above equation.
\subsubsection{Untwisting the twisted DWPF}
We give the following necessary results:
\begin{proposition}\label{NEEDEDSEC}
 \begin{equation}\begin{array}{llll}
  \langle 0| \mathcal{F}_{1\dots L} = \langle 0 |, &  \mathcal{F}_{1\dots L} | 0 \rangle = | 0 \rangle, & \langle 0| \mathcal{F}^{-1}_{1\dots L} = \langle 0 |, &  \mathcal{F}^{-1}_{1\dots L} | 0 \rangle = | 0 \rangle.
\end{array} 
 \label{Fwith1}\end{equation}
\end{proposition}
As with Proposition (\ref{gurg}), the verification of the above relations can be obtained through careful consideration of the form of $\mathcal{F}_{1\dots L}$ given by Eq. (\ref{ansatz1a}). We give the proof in Appendix \ref{APPSECOND}.

Through the application of Eqs. (\ref{Fwith0}) and (\ref{Fwith1}), and labeling the untwisted DWPF as follows,
\begin{equation}\begin{array}{lll}
\,_{(l_1 \dots l_M)}\langle 1 |{B}_{l_1 \dots l_M}(\mu_{1}) \dots {B}_{l_1 \dots l_M}(\mu_M)  |0\rangle_{(l_1 \dots l_M)} &=& Z^{(B)}_M(\{\mu\},\{\xi_l\})\\
\,_{(\genfrac{}{}{0pt}{}{l_1 \dots l_M}{\ne l_q})}\langle 1 | B_{\genfrac{}{}{0pt}{}{l_1 \dots l_M}{\ne l_q}}(\mu_{2})\dots B_{\genfrac{}{}{0pt}{}{l_1 \dots l_M}{\ne l_q}}(\mu_{M})| 0\rangle_{(\genfrac{}{}{0pt}{}{l_1 \dots l_M}{\ne l_q})} &=& Z^{(B)}_{M-1}(\{\mu\},\{\xi_l\}|\hat{\mu}_1,\hat{\xi}_{l_q}),
\end{array}\label{DWPFun}\end{equation}
Eq. (\ref{twistedrecursion}) becomes,
\begin{equation}\begin{array}{lll}
Z^{(B)}_M(\{\mu\},\{\xi_l\}) &=& {\displaystyle \sum^M_{q=1}}c_+(\mu_1,\xi_{l_{q}})  
 {\displaystyle \prod^{M}_{\genfrac{}{}{0pt}{}{j=1}{\ne q} } } \frac{a_-(\mu_1,\xi_{l_j})}{b_-(\xi_{l_j},\xi_{l_{q}})\theta(\xi_{l_{q}},\xi_{l_j})} \\
&&  \times  {\displaystyle \prod^{M}_{m=2}}b_-(\mu_m,\xi_{l_q})   Z^{(B)}_{M-1}(\{\mu\},\{\xi_l\}|\hat{\mu}_1,\hat{\xi}_{l_q}).
\end{array}
\label{untwistedrecursion}\end{equation}
\subsubsection{Explicit solution}\label{explicitI} For $M=\{1,2\}$ we obtain,
\begin{equation*}
 \begin{array}{lll}
  Z^{(B)}_1(\mu_1,\xi_{l_1}) &=& c_+(\mu_1,\xi_{l_1})\\
 Z^{(B)}_2(\{\mu\},\{\xi_{l}\})&=&\frac{c_+(\mu_1,\xi_{l_1})c_+(\mu_2,\xi_{l_2})b_-(\mu_2,\xi_{l_1})a_-(\mu_1,\xi_{l_2})}{b_-(\xi_{l_2},\xi_{l_1}) a_-(\xi_{l_1},\xi_{l_2})}\\
&&+ \frac{c_+(\mu_1,\xi_{l_2})c_+(\mu_2,\xi_{l_1})b_-(\mu_2,\xi_{l_2})a_-(\mu_1,\xi_{l_1})}{b_-(\xi_{l_1},\xi_{l_2})}.
\end{array}\end{equation*}
We now offer the following general result. 
\begin{proposition}\label{propexpDWPF}
 \begin{equation}\begin{array}{l}
 Z^{(B)}_M(\{\mu\},\{\xi_{l}\})=  {\displaystyle \sum_{\sigma \in S_M}} {\displaystyle \prod^M_{i=1}}c_+(\mu_i,\xi_{l_{\sigma_i}}) {\displaystyle  \prod_{1 \le j < k \le M}} \frac{b_-(\mu_k,\xi_{l_{\sigma_j}})a_-(\mu_j,\xi_{l_{\sigma_k}}) }{b_-(\xi_{l_{\sigma_k}},\xi_{l_{\sigma_j}})\theta(\xi_{l_{\sigma_j}},\xi_{l_{\sigma_k}})}
\end{array}\label{explicitZBM} \end{equation}
\end{proposition}
\textbf{Proof.} Noting that the above formula is correct for $M = \{1,2\}$, we assume that it holds for some $M$, and focus on the $M+1$ case of Eq. (\ref{untwistedrecursion}),
\begin{equation*}
\begin{array}{lll}
Z^{(B)}_{M+1}&=& {\displaystyle \sum^{M+1}_{q=1}}c_+(\mu_1,\xi_{l_{q}})  
 {\displaystyle \prod^{M+1}_{\genfrac{}{}{0pt}{}{n=1}{\ne q} } } \frac{a_-(\mu_1,\xi_{l_n})}{b_-(\xi_{l_n},\xi_{l_{q}})\theta(\xi_{l_{q}},\xi_{l_n})} 
{\displaystyle \prod^{M+1}_{m=2}}b_-(\mu_m,\xi_{l_q})  \\
&& \times   {\displaystyle \sum_{\sigma \in S^{(q)}_M}}{\displaystyle \prod^{M+1}_{i=2}}c_+(\mu_i,\xi_{l_{\sigma_i}}) {\displaystyle  \prod_{2 \le j < k \le M+1}} \frac{b_-(\mu_k,\xi_{l_{\sigma_j}})a_-(\mu_j,\xi_{l_{\sigma_k}}) }{b_-(\xi_{l_{\sigma_k}},\xi_{l_{\sigma_j}})\theta(\xi_{l_{\sigma_j}},\xi_{l_{\sigma_k}})},
\end{array}\end{equation*}
where the sum over the permutations with superscript $q$ is given by,
\begin{equation*}\begin{array}{lll}
 {\displaystyle \sum_{\sigma \in S^{(q)}_M} \equiv \sum^{M+1}_{\genfrac{}{}{0pt}{}{\sigma_2=1}{\ne q} } \sum^{M+1}_{\genfrac{}{}{0pt}{}{\sigma_3=1}{\ne q} } \dots \sum^{M+1}_{\genfrac{}{}{0pt}{}{\sigma_{M+1}=1}{\ne q} }} &\textrm{for}& \sigma_2 \ne \sigma_3 \ne \dots \ne \sigma_{M+1}  .
\end{array}\end{equation*}
The verification of the proposition follows immediately through the change in label, $ q \rightarrow \sigma_{1}$. $\square$
\subsection{DWPF of type C}
\subsubsection{Twisted recurrence relation}
Focusing on $\,_{(l_1\dots l_M)}\langle 0 | \tilde{C}_{l_1 \dots l_M}(\nu_M) \dots \tilde{C}_{l_1 \dots l_M}(\nu_1)  |1\rangle_{(l_1\dots l_M)}$, we insert a complete set of states between the operators $\tilde{C}_{l_1 \dots l_M}(\nu_{2})$ and $\tilde{C}_{l_1 \dots l_M}(\nu_{1})$ to obtain,
\begin{equation*}
 \begin{array}{ll}
 & \,_{(l_1\dots l_M)}\langle 0 | \tilde{C}_{l_1 \dots l_M}(\nu_M) \dots \tilde{C}_{l_1 \dots l_M}(\nu_1)  |1\rangle_{(l_1\dots l_M)}\\
=&  \sum^M_{q=1} \,_{(l_1\dots l_M)}\langle 0 |\tilde{C}_{l_1 \dots l_M}(\nu_{M}) \dots \tilde{C}_{l_1 \dots l_M}(\nu_{2})  | q \rangle_{(l_1\dots l_M)} \,_{(l_1\dots l_M)}\langle q | \tilde{C}_{l_1 \dots l_M}(\nu_1) |1\rangle_{(l_1\dots l_M)},
\end{array}
\end{equation*}
where we use the same notation for $| q \rangle_{(l_1\dots l_M)} $ and $ \,_{(l_1\dots l_M)}\langle q |$ provided in Eq. (\ref{style}). Since the action of $e^{(12)}$ on the basis vectors of the transpose ferromagnetic states (\ref{ferroT}) is zero, one can use matrix multiplication to obtain the following expression for $\,_{(l_1\dots l_M)}\langle q | \tilde{C}_{l_1 \dots l_M}(\nu_1) |1\rangle_{(l_1\dots l_M)}$,
\begin{equation*}
\,_{(l_1\dots l_M)}\langle q | \tilde{C}_{l_1 \dots l_M}(\nu_1) |1\rangle_{(l_1\dots l_M)}\\ =c_-(\nu_1,\xi_{l_{q}})  {\displaystyle \prod^{M}_{\genfrac{}{}{0pt}{}{j=1}{\ne q} }} \frac{a_-(\nu_1,\xi_{l_j})}{\theta(\xi_{l_q},\xi_{l_j})} .
\end{equation*}
Concentrating on $\,_{(l_1\dots l_M)}\langle 0 |\tilde{C}_{l_1 \dots l_M}(\nu_{M}) \dots \tilde{C}_{l_1 \dots l_M}(\nu_{2})  | q \rangle_{(l_1\dots l_M)}$, we apply a similar method to that detailed in Section (\ref{simp2}) to decrease the relevant number of vector spaces in the reference states from $M$ to $M-1$ to obtain,
\begin{equation*}
\begin{array}{ll}
&\,_{(l_1\dots l_M)}\langle 0 |\tilde{C}_{l_1 \dots l_M}(\nu_{M}) \dots \tilde{C}_{l_1 \dots l_M}(\nu_{2})  | q \rangle_{(l_1\dots l_M)} \\
=&  {\displaystyle\prod^{M}_{i=2}} b_-(\nu_{i},\xi_{l_q}){\displaystyle\prod^{M}_{\genfrac{}{}{0pt}{}{j=1}{\ne q}  } } \frac{ 1}{ b_-(\xi_{l_{j}},\xi_{l_q})}  \,_{(\genfrac{}{}{0pt}{}{l_1 \dots l_M}{\ne l_q})}\langle 0 | \tilde{C}_{\genfrac{}{}{0pt}{}{l_1 \dots l_M}{\ne l_q}}(\nu_{M})\dots \tilde{C}_{\genfrac{}{}{0pt}{}{l_1 \dots l_M}{\ne l_q}}(\nu_{2}) | 1 \rangle_{(\genfrac{}{}{0pt}{}{l_1 \dots l_M}{\ne l_q})} ,
\end{array}
\end{equation*}
where,
\begin{equation*}
 \tilde{C}_{\genfrac{}{}{0pt}{}{l_1 \dots l_M}{\ne l_q}}(\nu)= \sum^M_{\genfrac{}{}{0pt}{}{p=1}{\ne q}} c_-(\nu,\xi_{l_p}) e^{(12)}_{l_p} \otimes^{M}_{\genfrac{}{}{0pt}{}{i=1}{\ne p,q}} \left(  \begin{array}{cc} 
\frac{b_-(\nu,\xi_{l_i})}{b_-(\xi_{l_p},\xi_{l_i})}  & 0  \\
 0  & \frac{a_-(\nu,\xi_{l_i})}{\theta(\xi_{l_p},\xi_{l_i})}
\end{array}\right)_{l_i}  
\end{equation*}
and, 
\begin{equation*}\begin{array}{lll}
  \,_{(\genfrac{}{}{0pt}{}{l_1 \dots l_M}{\ne l_q})}\langle 0 | = \otimes^M_{\genfrac{}{}{0pt}{}{i=1}{\ne q}}\left( 1,0 \right)_{l_i} & , & |1 \rangle_{(\genfrac{}{}{0pt}{}{l_1 \dots l_M}{\ne l_q})} = \otimes^M_{\genfrac{}{}{0pt}{}{i=1}{\ne q}}\left( \begin{array}{l} 0\\  1  \end{array} \right)_{l_i}.
\end{array}\end{equation*}
Thus we obtain the following twisted recursion relation,
\begin{equation}
 \begin{array}{r}
\,_{(l_1 \dots l_M)}\langle 0 |\tilde{C}_{l_1 \dots l_M}(\nu_{M}) \dots \tilde{C}_{l_1 \dots l_M}(\nu_{1})  | 1 \rangle_{(l_1 \dots l_M)}
={\displaystyle \sum^M_{q=1}} c_-(\nu_1,\xi_{l_{q}})   
 {\displaystyle \prod^{M}_{\genfrac{}{}{0pt}{}{j=1}{\ne q} }} \frac{a_-(\nu_1,\xi_{l_j})}{b_-(\xi_{l_{j}},\xi_{l_q})\theta(\xi_{l_q},\xi_{l_j})} \\
{\displaystyle \prod^{M}_{m=2}} b_-(\nu_{m},\xi_{l_q}) 
\,_{(\genfrac{}{}{0pt}{}{l_1 \dots l_M}{\ne l_q})}\langle 0 | \tilde{C}_{\genfrac{}{}{0pt}{}{l_1 \dots l_M}{\ne l_q}}(\nu_{M})\dots \tilde{C}_{\genfrac{}{}{0pt}{}{l_1 \dots l_M}{\ne l_q}}(\nu_{2}) | 1 \rangle_{(\genfrac{}{}{0pt}{}{l_1 \dots l_M}{\ne l_q})}.
\end{array}
\label{twistedrecursionC}\end{equation}
We now focus on untwisting the above equation.
\subsubsection{Untwisting the twisted DWPF} Applying Eqs. (\ref{Fwith0}) and (\ref{Fwith1}) and using the labels,
\begin{equation*}\begin{array}{lll}
 \,_{(l_1 \dots l_M)}\langle 0 |{C}_{l_1 \dots l_M}(\nu_{M}) \dots {C}_{l_1 \dots l_M}(\nu_{1})  | 1 \rangle_{(l_1 \dots l_M)} &= & Z^{(C)}_M(\{\nu\},\{\xi_{l}\})\\
\,_{(\genfrac{}{}{0pt}{}{l_1 \dots l_M}{\ne l_q})}\langle 0 | C_{\genfrac{}{}{0pt}{}{l_1 \dots l_M}{\ne l_q}}(\nu_{M})\dots C_{\genfrac{}{}{0pt}{}{l_1 \dots l_M}{\ne l_q}}(\nu_{2}) | 1 \rangle_{(\genfrac{}{}{0pt}{}{l_1 \dots l_M}{\ne l_q})} &=& Z^{(C)}_{M-1}(\{\nu\},\{\xi_{l}\}|\hat{\nu}_1,\hat{\xi}_{l_q}),
\end{array}\end{equation*}
Eq. (\ref{twistedrecursionC}) becomes the following recurrence relation,
\begin{equation}\begin{array}{lll}
Z^{(C)}_M(\{\nu\},\{\xi_{l}\})&=& {\displaystyle \sum^M_{q=1}} c_-(\nu_1,\xi_{l_{q}})   {\displaystyle \prod^{M}_{\genfrac{}{}{0pt}{}{j=1}{\ne q} }} \frac{a_-(\nu_1,\xi_{l_j})\theta(\xi_{l_j},\xi_{l_q})}{b_-(\xi_{l_{j}},\xi_{l_q})}   \\
&&   { {\displaystyle \prod^{M}_{m=2}}} b_-(\nu_{m},\xi_{l_q})   Z^{(C)}_{M-1}(\{\nu\},\{\xi_{l}\}|\hat{\nu}_1,\hat{\xi}_{l_q}).
\end{array}
\label{untwistedrecC}\end{equation}
\subsubsection{Explicit solution} For $M = \{1,2\}$ we obtain,
\begin{equation*}
 \begin{array}{lll}
  Z^{(C)}_1(\nu_1,{\xi}_{l_1}) &=& c_-(\nu_1,\xi_{l_1})\\
 Z^{(C)}_2(\{\nu\},\{\xi_{l}\}) &=& \frac{c_-(\nu_1,\xi_{l_1})c_-(\nu_2,\xi_{l_2})b_-(\nu_2,\xi_{l_1})a_-(\nu_1,\xi_{l_2})}{b_-(\xi_{l_2},\xi_{l_1})}\\
&& + \frac{c_-(\nu_1,\xi_{l_2})c_-(\nu_2,\xi_{l_1})b_-(\nu_2,\xi_{l_2})a_-(\nu_1,\xi_{l_1})a_-(\xi_{l_1},\xi_{l_2})}{b_-(\xi_{l_1},\xi_{l_2})}.
 \end{array}
\end{equation*}
We now offer the following general result.
\begin{proposition}
 \begin{equation}
 Z^{(C)}_M(\{\nu\},\{\xi_{l}\})= {\displaystyle \sum_{\sigma \in S_M}} {\displaystyle \prod^M_{i=1}}c_-(\nu_i,\xi_{l_{\sigma_i}})  {\displaystyle \prod_{1 \le j < k \le M}} \frac{b_-(\nu_k,\xi_{l_{\sigma_j}})a_-(\nu_j,\xi_{l_{\sigma_k}})\theta(\xi_{l_{\sigma_k}},\xi_{l_{\sigma_j}})}{b_-(\xi_{l_{\sigma_k}},\xi_{l_{\sigma_j}})} .
\label{explicitZCM} \end{equation}
\end{proposition}
\textbf{Proof.} The verification of the above expression follows exactly from the proof of Proposition \ref{propexpDWPF}. $\square$
\subsection{Explicit form for the scalar product} Applying the results of this section, we obtain the following explicit form for the scalar product,
\begin{equation}
  \begin{array}{lll}
 \mathcal{S}( \{\mu\}, \{\nu\} )
&=&  \sum_{1 \le l_1 < \dots < l_M \le L}   \prod^M_{m=1} \prod^L_{\genfrac{}{}{0pt}{}{p = 1}{\ne l_1, \dots, l_M}} \frac{a_-(\mu_m,\xi_p)a_-(\nu_m,\xi_p)}{b_-(\xi_{p},\xi_{l_m})a_-(\xi_{l_m},\xi_p)}\\
&& \times  Z^{(C)}_M(\{\nu\},\{\xi_{l}\}) Z^{(B)}_M(\{\mu\},\{\xi_{l}\}) \\
&=& \sum_{1 \le l_1 < \dots < l_M \le L} \sum_{\{\sigma, \tau\} \in S_M}   \prod^M_{m=1} \prod^L_{\genfrac{}{}{0pt}{}{p = 1}{\ne l_1, \dots, l_M}} \frac{a_-(\mu_m,\xi_p)a_-(\nu_m,\xi_p)}{b_-(\xi_{p},\xi_{l_m})a_-(\xi_{l_m},\xi_p)} \\
&& \times \prod^M_{i=1}c_-(\nu_i,\xi_{l_{\sigma_i}})c_+(\mu_i,\xi_{l_{\tau_i}}) \prod_{1 \le j < k \le M} \frac{\theta(\xi_{l_{\sigma_k}},\xi_{l_{\sigma_j}})}{\theta(\xi_{l_{\tau_j}},\xi_{l_{\tau_k}})} \\
&& \times \frac{b_-(\nu_k,\xi_{l_{\sigma_j}})b_-(\mu_k,\xi_{l_{\tau_j}})a_-(\nu_j,\xi_{l_{\sigma_k}})a_-(\mu_j,\xi_{l_{\tau_k}})}{ b_-(\xi_{l_{\sigma_k}},\xi_{l_{\sigma_j}}) b_-(\xi_{l_{\tau_k}},\xi_{l_{\tau_j}})}  .
 \end{array}
\label{scalarexpl}
\end{equation}
\section{Application of results}\label{7} In this section we would like to offer an explicit example of our results in the context of an asymmetric six-vertex model that does not satisfy the free-fermion condition. As discussed in \cite{BA,BA2}, such a model contains not only the azimuthal anisotropy but also the presence of electric fields in both the horizontal and vertical directions. From the Yang-Baxter relations one concludes that while the anisotropy parameter is fixed, the electric fields can be site dependent. In this general situation the corresponding $R$-matrix in terms of rational spectral parameters can be written as,
\begin{equation}
 R\left(\begin{array}{cc} \mu & \xi \\ x & z \end{array} \right) = \left( \begin{array}{cccc}
                                  1 & 0 & 0 & 0 \\
0 & \frac{\sqrt{\rho}}{x}\frac{\mu-\xi}{\mu \rho - \xi} & (\rho-1)\sqrt{\frac{z}{x}}\frac{\sqrt{\mu \xi}}{\mu \rho - \xi} & 0\\
0 &  (\rho-1)\sqrt{\frac{z}{x}}\frac{\sqrt{\mu \xi}}{\mu \rho - \xi} & \sqrt{\rho}z \frac{\mu-\xi}{\mu \rho - \xi} & 0 \\
0 & 0 & 0 & \frac{z}{x}
                                 \end{array}\right),
\label{Rodrigo}\end{equation}
where the global complex parameter $\rho$ denotes the azimuthal anisotropy.

The parameters $\mu$ and $\xi$ are the site dependent additive spectral parameters present in the symmetrical six-vertex model. The parameters $x$ and $z$ represent the site dependent electric fields on the vertical and horizontal directions and they are clearly non-additive. Here we stress that the above $R$-matrix satisfies the following Yang-Baxter relation,
\begin{equation*}\begin{array}{ll}
& R_{12}\left(\begin{array}{cc} \mu_1 & \mu_2 \\ x_1 & x_2 \end{array} \right)R_{13}\left(\begin{array}{cc} \mu_1 & \mu_3 \\ x_1 & x_3 \end{array} \right)R_{23}\left(\begin{array}{cc} \mu_2 & \mu_3 \\ x_2 & x_3 \end{array} \right)\\
= & R_{23}\left(\begin{array}{cc} \mu_2 & \mu_3 \\ x_2 & x_3 \end{array} \right)R_{13}\left(\begin{array}{cc} \mu_1 & \mu_3 \\ x_1 & x_3 \end{array} \right)R_{12}\left(\begin{array}{cc} \mu_1 & \mu_2 \\ x_1 & x_2 \end{array} \right),
\end{array}\end{equation*}
in both the additive and non-additive spectral parameters. 

The above relation allows us to build up the following transfer matrix,
\begin{equation*}\begin{array}{lll}
 \mathcal{T}_{a,1\dots L}\left(\begin{array}{c} \mu \\ x \end{array} \right) &=& R_{aL}\left(\begin{array}{cc} \mu & \xi_{L} \\ x & z_{L} \end{array} \right)\dots R_{a 1}\left(\begin{array}{cc} \mu & \xi_{1} \\ x & z_{1} \end{array} \right)\\
&=& \left( \begin{array}{cc}
            A_{1\dots L}\left(\begin{array}{c} \mu \\ x \end{array} \right) & B_{1\dots L}\left(\begin{array}{c} \mu \\ x \end{array} \right)\\
	   C_{1\dots L}\left(\begin{array}{c} \mu \\ x \end{array} \right) & D_{1\dots L}\left(\begin{array}{c} \mu \\ x \end{array} \right)
           \end{array}\right)_a,
\end{array}\end{equation*}
with two types of inhomogeneities $\{\xi_1,\dots,\xi_L\}$ and $\{z_1,\dots,z_L\}$ associated to 
the additive and non-additive rapidities, respectively. 

We recall that much work done in this model as far as DWPF and scalar products are concerned
has been restricted to consider only the additive inhomogeneities \cite{KO1,IK,IZ,KMT,PRO}. In what follows we shall consider 
the most general possible situation which includes electric fields and inhomogeneities. We shall present the corresponding explicit expressions for the DWPF's and scalar product.
\subsection{Determinant formula for the DWPF of type B} Using the results of the past sections of this work, the DWPF of type $B$ is defined as,
\begin{equation*}
Z^{(B)}_M \left(\begin{array}{ll}\{\mu\} &\{\xi\} \\ \{x\} & \{z\} \end{array} \right) = \,_{(1\dots M)}\langle 1 | B_{1\dots M}\left(\begin{array}{c} \mu_1 \\ x_1 \end{array} \right) \dots B_{1\dots M}\left(\begin{array}{c} \mu_M \\ x_M \end{array} \right) |0\rangle_{(1\dots M)},
\end{equation*}
for which we offer the following determinant formula,
\begin{equation}\begin{array}{lll}
Z^{(B)}_M \left(\begin{array}{ll}\{\mu\} &\{\xi\} \\ \{x\} & \{z\} \end{array} \right) &=&  \left( \frac{z_1}{x_1} \right)^{\frac{2M-1}{2}}  \left( \frac{z_2}{x_{2}} \right)^{ \frac{2M-3}{2}} \dots \left( \frac{z_M}{x_M} \right)^{\frac{1}{2}}  \\
&& \times \frac{\prod^M_{k=1}\sqrt{\mu_k \xi_k}{ \prod^M_{mn=1}} (\mu_m - \xi_n)}{{ \prod_{1 \le p < q \le M}}(\mu_p-\mu_q)(\xi_q-\xi_p)}\textrm{det} \left[\frac{\rho-1}{(\mu_i \rho - \xi_j)(\mu_i-\xi_j)} \right]_{ij=1\dots M}\\
&=&\left( \frac{z_1}{x_1} \right)^{\frac{2M-1}{2}}  \left( \frac{z_2}{x_{2}} \right)^{ \frac{2M-3}{2}} \dots \left( \frac{z_M}{x_M} \right)^{\frac{1}{2}} Z^{(Izer)}_M(\{\mu\},\{\xi\}).
\end{array}
\label{RodrigoDWPFB}\end{equation}
We recognize that,
\begin{equation*}\begin{array}{l}
Z^{(Izer)}_M(\{\mu\},\{\xi\}) =   \frac{\prod^M_{k=1}\sqrt{\mu_k \xi_k}{ \prod^M_{mn=1}} (\mu_m - \xi_n)}{{ \prod_{1 \le p < q \le M}}(\mu_p-\mu_q)(\xi_q-\xi_p)}\textrm{det} \left[\frac{\rho-1}{(\mu_i \rho - \xi_j)(\mu_i-\xi_j)} \right]_{ij=1\dots M},
\end{array}\end{equation*}
is the usual Izergin-type DWPF of the symmetric six-vertex model \cite{KO1,IZ,KMT}.

To show the validity of Eq. (\ref{RodrigoDWPFB}), we substitute it into the recursion relation given by Eq. (\ref{untwistedrecursion}),
\begin{equation*}
\begin{array}{ll}
& Z^{(B)}_M\left(\begin{array}{ll}\{\mu\} & \{\xi\} \\ \{x\} & \{z\} \end{array} \right) =  \sum^M_{q=1}c_+ \left(\begin{array}{ll}  \mu_1 & \xi_{{q}} \\ x_1 & z_q \end{array} \right) 
  \prod^{M}_{\genfrac{}{}{0pt}{}{j=1}{\ne q} } \frac{a_-\left( \begin{array}{ll} \mu_1 & \xi_{j} \\ x_1 & z_j \end{array} \right)  }{b_- \left( \begin{array}{ll} \xi_{j} & \xi_{{q}} \\ z_j & z_q \end{array} \right)}   \\
& \times  \prod^{M}_{i=q+1}a_- \left( \begin{array}{ll} \xi_{i} & \xi_{{q}} \\ z_i & z_q \end{array} \right)  \prod^{M}_{m=2}b_- \left( \begin{array}{ll} \mu_m & \xi_{q} \\ x_m & z_q \end{array} \right)   Z^{(B)}_{M-1}\left( \begin{array}{ll} \hat{\mu}_1 & \hat{\xi}_{q} \\ \hat{x}_1 & \hat{z}_q \end{array} \right).
\end{array}
\end{equation*}
In doing so, the rapidities $\{x\}$ and inhomogeneities $\{z\}$ on either side of the expression cancel. Thus we are left to prove the validity of the following expression,
\begin{equation} 
 \begin{array}{lll}
  \textrm{det} \left[\phi(\mu_i, \xi_j) \right]_{ij=1\dots M} &=& \frac{1}{\Phi_1}\sum^M_{p=1} (-1)^{p-1} g_p \textrm{det} \left[\phi(\mu_i, \xi_j) \right]_{\genfrac{}{}{0pt}{}{i=2\dots M}{j=1\dots (p-1)(p+1) \dots M} },
 \end{array}
\label{polyeq}\end{equation}
where we have used the labels,
\begin{equation*}
 \begin{array}{lll}
  \Phi_k = \frac{\prod^M_{m=1}(\mu_k-\xi_m)}{\prod^M_{\genfrac{}{}{0pt}{}{n=1}{\ne k}}(\mu_k-\mu_n)}, & g_p = \frac{\prod^M_{\genfrac{}{}{0pt}{}{m=1}{\ne p}}(\xi_m \rho-\xi_p)}{\prod^M_{n=1}(\mu_n \rho-\xi_p)},&\phi(\mu_i, \xi_j) = \frac{1}{(\mu_i q - \xi_j)(\mu_i-\xi_j)}.
 \end{array}
\end{equation*}
An equivalent expression was considered in \cite{PRO} to show the validity of the determinant formula of the DWPF of the symmetric six vertex model.

To verify Eq. (\ref{polyeq}) we apply the polynomial identity,
\begin{proposition}\label{POLLY1}
\begin{equation}\begin{array}{lll}
{\displaystyle g_i = \sum^M_{k=1}\Phi_k \phi(\mu_k,\xi_i)} &\textrm{for}& i = (1,\dots,M).
\end{array}\label{polyiden}\end{equation}
\end{proposition}
The verification of the above identity is possible through the application of \textit{Liouville's theorem}. We offer a complete proof in Appendix \ref{APPPOLLY1}. 

To apply Eq. (\ref{polyiden}) to verify Eq. (\ref{polyeq}), we express Eq. (\ref{polyiden}) as the following matrix equation,
\begin{equation}
 \left( \begin{array}{c} g_1 \\ \vdots \\ g_M \end{array} \right) =  \left( 
\begin{array}{ccc} \phi(\mu_1,\xi_1) & \dots & \phi(\mu_M,\xi_1) \\
\vdots & \ddots & \vdots \\
\phi(\mu_1,\xi_M) & \dots & \phi(\mu_M,\xi_M) \end{array}
 \right) \left( \begin{array}{c}  \Phi_1 \\   \vdots \\  \Phi_M \end{array} \right).
\label{matrixforC}\end{equation}
Using Cramer's rule to find the expression for $\Phi_1$, we obtain exactly Eq. (\ref{polyeq}), hence verifying the determinant form for the DWPF of type $B$.
\subsection{Determinant formula for the DWPF of type C} In the notation employed in this section, the DWPF of type $C$ is defined as,
\begin{equation*}
 Z^{(C)}_M \left(\begin{array}{ll}\{\nu\} &\{\xi\} \\ \{y\} & \{z\} \end{array} \right) = \,_{(1\dots M)}\langle 0 | C_{1\dots M}\left(\begin{array}{c} \nu_M \\ y_M \end{array} \right) \dots C_{1\dots M}\left(\begin{array}{c} \nu_1 \\ y_1 \end{array} \right) |1\rangle_{(1\dots M)}
\end{equation*}
where we offer the following determinant formula,
\begin{equation}\begin{array}{l}
 Z^{(C)}_M \left(\begin{array}{ll} \{\nu\} & \{\xi\} \\ \{y\} & \{z\} \end{array} \right) =\left( \frac{z_M}{y_1} \right)^{\frac{2M-1}{2}}  \left( \frac{z_{M-1}}{y_{2}} \right)^{ \frac{2M-3}{2}} \dots \left( \frac{z_1}{y_M} \right)^{\frac{1}{2}} Z^{(Izer)}_M(\{\nu\},\{\xi\}).
\end{array}
\label{RodrigoDWPFC}\end{equation}
As with $Z^{(B)}$, to show that the above determinant form of $Z^{(C)}$ is valid we substitute it into the corresponding recursion relation given by Eq. (\ref{untwistedrecC}),
\begin{equation*}
\begin{array}{ll}
& Z^{(C)}_M\left(\begin{array}{ll} \{\nu\} & \{\xi\} \\ \{y\} & \{z\} \end{array} \right) =  \sum^M_{q=1} c_- \left(\begin{array}{ll}  \nu_1 & \xi_{{q}} \\ y_1 & z_q \end{array} \right) 
  \prod^{M}_{\genfrac{}{}{0pt}{}{j=1}{\ne q} } \frac{a_-\left( \begin{array}{ll} \nu_1 & \xi_{j} \\ y_1 & z_j \end{array} \right)  }{b_- \left( \begin{array}{ll} \xi_{j} & \xi_{{q}} \\ z_j & z_q \end{array} \right)}   \\
& \times   \prod^{q-1}_{i=1} a_- \left( \begin{array}{ll} \xi_{i} & \xi_{{q}} \\ z_i & z_q \end{array} \right)    \prod^{M}_{m=2}b_- \left( \begin{array}{ll} \nu_m & \xi_{q} \\ y_m & z_q \end{array} \right) Z^{(C)}_{M-1}\left( \begin{array}{ll} \hat{\nu}_1 & \hat{\xi}_{q} \\ \hat{y}_1 & \hat{z}_q \end{array} \right).
\end{array}
\end{equation*}
In doing so, the rapidities $\{y\}$ and inhomogeneities $\{z\}$ on either side of the expression cancel and we obtain exactly Eq. (\ref{polyeq}) with rapidities $\mu_j$ replaced by $\nu_j$, hence proving the validity of the determinant form of DWPF of type $C$.
\subsection{Scalar product I - all rapidities are free} Using the notation established in this section, we define the scalar product with rapidities $\{\mu\}$, $\{x\}$, $\{\nu\}$ and $\{y\}$, and inhomogeneities $\{\xi\}$ and $\{z\}$ as,
\begin{equation*}\begin{array}{l}
 \mathcal{S}\left( \begin{array}{cc} \{\mu\} & \{\nu\} \\ \{x\} & \{y\} \end{array} \right) 
= \langle 1| B_{1\dots L} \left( \begin{array}{c} \mu_1 \\ x_1 \end{array} \right) \dots B_{1\dots L} \left( \begin{array}{c} \mu_M \\ x_M \end{array} \right) C_{1\dots L} \left( \begin{array}{c} \nu_M \\ y_M \end{array} \right) \dots C_{1\dots L} \left( \begin{array}{c} \nu_1 \\ y_1 \end{array} \right)|1 \rangle
\end{array}\end{equation*}
Substituting the expressions for $Z^{(B)}_M$ (\ref{RodrigoDWPFB}) and $Z^{(C)}_M$ (\ref{RodrigoDWPFC}) into Eq. (\ref{scalarexpl}) we obtain,
\begin{equation}
  \mathcal{S}\left( \begin{array}{cc} \{\mu\} & \{\nu\} \\ \{x\} & \{y\} \end{array} \right) =
 \frac{ \prod^L_{l=1}z^M_l }{\prod^M_{m=1} (x_m y_m)^{L+\frac{1}{2}-m}} \mathcal{S}^{(sym)} (\{\mu\} , \{\nu\}) ,
\label{NON-SYM}\end{equation}
where we recognize that, 
\begin{equation*}\begin{array}{lll}
\mathcal{S}^{(sym)} (\{\mu\} , \{\nu\}) &=& \rho^{-\frac{M(L-M)}{2}}{\displaystyle \sum_{1 \le l_1 < \dots < l_M \le L}}  {\displaystyle \prod^M_{m=1} \prod^L_{\genfrac{}{}{0pt}{}{n = 1}{\ne l_1, \dots, l_M}}} \frac{\xi_{n}\rho - \xi_{l_m}}{\xi_{n} -\xi_{l_m}}\\
&&Z^{({Izer})}_M(\{\mu\},\{\xi_l\}) Z^{({Izer})}_M(\{\nu\},\{\xi_l\}) ,
\end{array}\end{equation*}
is the scalar product of the trigonometric symmetric six-vertex model where both sets of rapidities, $\{\nu\}$ and $\{\mu\}$ are free. We remark that $\mathcal{S}^{(sym)}$ can be constructed from the following expression,
\begin{equation}
\mathcal{S}^{(sym)} (\{\mu\} , \{\nu\}) = \langle 1| \tilde{\mathcal{B}}_{1\dots L} (  \mu_1 ) \dots \tilde{\mathcal{B}}_{1\dots L} (  \mu_M ) \tilde{\mathcal{C}}_{1\dots L} ( \nu_M ) \dots \tilde{\mathcal{C}}_{1\dots L} (\nu_1 )|1 \rangle,
\label{SYM-SYM}\end{equation}
where the twisted monodromy operators $\tilde{\mathcal{B}}_{1\dots L} (  \mu )$ and $\tilde{\mathcal{C}}_{1\dots L} (  \nu )$ are the operators associated with the symmetric six-vertex model \cite{MS}, given explicitly by,
\begin{eqnarray}
\tilde{\mathcal{B}}_{1\dots L}(\mu) &=& \sum^L_{l=1}c(\mu,\xi_l) e^{(21)}_l \otimes^{L}_{\genfrac{}{}{0pt}{}{i=1}{\ne l}} \left(  \begin{array}{cc} 
b(\mu,\xi_i)  & 0  \\
 0  & \frac{1}{b(\xi_i,\xi_l)}
\end{array}\right)_i   \label{Btil}\\
 \tilde{\mathcal{C}}_{1\dots L}(\nu) &=& \sum^L_{l=1}c(\nu,\xi_l) e^{(12)}_l\otimes^{L}_{\genfrac{}{}{0pt}{}{i=1}{\ne l}} \left(  \begin{array}{cc} 
\frac{b(\nu,\xi_i)}{b(\xi_l,\xi_i)}  & 0  \\
 0  & 1
\end{array}\right)_i   \label{Ctil},
\end{eqnarray}
where the weights $b$ and $c$ (without $+$ or $-$ subscripts) are,
\begin{equation}
b(\mu,\xi) = \sqrt{\rho} \frac{\mu-\xi}{\mu \rho - \xi} \textrm{    ,    } c(\mu,\xi) =  (\rho-1)\frac{\sqrt{\mu \xi}}{\mu \rho - \xi}.
\end{equation}
In the next section we impose that the rapidities $\{\nu\}$ satisfy the Bethe equations, given by Eq. (\ref{BETHEGEN}), and obtain a Slavnov-type \cite{Slav} determinant expression for the scalar product.
\subsection{Scalar product II - on-shell Slavnov-type determinant expression}
\subsubsection{Bethe equations} Expressing the Bethe equations (\ref{BETHEGEN}) in the notation employed in this section we obtain,
\begin{eqnarray*}
\prod^L_{k=1}  \frac{a_-\left( \begin{array}{cc} \nu_j & \xi_k \\ y_j & z_k \end{array} \right)}{b_+ \left( \begin{array}{cc} \nu_j & \xi_k \\ y_j & z_k \end{array} \right)}
=  \prod^M_{\genfrac{}{}{0pt}{}{m=1}{ \ne j}}\frac{b_+\left( \begin{array}{cc} \nu_m & \nu_j \\ y_m & y_j \end{array} \right)a_-\left( \begin{array}{cc} \nu_j & \nu_m \\ y_j & y_m \end{array} \right)}{b_+\left( \begin{array}{cc} \nu_j & \nu_m \\ y_j & y_m \end{array} \right)} &\textrm{for}& j=(1,\dots,M) .
\end{eqnarray*}
By using the explicit weights given in Eq. (\ref{Rodrigo}) the above expression becomes,
\begin{eqnarray}
\prod^L_{k=1} \frac{ \xi_k-\nu_j}{ \xi_k- \nu_j \rho} \prod^M_{\genfrac{}{}{0pt}{}{m=1}{ \ne j}} (\nu_m  - \nu_j \rho) = \rho^{-\frac{L}{2}} \prod^L_{k=1}z_k \prod^M_{\genfrac{}{}{0pt}{}{m=1}{ \ne j}}(\nu_m \rho - \nu_j ) &\textrm{for}& j=(1,\dots,M).
\label{EXPBETHE}\end{eqnarray}
We now proceed to show that by imposing that the rapidities $\{\nu\}$ satisfy Eq. (\ref{EXPBETHE}) the scalar product can be represented in a determinant form similar to that derived by Slavnov \cite{Slav} for the scalar product of the symmetric six-vertex model.
\subsubsection{Determinant expression} 
\begin{equation}
 \mathcal{S}\left( \begin{array}{cc} \{\mu\} & \{\nu\}_{\beta} \\ \{x\} & \{y\} \end{array} \right) =   \frac{\prod^L_{n=1}z^M_n}{\prod^M_{m=1} (x_m y_m)^{L+\frac{1}{2}-m}}\frac{\rho^{\frac{M(L-M)}{2}}\prod^M_{k=1}\sqrt{\mu_k \nu_k} }{\prod_{1 \le q < l \le M}(\mu_l-\mu_q)(\nu_q - \nu_l)}  \textrm{det} \left[\mathcal{H}_{ij} \right]_{ij=1 \dots M},
\label{sui6}\end{equation}
where,
\begin{equation}
  \mathcal{H}_{ij} = \frac{\rho-1}{\nu_i-\mu_j}\left\{\rho^{-\frac{L}{2}}  \prod^L_{k=1}z_k {\displaystyle \prod^M_{\genfrac{}{}{0pt}{}{m=1}{\ne i }}} (\nu_m \rho - \mu_{j}) - {\displaystyle \prod^L_{k=1}} \frac{\mu_j  - \xi_k}{\mu_j\rho - \xi_k} {\displaystyle  \prod^M_{\genfrac{}{}{0pt}{}{m=1}{\ne i }} } (\nu_m  - \mu_j\rho) \right\}.
\label{theH}\end{equation}
We briefly remark that the above expression for the matrix elements $ \mathcal{H}_{ij}$ offers a slight generalization to the original determinant formula devised by Slavnov \cite{Slav} due to the explicit dependence on the inhomogeneous electric fields - entering through the term $\prod^L_{k=1}z_k$. We also reiterate that in the above expression we are using the reference states $\langle 1 |$ and $|1\rangle$ as opposed to $\langle 0 |$ and $|0\rangle$. 

In order to offer a verification of the above expression we employ the same method used in \cite{Kita} and consider the so-called \textit{intermediate functions}, labeled $G^{(M-k)}_{p_1 \dots p_k}$. We note that since the details are almost the same in both cases, save for the Bethe equations themselves, we keep them to a minimum.
\subsubsection{Intermediate functions} 
The general idea of this procedure is to insert a complete set of states in between each $\tilde{\mathcal{B}}_{1\dots L}(\mu)$ operator in the scalar product (\ref{SYM-SYM}). The intermediate functions $G^{(M-k)}_{p_1 \dots p_k}$ are defined as,
\begin{equation}
G^{(M-k)}_{p_1 \dots p_k} = \langle p_1 \dots p_k | \tilde{\mathcal{B}}_{1\dots L}(\mu_{k+1}) \dots \tilde{\mathcal{B}}_{1\dots L}(\mu_M)  \tilde{\mathcal{C}}_{1\dots L}(\nu_{M}) \dots \tilde{\mathcal{C}}_{1\dots L}(\nu_1) |1 \rangle,
\label{labely}\end{equation}
where the bra $\langle p_1 \dots p_k |$ is given in Eq. (\ref{NICE}). 

For the value $k=0$ the intermediate function becomes the symmetric scalar product (\ref{SYM-SYM}). Additionally, for $k=M$ we obtain the Izergin-DWPF,
\begin{equation*}\begin{array}{lll}
G^{(0)}_{p_1 \dots p_{M}} &=& \langle p_1 \dots p_M | \tilde{\mathcal{C}}_{1\dots L}(\nu_{M}) \dots \tilde{\mathcal{C}}_{1\dots L}(\nu_1) |1 \rangle\\
&=& Z^{(Izer)}_M(\{\nu\},\{\xi_p\}).
\end{array}
\end{equation*}
Since the action of $e^{(21)}$ on the basis vectors of the ferromagnetic state (\ref{ferro}) produces zero, we can use elementary matrix multiplication to show that the intermediate functions satisfy the following recurrence relation,
\begin{equation}\begin{array}{lll}
G^{(M-k)}_{p_1\dots p_k} &=&  \sum^L_{p_{k+1}=1}\langle p_1\dots p_k| \tilde{\mathcal{B}}_{1\dots L}(\mu_{k+1}) | p_1 \dots p_{k+1} \rangle G^{(M-k-1)}_{p_1 \dots p_{k+1}}  \\
&=& \sum^L_{p_{k+1}=1}c(\mu_{k+1},\xi_{p_{k+1}})  \frac{\prod^k_{i=1}b(\mu_{k+1},\xi_{p_i})}{\prod^L_{\genfrac{}{}{0pt}{}{j=1}{\ne p_1 \dots p_{k+1}}} b(\xi_j,\xi_{p_{k+1}})} G^{(M-k-1)}_{p_1 \dots p_{k+1}}.
\end{array} \label{generalono}
\end{equation}
Beginning the procedure to construct the determinant expression, we set $k=M-1$ in Eq. (\ref{generalono}) to obtain, 
\begin{equation*} \begin{array}{lll}
G^{(1)}_{p_1\dots p_{M-1}}
&=&\rho^{M-\frac{L}{2}-\frac{1}{2}} \frac{\sqrt{\mu_M}\prod^M_{j_1=1}\sqrt{\nu_{j_1}}\prod^{M-1}_{j_2=1}\sqrt{\xi_{j_2}} \prod^M_{l_1=1}\prod^{M-1}_{l_2=1}(\nu_{l_1}-\xi_{p_{l_2}}) }{\prod_{1 \le r_1 < r_2 \le M}(\nu_{r_2}-\nu_{r_1})\prod_{1 \le s_1 < s_2 \le M-1}(\xi_{p_{s_1}}-\xi_{p_{s_2}}) }\prod^{M-1}_{h=1}\frac{\mu_M-\xi_{p_h}}{\mu_M \rho -\xi_{p_h}}\\
&&\times {\displaystyle \sum^L_{p_M =1}}\frac{(\rho-1) \xi_{p_M}}{\mu_M \rho-\xi_{p_M}} {\displaystyle \prod^L_{\genfrac{}{}{0pt}{}{k=1}{\ne p_M }}}\frac{\xi_{k} \rho - \xi_{p_M}}{\xi_{k} - \xi_{p_M}}\frac{\prod^M_{m=1} (\nu_m - \xi_{p_M})}{\prod^{M-1}_{n=1} (\xi_{p_n} \rho - \xi_{p_M} )} \textrm{det} \left[ \frac{\rho-1}{(\nu_i \rho - \xi_{p_j})(\nu_i - \xi_{p_j})} \right]_{ij=1 \dots M},
\end{array}\end{equation*}
where we have clearly differentiated the terms which do and do not contribute to the sum over $p_M$. 

Focusing solely on the sum $\sum^L_{p_M =1}$, we observe that only the final column in the determinant has terms containing the summation variable $p_M$. Hence we absorb the sum into the final column of the determinant and the corresponding entry $(i,M)$ of the determinant is given by,
\begin{equation}
\sum^L_{p_M =1}\frac{(\rho-1)^2 \xi_{p_M}}{(\mu_M \rho -\xi_{p_M})(\nu_i \rho - \xi_{p_M})} 
{\displaystyle \prod^L_{\genfrac{}{}{0pt}{}{k=1}{\ne p_M }}}\frac{\xi_{k} \rho - \xi_{p_M}}{\xi_{k} - \xi_{p_M}}\frac{\prod^M_{\genfrac{}{}{0pt}{}{m=1}{\ne i }} (\nu_m  - \xi_{p_M})}{\prod^{M-1}_{n=1} (\xi_{p_n} \rho - \xi_{p_M} )}.
\label{nasty1.1}\end{equation}
The strategy is to apply the Bethe equations to the above expression which has the effect transforming Eq. (\ref{nasty1.1}) into a single relevant term (proportional to $\mathcal{H}_{iM}$) and other irrelevant terms which are proportional to the remaining entries of the determinant, allowing us to discard them. To do achieve this goal we offer the following identity:
\begin{proposition}\label{SECONDPOLLY}
\begin{equation}\begin{array}{ll}
& {\displaystyle \sum^L_{p_{q} =1}}\frac{(\rho-1)^2}{\prod^M_{j=q}(\mu_j \rho-\xi_{p_{q}})} \frac{\xi_{p_{q}}}{\nu_i \rho - \xi_{p_{q}}}
{\displaystyle \prod^L_{\genfrac{}{}{0pt}{}{k=1}{\ne p_q }}}\frac{\xi_k \rho - \xi_{p_{q}} }{\xi_{k} - \xi_{p_q}}\frac{\prod^M_{\genfrac{}{}{0pt}{}{m=1}{\ne i }} (\nu_m  - \xi_{p_{q}})}{\prod^{q-1}_{n=1} (\xi_{p_{n}} \rho - \xi_{p_q} )}\\
=& {\displaystyle \sum^M_{l=q}}\frac{\rho^{L-M}\mathcal{H}_{i l}}{\prod^M_{\genfrac{}{}{0pt}{}{m=q}{\ne l }}(\mu_m-\mu_l) \prod^{q-1}_{n=1}( \xi_{p_n}-\mu_l)} + {\displaystyle  \sum^{q-1}_{l=1}}\kappa^{(q)}_l \frac{1}{(\nu_i \rho - \xi_{p_l})(\nu_i - \xi_{p_l})},
\end{array}
\label{gensui2}\end{equation}
where $1 \le q \le M$, $\mathcal{H}_{i l}$ is given in Eq. (\ref{theH}) and the co-efficients $\kappa^{(q)}_l $ are given by,
\begin{equation*}
 \kappa^{(q)}_l = \frac{(\rho-1)\rho^{\frac{L}{2}-M}}{\prod^M_{s=q}(\mu_s-\xi_{p_l})}\frac{\prod^L_{k=1}z_k \prod^M_{n=1}(\nu_n \rho - \xi_{p_l}) }{\prod^{q-1}_{\genfrac{}{}{0pt}{}{m=1}{\ne l }} (\xi_{p_m}-\xi_{p_l})}.
\end{equation*}
\end{proposition}
To verify the above identity, we notice that each term is a rational function in $\mu_l$ of order $\mu^{-1}_l$. With the application of the Bethe equations (\ref{EXPBETHE}) we can show that both sides of the identity have the same residue values, and hence by Liouville's theorem, they are equal. We offer a complete proof in Appendix \ref{APPPOLLY2}.

Applying the $q=M$ case of Eq. (\ref{gensui2}) to the sum in Eq. (\ref{nasty1.1}), we obtain the following simplified expression for the intermediate function $G^{(1)}_{p_1\dots p_{M-1}}$,
\begin{equation} 
 G^{(1)}_{p_1\dots p_{M-1}} = \frac{\rho^{\frac{L-1}{2}} \prod^M_{k=1}\sqrt{\nu_k}  }{     \prod_{1\le m < n \le M } (\nu_n - \nu_m)     }   \mathcal{M}^{(1)} \textrm{det} \left[ \frac{\rho-1}{(\nu_i \rho - \xi_{p_j})(\nu_i - \xi_{p_j})}, \mathcal{H}_{i M} \right]_{\genfrac{}{}{0pt}{}{i=1 \dots M}{j=1\dots M-1}},
\label{eqnono1.1}\end{equation}
where,
\begin{equation}\begin{array}{lll}
\mathcal{M}^{(l)} &=& (-1)^{\sum^{M-1}_{r=M-l}r} \frac{\prod^M_{j_1=M-(l-1)}\sqrt{\mu_{j_1}}  }{\prod_{M-(l-1) \le m_1 < m_2 \le M}(\mu_{m_2}-\mu_{m_1}) }  \\
&&\times \frac{\prod^{M-l}_{j_2=1}\sqrt{\xi_{p_{j_2}}} \prod^M_{j_3=1}\prod^{M-l}_{j_4=1} (\nu_{j_3}-\xi_{p_{j_4}})}{\prod_{1\le n_1 < n_2 \le M-l } (\xi_{p_{n_1}}- \xi_{p_{n_2}})\prod^M_{k_1=M-(l-1)} \prod^{M-l}_{k_2=1} (\mu_{k_1} \rho - \xi_{k_2})}   .
\end{array}\end{equation}
Following the procedure detailed above, we can perform an inductive proof to verify the following general result,
\begin{equation}
\begin{array}{l}
 G^{(l)}_{p_1\dots p_{M-l}} = \frac{\rho^{\frac{l(L-l)}{2}} \prod^M_{k=1}\sqrt{\nu_k}    }{      \prod_{1\le m < n \le M } (\nu_n - \nu_m)    }\mathcal{M}^{(l)}\textrm{det} \left[ \frac{\rho-1}{(\nu_i \rho - \xi_{p_j})(\nu_i - \xi_{p_j})}, \mathcal{H}_{i (M-l+1)},\dots,\mathcal{H}_{i M} \right]_{\genfrac{}{}{0pt}{}{i=1 \dots M}{j=1\dots M-l}}.
\end{array}
\label{sui5}\end{equation}
Substituting $l=M$ in Eq. (\ref{sui5}) we obtain the determinant form for symmetric scalar product,
\begin{equation}
\mathcal{S}^{(sym)}(\{\mu\}, \{\nu\}_{\beta}) = \frac{\rho^{\frac{M(L-M)}{2}}\prod^M_{k=1}\sqrt{\mu_k \nu_k} }{\prod_{1 \le q < l \le M}(\mu_l-\mu_q)(\nu_q - \nu_l)}\textrm{det} \left[\mathcal{H}_{ij} \right]_{ij=1 \dots M},
\label{sui56}\end{equation}
and substituting this expression into Eq. (\ref{NON-SYM}) we obtain Eq. (\ref{sui6}).
\subsubsection{The KP hierarchy} We comment briefly that replacing the term $\rho^{-\frac{L}{2}} \prod^L_{k=1}z_k$ in the determinant expression (\ref{theH}), with the corresponding Bethe equation expression, i.e.
\begin{equation*}
 \rho^{-\frac{L}{2}}  \prod^L_{k=1}z_k = \prod^L_{k=1} \frac{\nu_i   - \xi_k}{\nu_i \rho - \xi_k}\prod^M_{\genfrac{}{}{0pt}{}{j=1}{ \ne i}} \frac{\nu_j  - \nu_i \rho}{\nu_j \rho  - \nu_i } ,
\end{equation*}
we eliminate the explicit dependence of the inhomogeneities $\{z\}$ from the determinant and obtain an expression for the scalar product which is proportional to that considered in \cite{KP}. Hence we can immediately see that the scalar product (\ref{sui6}) is proportional to a restricted KP $\tau$-function in the rapidities $\{\mu\}$. 
\section{Discussion}
In this paper we have shown how to derive the expressions of the monodromy matrix of a general six-vertex model in the basis that factorizes the respective $R$-matrix. We have been able to carry out a strictly algebraic analysis relying solely on the unitary and Yang-Baxter relations (\ref{one}-\ref{three}). As a result we obtained the explicit expressions of the DWPF's and the off-shell scalar product for arbitrary Boltzmann weights. We have applied our results for an asymmetrical six-vertex model in the presence of inhomogeneous electric fields. The corresponding parameterized weights contain both additive and non-additive spectral variables. By applying certain polynomial identities amongst the weights we have been able to obtain a determinant representation for the DWPF's and the on-shell scalar product. It is only at this stage that we required the use of explicit Boltzmann weight parameterizations. It is still an open question whether or not such determinant formulas, or some other equivalently useful forms can be derived using only the information granted to us from relations (\ref{one}-\ref{three}). A positive answer would be an extra indication that the use of weight parameterization to determine the eigenspectrum properties of the transfer matrix could be deferred until the very end. We expect that the approach pursued in this paper is not exclusive to the six-vertex model. We hope that a similar analysis can be carried out to other integrable models in which the $F$-basis has been studied on the basis of explicit weight parameterizations. Of particular interest are the integrable models solved by the nested Bethe ansatz \cite{BOS1,CH1,CH2}. 
\appendix
\section{Proof of Proposition \ref{propABCD}}\label{APP1}
We proceed by induction. Noticing that the proposition is true by inspection for $L=1$ due to the form of the $R$-matrix, we assume that it is true up to some $L-1$. We then decompose the monodromy matrix as follows,
\begin{equation*}
\mathcal{T}_{a,1\dots L}(\mu) = R_{a L}(\mu) R_{a, 1 \dots (L-1)}(\mu),
\end{equation*}
and express the above relation as a matrix equation in auxiliary space $\mathcal{A}_a$ to obtain,
\begin{equation*}\begin{array}{ll}
& \left( \begin{array}{cc}
A_{1\dots L}(\mu)&  B_{1\dots L}(\mu) \\
C_{1\dots L}(\mu)&  D_{1\dots L}(\mu)
\end{array}\right)_{a}\\
=& \left( \begin{array}{cc}
\mathcal{I}_{1\dots (L-1)}\otimes \left(\begin{array}{cc} 1 & 0 \\ 0 & b_+(\mu,\xi_L) \end{array}\right)_L &  
\mathcal{I}_{1\dots (L-1)} \otimes c_+(\mu,\xi_L)e^{(21)}_L \\
\mathcal{I}_{1\dots (L-1)} \otimes c_-(\mu,\xi_L) e^{(12)}_L&  \mathcal{I}_{1\dots (L-1)}\otimes \left(\begin{array}{cc} b_-(\mu,\xi_L) & 0 \\ 0 & a_-(\mu,\xi_L) \end{array}\right)_L
\end{array}\right)_{a}\\
& \times \left(  \begin{array}{cc} 
A_{1\dots (L-1)}(\mu)\otimes \mathcal{I}_L&  B_{1\dots (L-1)}(\mu) \otimes \mathcal{I}_L\\
C_{1\dots (L-1)}(\mu)\otimes \mathcal{I}_L&  D_{1\dots (L-1)}(\mu)\otimes \mathcal{I}_L
\end{array}\right)_a.
\end{array}
\end{equation*}
Concentrating on the entry $A_{1\dots L}(\mu)$ we have,
\begin{equation*}
A_{1\dots L}(\mu) = A_{1\dots (L-1)}(\mu) \otimes \left(\begin{array}{cc} 1 & 0 \\ 0 & b_+(\mu,\xi_L) \end{array}\right)_L + C_{1\dots (L-1)}(\mu)\otimes c_+(\mu,\xi_L)e^{(21)}_L.
\end{equation*}
Since $C_{1\dots (L-1)}(\mu)$ is \textit{strictly} upper-triangular (the diagonal entries are zero), the second entry in the above sum is strictly upper-triangular. From the first entry, since $A_{1\dots (L-1)}(\mu)$ is upper-triangular, so to is $A_{1\dots L}(\mu)$. Concentrating on the entry $B_{1\dots L}(\mu)$ we have,
\begin{equation*}
B_{1\dots L}(\mu) = B_{1\dots (L-1)}(\mu) \otimes \left(\begin{array}{cc} 1 & 0 \\ 0 & b_+(\mu,\xi_L) \end{array}\right)_L + D_{1\dots (L-1)}(\mu)\otimes c_+(\mu,\xi_L)e^{(21)}_L,
\end{equation*}
which shows that since $B_{1\dots (L-1)}(\mu)$ and $D_{1\dots (L-1)}(\mu)$ are lower-triangular, so to is $B_{1\dots L}(\mu)$. Moreover, since the diagonal entries of $B_{1\dots (L-1)}(\mu)$ are given entirely by zeros, the same is also true for $B_{1\dots L}(\mu)$. Concentrating on the entry $C_{1\dots L}(\mu)$ we have,
\begin{equation*}\begin{array}{ll}
& C_{1\dots L}(\mu) \\
=& A_{1\dots (L-1)}(\mu) \otimes c_-(\mu,\xi_L)e^{(12)}_L + C_{1\dots (L-1)}(\mu)\otimes \left(\begin{array}{cc} b_-(\mu,\xi_L) & 0 \\ 0 & a_-(\mu,\xi_L) \end{array}\right)_L,
\end{array}\end{equation*}
which shows that since $A_{1\dots (L-1)}(\mu)$ and $C_{1\dots (L-1)}(\mu)$ are upper-triangular, so to is $C_{1\dots L}(\mu)$. Moreover, since the diagonal entries of $C_{1\dots (L-1)}(\mu)$ are given entirely by zeros, the same is also true for $C_{1\dots L}(\mu)$. Finally, concentrating on the entry $D_{1\dots L}(\mu)$, we have,
\begin{equation*}\begin{array}{ll}
& D_{1\dots L}(\mu) \\
=& B_{1\dots (L-1)}(\mu) \otimes c_-(\mu,\xi_L)e^{(12)}_L + D_{1\dots (L-1)}(\mu)\otimes \left(\begin{array}{cc} b_-(\mu,\xi_L) & 0 \\ 0 & a_-(\mu,\xi_L) \end{array}\right)_L.
\end{array}
\end{equation*}
Since $B_{1\dots (L-1)}(\mu)$ is strictly lower-triangular, the first entry in the above sum is strictly lower-triangular. Moreover, the presence of $e^{(12)}_L$ ensures that the first term does not contribute any entries in the bottom row. Focusing on the second entry, since $D_{1\dots (L-1)}(\mu)$ is lower-triangular, so to is $D_{1\dots L}(\mu)$, and since the only non zero entry in the final row of $D_{1\dots (L-1)}(\mu)$ is at the final column, so to is it the case for $D_{1\dots L}(\mu)$. Moreover, we can obtain the diagonal entries of $D_{1\dots L}(\mu)$ from the above relation since we have,
\begin{equation*}
\textrm{diag}\left\{ D_{1\dots L}(\mu) \right\} =\textrm{diag}\left\{ D_{1\dots (L-1)}(\mu) \right\} 
\otimes \left(\begin{array}{cc}
b_-(\mu,\xi_L) & 0\\
& a_-(\mu,\xi_L) \end{array} \right)_L. \textrm{  $\square$}
\end{equation*}
\section{The symmetric group and permutations}\label{APP3}
\subsection{Minimal decomposition and adjacent permutations} Let $\sigma$ be an element of the symmetric group $S_L$. $\sigma$ can be decomposed in a minimal way in terms of adjacent permutations, 
\begin{equation*}
\sigma = \sigma_{\alpha_p (\alpha_p+1)}\dots \sigma_{\alpha_2 (\alpha_2+1)}\sigma_{\alpha_1 (\alpha_1+1)},
\end{equation*}
where,
\begin{equation*}
\sigma_{\alpha(\alpha+1)}\left(1\dots (\alpha-1) \alpha (\alpha+1) (\alpha+2) \dots L\right) = \left(1 \dots (\alpha-1)(\alpha+1) \alpha(\alpha+2) \dots L\right).
\end{equation*}
\subsection{Adjacent generators and relations}
\begin{Theorem}
The symmetric group $S_L$ can be generated entirely from the following adjacent permutations,
\begin{equation}
\left\{ \sigma_{12}, \sigma_{23},\dots, \sigma_{(L-1)L} \right\},
\label{adjacentgen}
\end{equation}
which are equipped with the following relations,
\begin{eqnarray}
 \sigma^2_{\alpha(\alpha+1)} &=& 1 \label{rel1}\\
 \sigma_{\alpha(\alpha+1)} \sigma_{(\alpha+1)(\alpha+2)} \sigma_{\alpha(\alpha+1)} &=& \sigma_{(\alpha+1)(\alpha+2)} \sigma_{\alpha(\alpha+1)} \sigma_{(\alpha+1)(\alpha+2)}\label{rel2}\\
\sigma_{\alpha(\alpha+1)} \sigma_{\beta(\beta+1)} &=& \sigma_{\beta(\beta+1)} \sigma_{\alpha(\alpha+1)} \textrm{  for  } \beta \ne \alpha \pm 1.\label{rel3}
\end{eqnarray}
\end{Theorem}
\textbf{Proof.} See Chap. 9 of \cite{Humph}.
\subsection{The cyclic permutation} As a non-trivial example of decomposing a permutation into adjacent permutations, consider the cyclic permutation, labeled $ \sigma_{c}$ where,
\begin{equation*}
 \sigma_{c} (1 2 \dots L) = (2 3 \dots L 1).
\end{equation*}
The cyclic permutation can be minimally decomposed as the following $L-1$ adjacent permutations,
\begin{equation}
 \sigma_{c}  = \sigma_{(L-1)L} \sigma_{(L-2)(L-1)}\dots \sigma_{23} \sigma_{12}.
\label{CYCLE}\end{equation}
Additionally, since acting on any given element of $S_L$ with $\sigma_c$ $L$-times leaves the element invariant we have, 
\begin{equation}\begin{array}{llll}
& \sigma^{L}_{c} &=& 1 \\
\Rightarrow & \sigma^{-1}_{c}  &=&  \sigma^{L-1}_{c}.
\end{array}\label{cyclicinv}\end{equation} 
Hence one can obtain the inverse of $\sigma_c$ through simply taking various powers of $\sigma_c$. We use this fact to produce the following necessary alternative for the generation of $S_L$:
\begin{Proposition}
The symmetric group $S_L$ can be (alternatively) entirely generated from just two permutations, the cyclic permutation $\sigma_{c}$, and the adjacent permutation $\sigma_{12}$.
\end{Proposition}
\textbf{Proof.} One can generate the entire set of generators in Eq. (\ref{adjacentgen}) through the following relation,
\begin{eqnarray}
 \sigma_{\alpha (\alpha+1)} = \sigma^{-(\alpha-1)}_{c} \sigma_{12}\sigma^{\alpha-1}_{c}, & \alpha = \{1,\dots,L-1\}.
\label{cyclicN}
\end{eqnarray}
To see the validity of Eq. (\ref{cyclicN}), consider the operator $\sigma^{-(\alpha-1)}_{c} \sigma_{12}\sigma^{\alpha-1}_{c}$ acting on $(12\dots L)$,
\begin{eqnarray*}
 \sigma^{-(\alpha-1)}_{c} \sigma_{12}\sigma^{\alpha-1}_{c}(12 \dots L) &=&  \sigma^{-(\alpha-1)}_{c} \sigma_{12}  (\alpha (\alpha+1)\dots L 1 \dots (\alpha-1))\\
&=& \sigma^{-(\alpha-1)}_{c}   ( (\alpha+1)\alpha(\alpha+2)\dots L 1 \dots (\alpha-1))\\
&= &  (1\dots (\alpha-1)(\alpha+1)\alpha(\alpha+2)\dots L ) . \textrm{  $\square$}
\end{eqnarray*}
\section{Proof of proposition \ref{NandR}}\label{APP4}
We proceed in exactly the same method as we did for the verification of the factorization condition in Section (\ref{PROOF}). We note that in a similar method to Proposition (\ref{PROP2}), we can show that the auxiliary operators $\hat{\mathcal{R}}^{\{\sigma\}}_{1\dots L}$ also provide a valid representation of $S_L$. Applying Eq. (\ref{useful}) to Eq. (\ref{auxR}), Eq. (\ref{auxR}) becomes,
\begin{equation}
\mathcal{N}^{-1}_{1\dots L} P^{\{\sigma\}}_{1 \dots L}\mathcal{N}_{1\dots L} = \hat{\mathcal{R}}^{\{\sigma\}}_{1\dots L}.
\label{auxR2}\end{equation} 
Since both $P^{\{\sigma\}}_{1 \dots L}$ and $\hat{\mathcal{R}}^{\{\sigma\}}_{1\dots L}$ provide valid representations of $S_L$, we note that Eq. (\ref{auxR2}) is in a form where one can easily decompose the permutation. Explicitly illustrating this, consider the permutation $\sigma = \sigma_1 \sigma_2$ on the left-hand side and right-hand side of Eq. (\ref{auxR2}) respectively,
\begin{equation*}
\begin{array}{lll}
\mathcal{N}^{-1}_{1\dots L} P^{\{\sigma_1\sigma_2\}}_{1 \dots L}\mathcal{N}_{1\dots L} &=& \mathcal{N}^{-1}_{1\dots L} P^{\{\sigma_1\}}_{1 \dots L}\mathcal{N}_{1\dots L}\mathcal{N}^{-1}_{1\dots L} P^{\{\sigma_2\}}_{1 \dots L}\mathcal{N}_{1\dots L}\\
\hat{\mathcal{R}}^{\{\sigma_1\sigma_2\}}_{1\dots L}&=& \hat{\mathcal{R}}^{\{\sigma_1\}}_{1\dots L}\hat{\mathcal{R}}^{\{\sigma_2\}}_{1\dots L}
\end{array}
\end{equation*}
where $\{\sigma_1,\sigma_2\} \in S_L$. Since $S_L$ can be constructed entirely from the adjacent permutation $\sigma_{12}$ and the cyclic permutation $\sigma_c$, we have simply to verify Eq. (\ref{auxR}) for the aforementioned permutations. Beginning with $\sigma_{12}$ we obtain,
\begin{equation*}\begin{array}{lll}
 \mathcal{N}^{-1}_{213\dots L}\mathcal{N}_{1\dots L} &=&  \overbrace{\mathcal{N}^{-1}_{3\dots L}\mathcal{N}_{3\dots L}}^{\mathcal{I}_{3\dots L}} \overbrace{\mathcal{N}^{-1}_{1,3\dots L} \mathcal{N}_{1,2\dots L}}^{\mathcal{N}_{12}} \overbrace{\mathcal{N}^{-1}_{2,13\dots L}\mathcal{N}_{2,3\dots L}}^{\mathcal{N}^{-1}_{21}}\\
&=& \mathcal{R}_{12} .
\end{array}\end{equation*}
Focusing now on $\sigma_c$, through a simple induction argument one can obtain,
\begin{equation*}\begin{array}{lllll}
 \mathcal{N}^{-1}_{2\dots L1}\mathcal{N}_{1\dots L} &=&  \mathcal{N}^{-1}_{i\dots L1}\mathcal{N}_{i\dots L} \mathcal{N}^{-1}_{2\dots (i-1),1}  \mathcal{N}_{1,2\dots L} &,& 2 \le i \le L  \\
&=& \mathcal{N}^{-1}_{2\dots L,1}  \mathcal{N}_{1,2\dots L}\\
&=& \mathcal{R}_{1,2\dots L} .&\square
\end{array}\end{equation*}
\section{Explicit details for the case L=2}\label{APP5}
\begin{equation*}\begin{array}{lll}
\kappa^{(D)}_1 &=& 0 \textrm{  using Eq. (\ref{YB6})}\\
\kappa^{(C)}_1 &=& \overbrace{c_-(\mu,\xi_2)- \frac{c_-(\mu,\xi_1)b_-(\mu,\xi_2) c_-(\xi_1,\xi_2)}{b_-(\xi_1,\xi_2)}}^{\textrm{use Eq. (\ref{uni3}) and (\ref{YB2})}}\\
&=& \frac{b_-(\mu,\xi_1)c_-(\mu,\xi_2)}{b_-(\xi_2,\xi_1)}\\
\kappa^{(C)}_2 &=&\frac{\overbrace{c_-(\mu,\xi_1)a_-(\mu,\xi_2) c_-(\xi_1,\xi_2)+b_+(\mu,\xi_1)c_-(\mu,\xi_2) b_-(\xi_1,\xi_2) }^{\textrm{use Eq. (\ref{YB5})}}}{a_-(\xi_1,\xi_2)}\\
&=& a_-(\mu,\xi_1)c_-(\mu,\xi_2)\\
\kappa^{(B)}_1 &=&\overbrace{b_-(\mu,\xi_1)c_+(\mu,\xi_2)c_-(\xi_1,\xi_2)+c_+(\mu,\xi_1)b_-(\xi_1,\xi_2)}^{\textrm{use Eq. (\ref{YB1})}}\\
&=& c_+(\mu,\xi_1)b_-(\mu,\xi_2)\\
\kappa^{(B)}_2 &=& a_-(\xi_1,\xi_2)\overbrace{\left\{ c_+(\mu,\xi_1) b_+(\mu,\xi_2)  -\frac{a_-(\mu,\xi_1) c_+(\mu,\xi_2) c_-(\xi_1,\xi_2) }{b_-(\xi_1,\xi_2)} \right\}}^{\textrm{use Eq. (\ref{uni3}) and (\ref{YB4})}}\\
&=&\frac{c_+(\mu,\xi_1)a_-(\mu,\xi_2)\overbrace{a_-(\xi_1,\xi_2)a_-(\xi_2,\xi_1)}^{\textrm{use Eq. (\ref{uni5})}}}{b_-(\xi_2,\xi_1)}\\
&=& \frac{c_+(\mu,\xi_1)a_-(\mu,\xi_2)}{b_-(\xi_2,\xi_1)}
\end{array}\end{equation*}
\begin{equation*}\begin{array}{lll}
\kappa^{(A)}_1 &=& \overbrace{b_+(\mu,\xi_2)}^{\textrm{use Eq. (\ref{uni1}) and (\ref{uni3})}} - \overbrace{\frac{c_-(\mu,\xi_1)c_+(\mu,\xi_2) c_-(\xi_2,\xi_1)}{ b_-(\xi_2,\xi_1)}}^{\textrm{use Eq. (\ref{uni3})}}\\
&=&\frac{c_-(\mu,\xi_1)c_+(\mu,\xi_2) c_+(\xi_2,\xi_1)}{ b_-(\xi_2,\xi_1)} +\frac{c_-(\mu,\xi_2)c_+(\mu,\xi_2)}{ b_-(\xi_2,\mu)} + \frac{1}{b_-(\xi_2,\mu)}\\
&=& \frac{c_+(\mu,\xi_2)}{b_-(\mu,\xi_2)b_-(\xi_2,\xi_1)}\{\overbrace{c_-(\mu,\xi_2)b_-(\xi_2,\xi_1) + b_-(\mu,\xi_2)c_-(\mu,\xi_1)c_+(\xi_2,\xi_1) }^{\textrm{use Eq. (\ref{YB2})}}  \}+\frac{1}{b_-(\xi_2,\mu)}\\
&=& \frac{b(\mu,\xi_1)c_-(\mu,\xi_2)c_+(\mu,\xi_2)}{b_-(\mu,\xi_2)b_-(\xi_2,\xi_1)}+\frac{1}{b_-(\xi_2,\mu)}\\
\kappa^{(A)}_2 &=& b_+(\mu,\xi_2)c_-(\xi_1,\xi_2)-b_+(\mu,\xi_1)c_-(\xi_1,\xi_2)- \frac{\overbrace{c_-(\mu,\xi_1)c_+(\mu,\xi_2) c_-(\xi_1,\xi_2)}^{\textrm{use Eq. (\ref{YB7})}}\overbrace{c_-(\xi_1,\xi_2)}^{\textrm{use Eq. (\ref{uni3})}} }{ b_-(\xi_1,\xi_2)}\\
&=& b_+(\mu,\xi_2)c_-(\xi_1,\xi_2)-b_+(\mu,\xi_1)c_-(\xi_1,\xi_2)+ \frac{c_+(\mu,\xi_1)c_-(\mu,\xi_2) \overbrace{c_+(\xi_1,\xi_2)c_+(\xi_2,\xi_1)}^{\textrm{use Eq. (\ref{uni2})}}}{ b_-(\xi_2,\xi_1)}\\
&=& \overbrace{b_+(\mu,\xi_2)c_-(\xi_1,\xi_2)-b_+(\mu,\xi_1)c_-(\xi_1,\xi_2) - c_+(\mu,\xi_1)c_-(\mu,\xi_2)b_+(\xi_1,\xi_2)}^{\textrm{use Eq. (\ref{YB3})}} + \frac{c_+(\mu,\xi_1)c_-(\mu,\xi_2) }{ b_-(\xi_2,\xi_1)}\\
&=& \frac{c_+(\mu,\xi_1)c_-(\mu,\xi_2) }{ b_-(\xi_2,\xi_1)}\\
\kappa^{(A)}_3 &=& \overbrace{b_+(\mu,\xi_1)}^{\textrm{use Eq. (\ref{uni1}) and (\ref{uni3})}} + \frac{c_-(\mu,\xi_1)c_+(\mu,\xi_2) c_-(\xi_1,\xi_2)}{ b_-(\xi_1,\xi_2)}\\
&=& \frac{c_-(\mu,\xi_1)c_+(\mu,\xi_2) c_-(\xi_1,\xi_2)}{ b_-(\xi_1,\xi_2)} + \frac{c_-(\mu,\xi_1)c_+(\mu,\xi_1)}{b_-(\mu,\xi_1)}+\frac{1}{b_-(\xi_1,\mu)}\\
&=& \frac{c_-(\mu,\xi_1)}{b_-(\mu,\xi_1)b_-(\xi_1,\xi_2)}\{ \overbrace{b_-(\mu,\xi_1) c_+(\mu,\xi_2) c_-(\xi_1,\xi_2)+ c_+(\mu,\xi_1)b_-(\xi_1,\xi_2)}^{\textrm{use Eq. (\ref{YB1})}} \}+ \frac{1}{b_-(\xi_1,\mu)}\\
&=& \frac{c_-(\mu,\xi_1)c_+(\mu,\xi_1)b_-(\mu,\xi_2)}{b_-(\mu,\xi_1)b_-(\xi_1,\xi_2)} + \frac{1}{b_-(\xi_1,\mu)}\\
\kappa^{(A)}_4 &=& \overbrace{b_+(\mu,\xi_1) b_+(\mu,\xi_2)}^{\textrm{use Eq. (\ref{uni1}) and (\ref{uni3})}}\\
&=&\left\{ \frac{1}{b_-(\xi_1,\mu)}  + \frac{c_-(\mu,\xi_1)c_+(\mu,\xi_1)}{b_-(\mu,\xi_1)} \right\}\left\{ \frac{1}{b_-(\xi_2,\mu)}  + \frac{c_-(\mu,\xi_2)c_+(\mu,\xi_2)}{b_-(\mu,\xi_2)} \right\}\\
&=& \frac{1}{b_-(\xi_1,\mu)b_-(\xi_2,\mu)} + \frac{c_-(\mu,\xi_1)c_+(\mu,\xi_1)}{b_-(\mu,\xi_1)b_-(\xi_2,\mu)} + \frac{c_-(\mu,\xi_2)c_+(\mu,\xi_2) }{b_-(\mu,\xi_2)b_-(\xi_1,\mu)} + \frac{c_-(\mu,\xi_1)c_+(\mu,\xi_1)c_-(\mu,\xi_2)c_+(\mu,\xi_2)}{b_-(\mu,\xi_1)b_-(\mu,\xi_2)}\\
&&+ \frac{c_-(\mu,\xi_1)c_+(\mu,\xi_2)}{b_-(\mu,\xi_1)b_-(\mu,\xi_2)b_-(\xi_1,\xi_2)}\overbrace{\left\{ \begin{array}{l} c_+(\mu,\xi_1) c_-(\mu, \xi_2) b_-(\xi_1,\xi_2) + b_-(\mu,\xi_1) a_-(\mu, \xi_2) c_-(\xi_1,\xi_2)  \\
- a_-(\mu,\xi_1) b_-(\mu, \xi_2) c_-(\xi_1,\xi_2)
\end{array}\right\}}^{\textrm{= 0 using Eq. (\ref{YB6})}}
\end{array}\end{equation*}
\begin{equation*}\begin{array}{lll}
&=&\frac{1}{b_-(\xi_1,\mu)b_-(\xi_2,\mu)}+ \frac{c_-(\mu,\xi_1)c_+(\mu,\xi_2)c_-(\xi_1,\xi_2)a_-(\mu,\xi_2)}{b_-(\xi_1,\xi_2)b_-(\mu,\xi_2)} -\frac{c_-(\mu,\xi_1)c_+(\mu,\xi_2)\overbrace{c_-(\xi_1,\xi_2)}^{\textrm{use Eq. (\ref{uni3})}} a_-(\mu,\xi_1)}{b_-(\mu,\xi_1)b_-(\xi_1,\xi_2)}\\
&&+\frac{c_-(\mu,\xi_1)c_+(\mu,\xi_1)}{b_-(\mu,\xi_1)}\overbrace{\left\{\frac{1}{b_-(\xi_2,\mu)}+\frac{c_-(\mu,\xi_2)c_+(\mu,\xi_2)}{b_-(\mu,\xi_2)}  \right\}}^{\textrm{use Eq. (\ref{uni1}) and (\ref{uni3})}}\\
&&+\frac{c_-(\mu,\xi_2)c_+(\mu,\xi_2)}{b_-(\mu,\xi_2)}\overbrace{\left\{ \frac{1}{b_-(\xi_1,\mu)}+\frac{c_-(\mu,\xi_1)c_+(\mu,\xi_1)}{b_-(\mu,\xi_1)} \right\}}^{\textrm{use Eq. (\ref{uni1}) and (\ref{uni3})}}\\
&=& \frac{1}{b_-(\xi_1,\mu)b_-(\xi_2,\mu)} \\
&&+ \frac{c_-(\mu,\xi_1)}{b_-(\mu,\xi_1)b_-(\xi_2,\xi_1)} \{ \overbrace{b_+(\mu,\xi_2) c_+(\mu, \xi_1) b_-(\xi_2,\xi_1) + c_+(\mu,\xi_2) a_-(\mu, \xi_1) c_+(\xi_2,\xi_1)}^{\textrm{use Eq. (\ref{YB4})}} \}\\
&&+ \frac{c_+(\mu,\xi_2)}{b_-(\mu,\xi_2)b_-(\xi_1,\xi_2)} \{ \overbrace{b_+(\mu,\xi_1) c_-(\mu, \xi_2) b_-(\xi_1,\xi_2) + c_-(\mu,\xi_1) a_-(\mu, \xi_2) c_-(\xi_1,\xi_2)}^{\textrm{use Eq. (\ref{YB5})}} \}
\end{array}\end{equation*}
\begin{equation*}\begin{array}{lll}
&=& \frac{1}{b_-(\xi_1,\mu)b_-(\xi_2,\mu)} +  \frac{c_-(\mu,\xi_1)c_+(\mu,\xi_1)a_-(\mu,\xi_2)\overbrace{a_-(\xi_2,\xi_1)}^{\textrm{use Eq. (\ref{uni5})}}}{b_-(\mu,\xi_1)b_-(\xi_2,\xi_1)}+  \frac{c_-(\mu,\xi_2)c_+(\mu,\xi_2)a_-(\mu,\xi_1)a_-(\xi_1,\xi_2)}{b_-(\mu,\xi_2)b_-(\xi_1,\xi_2)}\\
&=& \frac{1}{b_-(\xi_1,\mu)b_-(\xi_2,\mu)} +  \frac{c_-(\mu,\xi_1)c_+(\mu,\xi_1)a_-(\mu,\xi_2) }{b_-(\mu,\xi_1)b_-(\xi_2,\xi_1)a_-(\xi_1,\xi_2)}+  \frac{c_-(\mu,\xi_2)c_+(\mu,\xi_2)a_-(\mu,\xi_1)a_-(\xi_1,\xi_2)}{b_-(\mu,\xi_2)b_-(\xi_1,\xi_2)}.
\end{array}
\end{equation*}
\section{Expanding the twisted monodromy operators in vector space $V_1$}
\subsection{Expanding $\tilde{C}_{1 \dots L}$}\label{APP6}
To expand the assumed form of $\tilde{C}_{1\dots L}$ (\ref{C}) in the vector space $V_1$ we must look at the two distinct cases of $l=1$ and $l \ne 1$.\\
$l=1$
\begin{equation*}
c_-(\mu,\xi_1)  e^{(12)}_1 \otimes^L_{j=2}\left(  \begin{array}{cc} 
\frac{b_-(\mu,\xi_j)}{b_-(\xi_1,\xi_j)}  & 0  \\
 0  & \frac{a_-(\mu,\xi_j)}{a_-(\xi_1,\xi_j)} 
\end{array}\right)_j
\end{equation*}
$l \ne 1$
\begin{equation*}\begin{array}{l}
\sum^L_{l=2}c_-(\mu,\xi_l)\left(  \begin{array}{cc} 
\frac{b_-(\mu,\xi_1)}{b_-(\xi_l,\xi_1)}  & 0  \\
 0  & a_-(\mu,\xi_1)
\end{array}\right)_1\otimes^{l-1}_{i=2} \left(  \begin{array}{cc} 
\frac{b_-(\mu,\xi_i)}{b_-(\xi_l,\xi_i)}  & 0  \\
 0  & a_-(\mu,\xi_i)
\end{array}\right)_i  e^{(12)}_l\\
\otimes^L_{j=l+1}\left(  \begin{array}{cc} 
\frac{b_-(\mu,\xi_j)}{b_-(\xi_l,\xi_j)}  & 0  \\
 0  & \frac{a_-(\mu,\xi_j)}{a_-(\xi_l,\xi_j)} 
\end{array}\right)_j.
\end{array}
\end{equation*}
Collecting each of the four terms explicitly in $V_1$ we obtain the expressions (\ref{alphaCa}-\ref{deltaCa}).
\subsection{Expanding $\tilde{B}_{1 \dots L}$}\label{APP7}
As with $\tilde{C}_{1\dots L}$, to expand $\tilde{B}_{1\dots L}$ (\ref{B}) in vector space $V_1$ we must consider the cases $l=1$ and $l\ne 1$ of Eq. (\ref{B}) separately. \\
$l=1$
\begin{equation*}
c_+(\mu,\xi_1)  e^{(21)}_1 \otimes^L_{j=2}\left(  \begin{array}{cc} 
b_-(\mu,\xi_j)  & 0  \\
 0  & \frac{a_-(\mu,\xi_j)}{b_-(\xi_j,\xi_1)} 
\end{array}\right)_j
\end{equation*}
$l \ne 1$
\begin{equation*}\begin{array}{l}
\sum^L_{l=2}c_+(\mu,\xi_l)\left(  \begin{array}{cc} 
b_-(\mu,\xi_1)  & 0  \\
 0  & \frac{a_-(\mu,\xi_1)a_-(\xi_1,\xi_l)}{b_-(\xi_1,\xi_l)}
\end{array}\right)_1 \\
\otimes^{l-1}_{i=2} \left(  \begin{array}{cc} 
b_-(\mu,\xi_i)  & 0  \\
 0  & \frac{a_-(\mu,\xi_i)a_-(\xi_i,\xi_l)}{b_-(\xi_i,\xi_l)}
\end{array}\right)_i 
 e^{(21)}_l\otimes^L_{j=l+1}\left(  \begin{array}{cc} 
b_-(\mu,\xi_j)  & 0  \\
 0  & \frac{a_-(\mu,\xi_j)}{b_-(\xi_j,\xi_l)} 
\end{array}\right)_j
\end{array}
\end{equation*}
Collecting each of the three non-zero terms explicitly in $V_1$ we obtain the expressions (\ref{alphaBa}-\ref{gammaBa}).
\subsection{Expanding $\tilde{A}_{1 \dots L}$}\label{APP8}
We note that Eq. (\ref{A}) is explicitly given by,
\begin{equation}
 \begin{array}{l}
  \left( \begin{array}{cc}
1&  0\\
0&  \frac{1}{b_-(\xi_1,\mu)}
\end{array}\right)_{1}\otimes \Omega_0(\mu)+   \sum^L_{l_1 l_2=1}\frac{c_+(\mu,\xi_{l_1})c_-(\mu,\xi_{l_2})}{b_-(\mu,\xi_{l_1})}\\
\times \left\{ \otimes^{l_1-1}_{i_1=1} \left(  \begin{array}{cc} 
1  & 0  \\
 0  & \frac{a_-(\xi_{i_1},\xi_{l_1})}{b_-(\xi_{i_1},\xi_{l_1})}
\end{array}\right)_{i_1} e^{(21)}_{l_1} \otimes^L_{j_1=l_1+1}\left(  \begin{array}{cc} 
1  & 0  \\
 0  & \frac{1}{b_-(\xi_{j_1},\xi_{l_1})} 
\end{array}\right)_{j_1}\right\}\\
\times \left\{\otimes^{l_2-1}_{i_2=1} \left(  \begin{array}{cc} 
\frac{b_-(\mu,\xi_{i_2})}{b_-(\xi_{l_2},\xi_{i_2})}  & 0  \\
 0  & a_-(\mu,\xi_{i_2})
\end{array}\right)_{i_2}  e^{(12)}_{l_2}  \otimes^L_{j_2=l_2+1}\left(  \begin{array}{cc} 
\frac{b_-(\mu,\xi_{j_2})}{b_-(\xi_{l_2},\xi_{j_2})}  & 0  \\
 0  & \frac{a_-(\mu,\xi_{j_2})}{a_-(\xi_{l_2},\xi_{j_2})} 
\end{array}\right)_{j_2}\right\}.
 \end{array}
\label{expA}\end{equation}
The double summation in Eq. (\ref{expA}) has three relevant sets of values, $l_1=l_2=l$, $l_1>l_2$ and $l_1<l_2$. We expand Eq. (\ref{expA}) in those three sets in vector space $V_1$ below.\\
$l_1=l_2=l$
\begin{equation}\begin{array}{l}
\frac{c_+(\mu,\xi_1)c_-(\mu,\xi_1)}{b_-(\mu,\xi_1)}  e^{(22)}_{1}\otimes \Omega_1(\mu)\\
 +  \sum^L_{l=2}\frac{c_+(\mu,\xi_l)c_-(\mu,\xi_l)}{b_-(\mu,\xi_l)}\left(  \begin{array}{cc} 
 \frac{b_-(\mu,\xi_{1})}{b_-(\xi_{l},\xi_{1})} & 0  \\
 0  & \frac{a_-(\mu,\xi_{1})a_-(\xi_{1},\xi_{l})}{b_-(\xi_{1},\xi_{l})}
\end{array}\right)_{1} \otimes \Omega^{(l)}_2(\mu).
\end{array}
\label{lequal}
\end{equation}
$l_1>l_2$
\begin{equation}
\begin{array}{l}
 \sum^N_{l=2}\frac{c_+(\mu,\xi_{l})c_-(\mu,\xi_{1})}{b_-(\xi_{1},\xi_{l})}e^{(12)}_{1} \otimes \Omega^{(l)}_3(\mu) \\
+  \sum_{2 \le l_2 < l_1 \le L}\frac{c_+(\mu,\xi_{l_1})c_-(\mu,\xi_{l_2})}{b_-(\xi_{l_2},\xi_{l_1})}\left(  \begin{array}{cc} 
\frac{b_-(\mu,\xi_1)}{b_-(\xi_{l_2},\xi_1)}  & 0  \\
 0  & \frac{a_-(\mu,\xi_1)a_-(\xi_{1},\xi_{l_1})}{b_-(\xi_{1},\xi_{l_1})}
\end{array}\right)_{1} \otimes \Omega^{(l_1,l_2)}_5(\mu).
\end{array}
\label{l1greater}\end{equation}
$l_2>l_1$
\begin{equation}\begin{array}{l}
  \sum^N_{l=2}\frac{c_+(\mu,\xi_{1})c_-(\mu,\xi_{l})}{b_-(\xi_{l},\xi_{1})} e^{(21)}_{1} \otimes \Omega^{(l)}_4(\mu)\\
+ \sum_{2 \le l_1 < l_2 \le L}\frac{c_+(\mu,\xi_{l_1})c_-(\mu,\xi_{l_2})}{b_-(\xi_{l_2},\xi_{l_1})} \left(  \begin{array}{cc} 
\frac{b_-(\mu,\xi_1)}{b_-(\xi_{l_2},\xi_1)}  & 0  \\
 0  & \frac{a_-(\mu,\xi_1)a_-(\xi_{1},\xi_{l_1})}{b_-(\xi_{1},\xi_{l_1})}
\end{array}\right)_{1}\otimes \Omega^{(l_1,l_2)}_6(\mu).
\end{array}
\label{l1lesser}\end{equation}
Collecting each of the four terms explicitly in $V_1$ we obtain the expressions (\ref{alphaAa}-\ref{deltaAa}).
\section{The action of the F-basis on reference states}
\subsection{Proof of proposition \ref{gurg}}\label{APPFIRST}
From the explicit form of $\langle 1 |$, we see that it only ever multiplies entries from the final row of $\mathcal{F}_{1\dots L}$. One can explicitly obtain the entries of the final row through the following elementary analysis. Recalling the following form of $\mathcal{F}_{1\dots L}$ for convenience,
\begin{equation*}
\mathcal{F}_{1\dots L} = \mathcal{F}_{(L-1)L}\mathcal{F}_{(L-2),(L-1)L}\dots \mathcal{F}_{1,2\dots L} 
\end{equation*}
where,
\begin{equation*}
\mathcal{F}_{i,(i+1) \dots L}= \left( \begin{array}{cc} \mathcal{I}_{(i+1)\dots L} & 0 \\ C_{(i+1)\dots L}(\xi_i) & D_{(i+1)\dots L}(\xi_i)
\end{array}\right)_i,
\end{equation*}
we remember two important properties for the $C$ and $D$ operators:
\begin{itemize}
\item{$C_{(i+1)\dots L}(\xi_{i})$ is strictly upper triangular.}\\
\item{The last row of $D_{(i+1)\dots L}(\xi_{i})$ only has one non zero entry at the last column, explicitly given by $\prod^L_{j=i+1}a_-(\xi_i,\xi_j)$.}
\end{itemize}
From the explicit form of $\mathcal{F}_{1\dots L}$ given above we see that the final row only has one non zero entry at the final column with value $\prod_{1 \le i < j \le L} a_-(\xi_i,\xi_j)$, hence verifying the first result of this proposition. For the second result, we notice that only entries from the final column of $\mathcal{F}^{-1}_{1\dots L}$ multiply $|1\rangle$. From this observation, and recalling that $D_{(i+1)\dots L}(\xi_i)$ is lower triangular, we have the following values for the co-factors of $\mathcal{F}_{1\dots L}$,
\begin{equation*}\begin{array}{llcl}
 \textrm{det}\left\{  \left( \mathcal{F}_{1\dots L} \right)_{kl} \right\}_{\genfrac{}{}{0pt}{}{k=1 \dots 2^L-1}{l=1 \dots 2^L-1}} &=&{\displaystyle\prod_{1 \le i < j \le L}} a_-(\xi_j,\xi_i) \textrm{det}\left\{  \mathcal{F}_{1\dots L}  \right\}\\
\textrm{det}\left\{  \left( \mathcal{F}_{1\dots L} \right)_{kl} \right\}_{\genfrac{}{}{0pt}{}{k=1 \dots 2^L-1}{l=1\dots (i-1)(i+1) \dots 2^L}} &=& 0 &\textrm{$i \ne 2^L$}\\
\textrm{det}\left\{  \left( \mathcal{F}_{1\dots L} \right)_{kl} \right\}_{\genfrac{}{}{0pt}{}{k=1\dots (j-1)(j+1) \dots 2^L}{l=1 \dots 2^L-1}} &=& 0 &\textrm{$j \ne 2^L$}.
\end{array}
\end{equation*}
Recalling Cramer's rule for the inverse of a matrix,
\begin{equation}
\left( \mathcal{F}^{-1}_{1\dots L} \right)_{ij} = \frac{(-1)^{i+j}}{\textrm{det}\left\{ \mathcal{F}_{1\dots L} \right\}}\textrm{det}\left\{  \left( \mathcal{F}_{1\dots L} \right)_{kl} \right\}_{\genfrac{}{}{0pt}{}{k=1\dots j-1,j+1 \dots 2^L}{l=1\dots i-1,i+1\dots 2^L} },
\label{Cramer}\end{equation}
we obtain the following values for $\mathcal{F}^{-1}_{1\dots L}$,
\begin{equation*}\begin{array}{llcl}
\left( \mathcal{F}^{-1}_{1\dots L} \right)_{(2^L)(2^L)} &=& {\displaystyle\prod_{1 \le i < j \le L}} a_-(\xi_j,\xi_i)\\
\left( \mathcal{F}^{-1}_{1\dots L} \right)_{i(2^L)} &=& 0 &\textrm{$i \in \{1,\dots, 2^L-1 \}$}\\
\left( \mathcal{F}^{-1}_{1\dots L} \right)_{(2^L)j} &=& 0 &\textrm{$j \in \{1,\dots, 2^L-1 \}$},
\end{array}
\end{equation*}
thus verifying the final result. $\square$
\subsection{Proof of proposition \ref{NEEDEDSEC}}\label{APPSECOND}
From the form of $\langle 0|$, we see that it only ever multiplies the entries of the top row of $\mathcal{F}_{1\dots L}$ or $\mathcal{F}^{-1}_{1\dots L}$. Recalling the explicit form of $\mathcal{F}_{1\dots L}$ given in Section (\ref{APPFIRST}), we see that the top row of $\mathcal{F}_{1\dots L}$ is given explicitly by,
\begin{equation*}\begin{array}{lll}
\left(\mathcal{F}_{1\dots L}\right)_{1j} &=& \left\{ \begin{array}{lll} 1 &\textrm{for}& j=1\\
							    0 & \textrm{for} &j \in \{ 2, \dots, 2^L\} \end{array}\right. ,
\end{array}
\end{equation*}
hence verifying $\langle 0 | \mathcal{F}_{1\dots L} = \langle 0 |$. From the form of $|0 \rangle$, we see that it only ever multiplies the entries of the first column of $\mathcal{F}_{1 \dots L}$ or $\mathcal{F}^{-1}_{1 \dots L}$. Recalling that the operator $C_{(i+1) \dots L}(\xi_i)$ is upper triangular, with the diagonal entries equal to zero, we find that the first column of $\mathcal{F}_{1 \dots L}$ is given explicitly by,
\begin{equation*}\begin{array}{lll}
\left(\mathcal{F}_{1\dots L}\right)_{i1} &=& \left\{ \begin{array}{lll} 1 &\textrm{for}& i=1\\
							    0 & \textrm{for} &i \in \{ 2, \dots, 2^L\} \end{array}\right. ,
\end{array}
\end{equation*}
thus verifying $\mathcal{F}_{1 \dots L} |0 \rangle = |0 \rangle$. For the remaining results we note the following cofactor values concerning the first row and column of $\mathcal{F}_{1\dots L}$,
\begin{equation*}\begin{array}{llcl}
\textrm{det}\left\{  \left( \mathcal{F}_{1\dots L} \right)_{kl} \right\}_{\genfrac{}{}{0pt}{}{k=2 \dots 2^L}{l=2 \dots 2^L}} &=& \textrm{det}\left\{  \mathcal{F}_{1\dots L}  \right\}\\
\textrm{det}\left\{  \left( \mathcal{F}_{1\dots L} \right)_{kl} \right\}_{\genfrac{}{}{0pt}{}{k=2 \dots 2^L}{l=1\dots (i-1)(i+1) \dots 2^L}} &=& 0 &\textrm{for $i \in \{2,\dots, 2^L \}$}\\
\textrm{det}\left\{  \left( \mathcal{F}_{1\dots L} \right)_{kl} \right\}_{\genfrac{}{}{0pt}{}{k=1\dots (j-1)(j+1) \dots 2^L}{l=2 \dots 2^L}} &=& 0 &\textrm{for $j \in \{2,\dots, 2^L \}$},
\end{array}
\end{equation*}
hence applying Cramer's rule for the inverse of a matrix (\ref{Cramer}) we have,
\begin{equation*}\begin{array}{llcl}
\left( \mathcal{F}^{-1}_{1\dots L} \right)_{11} &=& 1\\
\left( \mathcal{F}^{-1}_{1\dots L} \right)_{i1} &=& 0 &\textrm{for $i \in \{2,\dots, 2^L \}$}\\
\left( \mathcal{F}^{-1}_{1\dots L} \right)_{1j} &=& 0 &\textrm{for $j \in \{2,\dots, 2^L \}$},
\end{array}
\end{equation*}
thus verifying $\langle 0 | \mathcal{F}^{-1}_{1\dots L} = \langle 0 |$ and $\mathcal{F}^{-1}_{1 \dots L} |0 \rangle = |0 \rangle$. $\square$
\section{Necessary polynomial identities}
\subsection{Proof of proposition \ref{POLLY1}}\label{APPPOLLY1}
We begin by writing Eq. (\ref{polyiden}) in the following form,
\begin{equation}
 1 = \frac{1}{\prod^M_{\genfrac{}{}{0pt}{}{l=1}{\ne i}}(\xi_l \rho - \xi_i)} \sum^M_{k=1} \frac{\prod^M_{\genfrac{}{}{0pt}{}{n=1}{\ne i}}(\mu_k-\xi_n)  \prod^M_{\genfrac{}{}{0pt}{}{m=1}{\ne k}}(\mu_m \rho-\xi_i)}{\prod^M_{\genfrac{}{}{0pt}{}{p=1}{\ne k}}(\mu_k -\mu_p)},
\label{RHS}\end{equation}
and use Liouville's theorem to show that the right hand side is a constant. To show that the right hand side is an \textit{entire function}, we show that the poles, which are situated at $\mu_{k_1} = \mu_{k_2}$, $\genfrac{}{}{0pt}{}{1 \le k_1,k_2 \le M}{k_1 \ne k_2}$, and $\xi_j \rho = \xi_i$, $\genfrac{}{}{0pt}{}{1 \le j \le M}{\ne i}$, are removable. Beginning with the pole $\mu_{k_1} = \mu_{k_2}$, we rearrange the sum on the right hand side of Eq. (\ref{RHS}) as,
\begin{equation*}
 \begin{array}{l}
\frac{\prod^M_{\genfrac{}{}{0pt}{}{m=1}{\ne k_1,k_2}}(\mu_m \rho - \xi_i)}{\mu_{k_1}-\mu_{k_2}}\left\{ \frac{(\mu_{k_2} \rho-\xi_i)\prod^M_{\genfrac{}{}{0pt}{}{n=1}{\ne i}}(\mu_{k_1}-\xi_n)  }{\prod^M_{\genfrac{}{}{0pt}{}{p=1}{\ne k_1,k_2}}(\mu_{k_1} -\mu_{p})} - \frac{(\mu_{k_1} \rho-\xi_i)\prod^M_{\genfrac{}{}{0pt}{}{n=1}{\ne i}}(\mu_{k_2}-\xi_n)  }{\prod^M_{\genfrac{}{}{0pt}{}{p=1}{\ne k_1,k_2}}(\mu_{k_2} -\mu_{p})}\right\}\\
 +  \sum^M_{\genfrac{}{}{0pt}{}{k=1}{\ne k_1,k_2}} \frac{\prod^M_{\genfrac{}{}{0pt}{}{n=1}{\ne i}}(\mu_k-\xi_n)  \prod^M_{\genfrac{}{}{0pt}{}{m=1}{\ne k}}(\mu_m \rho-\xi_i)}{\prod^M_{\genfrac{}{}{0pt}{}{p=1}{\ne k}}(\mu_k -\mu_p)}.
 \end{array}
\end{equation*}
Hence we can see through inspection that the limit $\mu_{k_1} \rightarrow \mu_{k_2}$ of the above expression has a residue value of zero. We now consider the pole at $\xi_j \rho = \xi_i$. Focusing on the summation on the right hand side of Eq. (\ref{RHS}) we obtain,
\begin{equation*}
 \lim_{\xi_i \rightarrow \xi_j \rho}  {\displaystyle \sum^M_{\genfrac{}{}{0pt}{}{k=1}{}}} \frac{\prod^M_{\genfrac{}{}{0pt}{}{n=1}{\ne i}}(\mu_k-\xi_n)  \prod^M_{\genfrac{}{}{0pt}{}{m=1}{\ne k}}(\mu_m \rho-\xi_i)}{\prod^M_{\genfrac{}{}{0pt}{}{p=1}{\ne k}}(\mu_k -\mu_p)} = \rho^{M-1} \frac{ \prod^M_{\genfrac{}{}{0pt}{}{p=1}{}}(\mu_p -\xi_j)}{\prod_{1 \le l < p \le M}(\mu_p - \mu_l)} \mathcal{J}_{ij},
\end{equation*}
where,
\begin{equation*}\begin{array}{lll}
\mathcal{J}_{ij} &=& {\displaystyle \sum^M_{k=1}} (-1)^{M-k} \prod^M_{\genfrac{}{}{0pt}{}{p=1}{\ne i,j}}(\mu_k-\xi_p)  \prod_{\genfrac{}{}{0pt}{}{1 \le l < m \le M}{\ne k}}(\mu_m - \mu_l)\\
&=& {\displaystyle\sum^{M-2}_{p=0}} (-1)^{p}  e_{p}(\{\xi\}, \hat{\xi}_i,\hat{\xi}_j)\textrm{det} \left[ \begin{array}{ccccc}
1 & \mu_1 & \dots & \mu^{M-2}_1 & \mu^{M-2-p}_1 \\
\vdots & \vdots & \ddots & \vdots & \vdots \\
1 & \mu_M & \dots & \mu^{M-2}_M & \mu^{M-2-p}_M
\end{array} \right]\\
&=&0,
\end{array}\end{equation*}
and $e_{p}(\{\xi\}, \hat{\xi}_i,\hat{\xi}_j)$ are \textit{elementary symmetric polynomials}\footnote{For more information on symmetric polynomials refer to chapter I of \cite{MacD}.} of order $p$ and variables\\
$\{\xi_1, \dots, \xi_{i-1},\xi_{i+1}, \dots, \xi_{j-1},\xi_{j+1} \dots, \xi_M \}$. Hence we have shown that the in limit $\xi_i \rightarrow \xi_j \rho$, the expression of the right hand side of Eq. (\ref{RHS}) obtains a residue value of zero, verifying that the aforementioned expression is an entire function.

Through inspection we can show that the expression at the right hand side of Eq. (\ref{RHS}) is bounded for any value of the variables by considering the order the polynomials on the numerator and denominator. Since it is \textit{entire} and \textit{bounded}, Liouville's theorem asserts that it is \textit{constant}. Substituting the values,
\begin{equation*}
 \xi_1 = \mu_1, \dots, \xi_i = \mu_i,\dots, \xi_M = \mu_M,
\end{equation*}
we find that the value of that constant is 1, hence verifying Eq. (\ref{polyiden}). $\square$
\subsection{Proof of proposition \ref{SECONDPOLLY}}\label{APPPOLLY2}
We begin by noticing that the left hand side of Eq. (\ref{gensui2}) is a rational function in $\mu_s$, $q \le s \le M$. Both left and right hand sides of Eq.  (\ref{gensui2}) have simple poles at $\mu_s \rightarrow \xi_{p_{q}}\frac{1}{\rho}$, and it is elementary to show that both sides have the following residue values,
\begin{equation*}
\frac{\rho^{L-M}(\rho-1)^2  }{\prod^M_{\genfrac{}{}{0pt}{}{j=q}{\ne s }} (\mu_j-\mu_s)} \frac{\mu_s}{\nu_i - \mu_s}
{\displaystyle \prod^L_{\genfrac{}{}{0pt}{}{k=1}{\ne p_{q} }}}\frac{\mu_s - \xi_k}{\mu_s \rho - \xi_k}\frac{\prod^M_{\genfrac{}{}{0pt}{}{m=1}{\ne i }} (\nu_m  -  \mu_s \rho)}{\prod^{q-1}_{n=1} (\xi_{p_{n}}  - \mu_s )}.
\end{equation*}
We observe however that the right hand side of Eq. (\ref{gensui2}) contains additional poles at the following values,
\begin{equation*}\begin{array}{llll}
\mu_s \rightarrow \xi_{p_l} & l \in \{1, \dots, q-1\}\\
\mu_s \rightarrow \mu_m & m \in \{q, \dots, s-1,s+1,\dots, M\}\\
\mu_s \rightarrow \nu_i 
\end{array}\end{equation*}
Verifying that the first two sets of residue values are zero is an elementary process. To show that the residue value in the limit $\mu_s \rightarrow \nu_i$ is zero additionally requires the application of the Bethe equations (\ref{EXPBETHE}).

Hence the left hand side and right hand side of Eq. (\ref{gensui2}) contain the same residue values at the same simple poles. Additionally, all the terms on both sides of Eq. (\ref{gensui2}) are rational functions in $\mu_s$ of order $1$ on the denominator - thus they are equivalent. $\square$

\section*{Acknowledgments}
The authors thank the Brazilian Research Agencies FAPESP and CNPq for financial support.

\end{document}